\documentclass{aa}
\usepackage{txfonts}
\usepackage{multirow}
\usepackage{epsfig}
\usepackage{natbib}
\usepackage{lscape}
\usepackage{wasysym}
\usepackage{array}
\usepackage{amssymb}
\usepackage{subfigure}
\usepackage{longtable}
\usepackage[colorlinks=true,citecolor=blue]{hyperref}
\usepackage{rotating}
\usepackage{color}
\usepackage{colortbl}
\usepackage{soul}
\usepackage[normalem]{ulem}
\usepackage{fancyheadings}
\bibpunct{(}{)}{;}{a}{}{,}

\newcommand\T{\rule{0pt}{2.6ex}}

\usepackage{graphicx}
\usepackage{txfonts}
\usepackage{dblfloatfix}
\begin{document}

   \title{Metallicity of M dwarfs }

  \subtitle{IV. A high-precision [Fe/H] and $T_{eff}$ technique from high-resolution optical spectra for M dwarfs \thanks{Based on observations made with the HARPS instrument on the ESO 3.6-m telescope at La Silla Observatory under programme ID 072.C-0488(E)}}

\author{ V. Neves\inst{1,2,3} \and X. Bonfils\inst{3,4} \and
  N. C. Santos\inst{2} \and X. Delfosse\inst{3,4} \and
  T. Forveille\inst{3,4}  \and F. Allard\inst{5}  \and
  S. Udry\inst{6}}
\institute{ Departamento de F\'{i}sica, Universidade Federal do Rio Grande do Norte,
59072-970 Natal, RN, Brazil \\
email: {\tt vasco.neves@dfte.ufrn.br}
\and
Centro de Astrof\'{\i}sica and Faculdade de Ci\^encias, Universidade do Porto, Rua das Estrelas, 4150-762 Porto, Portugal 
\and
Univ. Grenoble Alpes, IPAG, F-38000 Grenoble, France
\and
CNRS, IPAG, F-38000 Grenoble, France
\and
Centre de Recherche Astrophysique de Lyon, UMR 5574: CNRS,
Universit\'e de Lyon, \'Ecole Normale Sup\'erieure de Lyon, 46 All\'ee
d'Italie, F-69364 Lyon Cedex 07, France
\and
Observatoire de Gen\`eve, Universit\'e de Gen\`eve, 51 Chemin des
Maillettes, 1290 Sauverny, Switzerland
}

   \date{Received XXX; accepted XXX}

 
  \abstract
{}
   {In this work we develop a technique to obtain high precision determinations of both metallicity and effective temperature of M dwarfs in the optical.}
   {A new method is presented that makes use of the information of 4104 lines in the 530-690 nm spectral region. It consists in the measurement of pseudo equivalent widths and their correlation with established scales of [Fe/H] and $T_{eff}$.}
   {Our technique achieves a $rms$ of 0.08$\pm$0.01 for [Fe/H], 91$\pm$13 K for $T_{eff}$, and is valid in the (-0.85, 0.26 dex), (2800, 4100 K), and (M0.0, M5.0) intervals for [Fe/H], $T_{eff}$ and spectral type respectively. We also calculated the RMSE$_{V}$ which estimates uncertainties of the order of 0.12 dex for the metallicity and of 293 K for the effective temperature. The technique has an activity limit and should only be used for stars with  $\log{L_{H_{\alpha}}/L_{bol}} < -4.0$. Our method is available online at \url{http://www.astro.up.pt/resources/mcal}.  }
   {}


   \maketitle

\section{Introduction}

The precise derivation of M dwarf atmospheric parameters is still very challenging today. Cool and intrinsically faint, M dwarfs are not easy to study. As the M subtype increases more molecules form in their atmosphere, making the spectral continuum very hard or impossible to identify, at least in the visible region of the spectrum. Therefore, methods such as atomic line analysis that are dependent on the knowledge of the continuum are suited only for the metal poor and earliest types of M dwarfs \citep[e.g.][]{Woolf-2005,Woolf-2006}. On the other hand, spectral synthesis techniques do not reach yet a high precision comparable to FGK dwarf methods, due to the fundamental lack of knowledge of billions of molecular line strengths and transitions, and most studies have reached modest results \citep[e.g.][]{Valenti-1998,Bean-2006a}. Despite that, some important progress have been made using spectral synthesis fitting to high-resolution spectra in the infrared \citep{Onehag-2012}, where the depression of the continuum in some regions is less intense than in the visible region of the spectrum \citep[e.g.][]{Rajpurohit-2013b}. However, only a few stars have been measured this way, and the technique lacks external confirmation. An alternative method, based on high-resolution template spectra, to calculate metallicity, distance, stellar mass, and radius, was presented by \citep[][]{Pineda-2013}. This new technique, with similarities with our own method \citep[][]{Neves-2013}, is very promising but its [Fe/H] precision is still limited to 0.15 dex.

In this context, most parameter determinations, especially metallicity and effective temperature are instead based on calibrations using colours \citep[e.g.][]{Bonfils-2005,Johnson-2009,Schlaufman-2010,Johnson-2012,Neves-2012} or spectroscopic indices \citep[e.g.][]{Rojas-Ayala-2010,Rojas-Ayala-2012, Mann-2013a,Mann-2013b,Newton-2013}.

Regarding metallicity, some progress has been made in the last few years. A steady improvement was achieved, bringing the typical uncertainties of $\pm$ 0.20 dex of the photometric calibrations, below $\sim 0.10$ dex in the most recent low-resolution spectroscopic scales in the infrared \cite[e.g.][]{Rojas-Ayala-2012,Mann-2013a,Newton-2013}, following the pioneering work of \citet{Rojas-Ayala-2010}. However a true high-precision determination with a rms of the order of 0.05 dex, on par with the ones obtained for FGK dwarfs \citep[e.g.][]{Santos-2004b,Sousa-2007} has not yet been reached (see \citet{Neves-2012} introduction).

For temperature, on the other hand, important uncertainties and systematics still persist today. Although internal precisions are reported to be lower than 100K \citep[e.g.][]{Casagrande-2008,Rojas-Ayala-2012,Boyajian-2012}, their suffer from systematics ranging from 150 to 300 K making the determination of accurate temperature for M dwarfs a priority. In this context \citet{Boyajian-2012} presented several calibrations of $T_{eff}$, based on the largest sample to date of high precision interferometric measurements of K and M dwarf radii and bolometric fluxes, that in principle allow a very precise measurement of the effective temperature (a technique pioneered by \citet{Segransan-2003} for M dwarfs). However, some doubts still arise regarding the accuracy of the determination of the total flux of the stars, based on templates from \citet{Pickles-1998}, as recently pointed out by \citet[][]{Mann-2013b}. They have, in turn, also recently presented their own effective temperature method that is very similar to the one of \citet{Boyajian-2012} but rely on a combination of their low resolution spectra with BT-SETTL synthetic spectra from \citet{Allard-2011,Allard-2013} to calculate the bolometric flux. From these high-precision effective temperatures they established four visual and infrared spectroscopic indices, with precisions (but not accuracies) between 62 and 100 K. A similar effort regarding $T_{eff}$ determination of M dwarfs using synthetic spectra came from \citet{Rajpurohit-2013a}, where they compare synthetic spectra from the latest BT-SETTL models \citep{Allard-2012} to low-resolution optical spectra. They obtain a better agreement between synthetic and observed spectra when compared with previous models, estimating uncertainties of $T_{eff}$ of the order of 100 K. 




In this work we present a new method to try to overcome the aforementioned hurdles and improve on the precision of both metallicity and effective temperature of M dwarfs. An early version of this new technique was briefly presented in the Appendix of \citet{Neves-2013} and used to investigate the planet-metallicity relation of the HARPS GTO M dwarf sample \citep{Bonfils-2013}. In Sect. \ref{sec:calib} we describe in detail our method, as well as the sample selection, uncertainty estimation and a test of the technique as a function of resolution and signal-to-noise ratio (S/N). Afterwards, in Sect. \ref{sec:comp}, we compare our results with other determinations from the literature. Finally, in Sect \ref{sec:discussion} we discuss our results. The instructions to use our method are described in the Appendix.

\section{The method}
\label{sec:calib}

Our new method is based on the measurement of pseudo equivalent widths (EWs) of most lines/features in the 530-690 nm region of the spectra from a 102 star sample from the HARPS GTO M dwarf program, described in detail in Sect. 2 of \citet{Bonfils-2013}. It is a volume-limited sample (11 pc), and contains stars with  $\delta< +20^{\circ}$, $V < 14$ mag and includes only targets with $vsini\le 6.5$ km/s. Spectroscopic binaries as well as visual pairs with separations lower than 5$\arcsec$ were removed \textit{a priori}.

The features are defined as regions of the spectra that are formed by more than one line. The EWs are then correlated with the reference photometric [Fe/H] and $T_{eff}$ scales from \citet{Neves-2012} and \citet{Casagrande-2008} respectively. The two scales are in turn based on [Fe/H] determinations from FGK primaries with a M dwarf secondary and on an adaptation of the IRFM technique \citep{Blackwell-1977} respectively.

This method achieves an increase in precision of both parameters whereas its accuracy is tied to the original calibrations. The methodology is detailed in Sect. \ref{sec:method}. 

The reference [Fe/H] was calculated using stellar parallaxes, $V$, and $K_{S}$ magnitudes following the procedure described in \citet{Neves-2012}. The reference $T_{eff}$ is the average value of the $V-J$, $V-H$, and $V-K$ photometric scales taken from \citet{Casagrande-2008}. Table \ref{table:params} lists the quantities used to calculate these parameters. Column 1 shows the star name, column 2 and 3 the right ascension and declination respectively, column 4 the parallax of each star and its associated error, and column 5 the source of the parallax measurement. Column 6 depicts the stellar type of the star taken from Simbad\footnote{\url{http://simbad.u-strasbg.fr/}} \citep{Wenger-2000}, except in the case of Gl438, where it was obtained from \citet{Hawley-1997}. The photometric stellar type presented in Column 7 was calculated with the color relation of \citet{Lepine-2013}, and columns 8 to 11 display the $V$, $J$, $H$, and $K_{S}$ photometry. Lastly, column 12 details the source of the photometry. 


\begin{table*}[h!]
\caption{List containing our sample and the quantities used to calculate the reference [Fe/H] and T$_{eff}$ . Sorted by right ascension.}
\label{table:params}

\scalebox{0.93}{
\begin{tabular}{ l r r r c r r r r r r r r r}
\hline
\hline
star & $\alpha(2000)$ & $\delta(2000)$ & $\pi$  & $\pi$  & Stype & Stype & V   & J   & H   & K  & V/J/H/K \\
     &    &     &[mas]&  source     & (S)  & Phot. &[mag]&[mag]&[mag]&[mag] & source \\
\hline
Gl1 & 00:05:25 & -37:21:23 & $230.4 \pm  0.9$ & H & M1.5 & M1.5 &  $8.56 \pm 0.02$ & $5.34 \pm 0.02$ & $4.73 \pm 0.02$ & $4.54 \pm 0.02$ & 1/1/1/1 \\
Gl54.1 & 01:12:31 & -17:00:00 & $271.0 \pm  8.4$ & H & M4Ve & M4.0 &  $12.07 \pm 0.02$ & $7.26 \pm 0.02$ & $6.75 \pm 0.03$ & $6.42 \pm 0.02$ & 1/8/8/8 \\
Gl87 & 02:12:21 & +03:34:30 & $96.0 \pm  1.7$ & H & M2.5V & M1.5 &  $10.04 \pm 0.02$ & $6.83 \pm 0.02$ & $6.32 \pm 0.03$ & $6.08 \pm 0.02$ & 1/8/8/8 \\
Gl105B & 02:36:16 & +06:52:12 & $139.3 \pm  0.5$ & H & M4.5V & M3.5 &  $11.66 \pm 0.02$ & $7.33 \pm 0.02$ & $6.79 \pm 0.04$ & $6.57 \pm 0.02$ & 4/8/8/8 \\
HIP12961 & 02:46:43 & -23:05:12 & $43.5 \pm  1.7$ & H & M0  & - &  $10.24 \pm 0.02$ & $7.56 \pm 0.02$ & $6.93 \pm 0.03$ & $6.74 \pm 0.02$ & 1/8/8/8 \\
LP771-95A & 03:01:51 & -16:35:36 & $146.4 \pm  2.9$ & H06 & M3.5 & M2.0 &  $10.59 \pm 0.05$ & $7.11 \pm 0.02$ & $6.56 \pm 0.02$ & $6.29 \pm 0.02$ & 2/8/8/8 \\
GJ163 & 04:09:16 & -53:22:25 & $66.7 \pm  1.8$ & H & M3.5 & M3.0 &  $11.81 \pm 0.02$ & $7.95 \pm 0.03$ & $7.43 \pm 0.04$ & $7.13 \pm 0.02$ & 1/8/8/8 \\
Gl176 & 04:42:56 & +18:57:29 & $106.2 \pm  2.5$ & H & M2 & M2.0 &  $9.95 \pm 0.02$ & $6.46 \pm 0.02$ & $5.82 \pm 0.03$ & $5.61 \pm 0.03$ & 1/8/8/8 \\
GJ179 & 04:52:06 & +06:28:36 & $81.4 \pm  4.0$ & H & M3.5 & M3.5 &  $12.02 \pm 0.04$ & $7.81 \pm 0.02$ & $7.21 \pm 0.05$ & $6.94 \pm 0.02$ & 1/8/8/8 \\
Gl191 & 05:11:40 & -45:01:06 & $255.3 \pm  0.9$ & H & sdM1.0 & M0.5 &  $8.85 \pm 0.02$ & $5.82 \pm 0.03$ & $5.32 \pm 0.03$ & $5.05 \pm 0.02$ & 1/8/8/8 \\
Gl205 & 05:31:27 & -03:40:42 & $176.8 \pm  1.2$ & H & M1.5V & M1.5 &  $7.97 \pm 0.02$ & $4.75 \pm 0.05$ & $4.07 \pm 0.05$ & $3.85 \pm 0.03$ & 1/9/9/9 \\
Gl213 & 05:42:09 & +12:29:23 & $171.6 \pm  4.0$ & H & M4.0V & M4.0 &  $11.56 \pm 0.01$ & $7.12 \pm 0.02$ & $6.63 \pm 0.02$ & $6.39 \pm 0.02$ & 3/8/8/8 \\
Gl229 & 06:10:34 & -21:51:53 & $173.8 \pm  1.0$ & H & M1/M2V & M1.0 &  $8.12 \pm 0.02$ & $5.06 \pm 0.02$ & $4.36 \pm 0.02$ & $4.16 \pm 0.02$ & 1/1/1/1 \\
HIP31293 & 06:33:43 & -75:37:47 & $110.9 \pm  2.2$ & H & M2V & M2.5 &  $10.35 \pm 0.01$ & $6.72 \pm 0.02$ & $6.15 \pm 0.03$ & $5.86 \pm 0.02$ & 3/8/8/8 \\
HIP31292 & 06:33:47 & -75:37:30 & $114.5 \pm  3.2$ & H & M3V & M3.0 &  $11.41 \pm 0.01$ & $7.41 \pm 0.03$ & $6.85 \pm 0.03$ & $6.56 \pm 0.02$ & 3/8/8/8 \\
Gl250B & 06:52:18 & -05:11:24 & $114.8 \pm  0.4$ & H & M2 & M2.0 &  $10.08 \pm 0.01$ & $6.58 \pm 0.03$ & $5.98 \pm 0.06$ & $5.72 \pm 0.04$ & 5/8/8/8 \\
Gl273 & 07:27:24 & +05:13:30 & $263.0 \pm  1.4$ & H & M3.5V & M3.5 &  $9.87 \pm 0.02$ & $5.71 \pm 0.03$ & $5.22 \pm 0.06$ & $4.86 \pm 0.02$ & 1/8/8/8 \\
Gl300 & 08:12:41 & -21:33:12 & $125.8 \pm  1.0$ & H & M4 & M4.0 &  $12.13 \pm 0.01$ & $7.60 \pm 0.02$ & $6.96 \pm 0.03$ & $6.71 \pm 0.03$ & 2/8/8/8 \\
GJ2066 & 08:16:08 & +01:18:11 & $109.6 \pm  1.5$ & H & M2.0V & M2.0 &  $10.09 \pm 0.02$ & $6.62 \pm 0.03$ & $6.04 \pm 0.03$ & $5.77 \pm 0.02$ & 1/8/8/8 \\
GJ317 & 08:40:59 & -23:27:23 & $65.3 \pm  0.4$ & A12 & M3.5 & M3.0 &  $11.97 \pm 0.04$ & $7.93 \pm 0.03$ & $7.32 \pm 0.07$ & $7.03 \pm 0.02$ & 2/8/8/8 \\
Gl341 & 09:21:38 & -60:16:53 & $95.6 \pm  0.9$ & H & M0.0 & M0.5 &  $9.46 \pm 0.02$ & $6.44 \pm 0.02$ & $5.79 \pm 0.03$ & $5.59 \pm 0.02$ & 1/8/8/8 \\
GJ1125 & 09:30:44 & +00:19:18 & $103.5 \pm  3.9$ & H & M3.5 & M3.0 &  $11.71 \pm 0.02$ & $7.70 \pm 0.02$ & $7.18 \pm 0.03$ & $6.87 \pm 0.02$ & 1/8/8/8 \\
Gl357 & 09:36:02 & -21:39:42 & $110.8 \pm  1.9$ & H & M2.5V & M2.5 &  $10.91 \pm 0.02$ & $7.34 \pm 0.03$ & $6.74 \pm 0.03$ & $6.47 \pm 0.02$ & 1/8/8/8 \\
Gl358 & 09:39:47 & -41:04:00 & $105.6 \pm  1.6$ & H & M3 & M3.0 &  $10.69 \pm 0.02$ & $6.90 \pm 0.03$ & $6.32 \pm 0.05$ & $6.06 \pm 0.02$ & 1/8/8/8 \\
Gl367 & 09:44:30 & -45:46:36 & $101.3 \pm  3.2$ & H & M1.0 & M1.5 &  $9.98 \pm 0.02$ & $6.63 \pm 0.02$ & $6.04 \pm 0.04$ & $5.78 \pm 0.02$ & 1/8/8/8 \\
Gl382 & 10:12:17 & -03:44:47 & $127.1 \pm  1.9$ & H & M2.0V & M2.0 &  $9.26 \pm 0.02$ & $5.89 \pm 0.02$ & $5.26 \pm 0.02$ & $5.01 \pm 0.02$ & 1/8/8/8 \\
Gl393 & 10:28:55 & +00:50:23 & $141.5 \pm  2.2$ & H & M2.5V & M2.0 &  $9.63 \pm 0.02$ & $6.18 \pm 0.02$ & $5.61 \pm 0.03$ & $5.31 \pm 0.02$ & 1/8/8/8 \\
GJ3634 & 10:58:35 & -31:08:38 & $50.5 \pm  1.6$ & R10 & M2.5 & M2.5 &  $11.93 \pm 0.02$ & $8.36 \pm 0.02$ & $7.76 \pm 0.05$ & $7.47 \pm 0.03$ & 2/8/8/8 \\
Gl413.1 & 11:09:31 & -24:36:00 & $93.0 \pm  1.7$ & H & M2 & M2.0 &  $10.45 \pm 0.02$ & $6.95 \pm 0.02$ & $6.36 \pm 0.04$ & $6.10 \pm 0.02$ & 1/8/8/8 \\
Gl433 & 11:35:27 & -32:32:23 & $112.6 \pm  1.4$ & H & M1.5 & M1.5 &  $9.81 \pm 0.02$ & $6.47 \pm 0.02$ & $5.86 \pm 0.04$ & $5.62 \pm 0.02$ & 1/8/8/8 \\
Gl436 & 11:42:11 & +26:42:23 & $98.6 \pm  2.3$ & H & M3.5V & M2.5 &  $10.61 \pm 0.01$ & $6.90 \pm 0.02$ & $6.32 \pm 0.02$ & $6.07 \pm 0.02$ & 2/8/8/8 \\
Gl438 & 11:43:20 & -51:50:23 & $91.7 \pm  2.0$ & R10 & M0.0$^\dagger$ & M1.5 &  $10.36 \pm 0.04$ & $7.14 \pm 0.02$ & $6.58 \pm 0.04$ & $6.32 \pm 0.02$ & 2/8/8/8 \\
Gl447 & 11:47:44 & +00:48:16 & $299.6 \pm  2.2$ & H & M4.5V & M4.0 &  $11.12 \pm 0.01$ & $6.50 \pm 0.02$ & $5.95 \pm 0.02$ & $5.65 \pm 0.02$ & 3/8/8/8 \\
Gl465 & 12:24:53 & -18:14:30 & $113.0 \pm  2.5$ & H & M2 & M2.0 &  $11.27 \pm 0.02$ & $7.73 \pm 0.02$ & $7.25 \pm 0.02$ & $6.95 \pm 0.02$ & 1/8/8/8 \\
Gl479 & 12:37:53 & -52:00:06 & $103.2 \pm  2.3$ & H & M3V(e) & M3.0 &  $10.66 \pm 0.02$ & $6.86 \pm 0.02$ & $6.29 \pm 0.03$ & $6.02 \pm 0.02$ & 1/8/8/8 \\
Gl514 & 13:30:00 & +10:22:36 & $130.6 \pm  1.1$ & H & M1.0V & M1.0 &  $9.03 \pm 0.02$ & $5.90 \pm 0.02$ & $5.30 \pm 0.03$ & $5.04 \pm 0.03$ & 1/8/8/8 \\
Gl526 & 13:45:44 & +14:53:30 & $185.5 \pm  1.1$ & H & M4.0V & M1.0 &  $8.43 \pm 0.02$ & $5.24 \pm 0.05$ & $4.65 \pm 0.05$ & $4.42 \pm 0.02$ & 1/9/9/8 \\
Gl536 & 14:01:03 & -02:39:18 & $98.3 \pm  1.6$ & H & M1.5V & M1.0 &  $9.71 \pm 0.02$ & $6.52 \pm 0.02$ & $5.93 \pm 0.04$ & $5.68 \pm 0.02$ & 1/8/8/8 \\
Gl555 & 14:34:17 & -12:31:06 & $165.0 \pm  3.3$ & H & M4.0V & M4.0 &  $11.32 \pm 0.02$ & $6.84 \pm 0.02$ & $6.26 \pm 0.04$ & $5.94 \pm 0.03$ & 1/8/8/8 \\
Gl569A & 14:54:29 & +16:06:04 & $101.9 \pm  1.7$ & H & M2.5V & M3.0 &  $10.41 \pm 0.05$ & $6.63 \pm 0.02$ & $5.99 \pm 0.02$ & $5.77 \pm 0.02$ & 6/8/8/8 \\
Gl581 & 15:19:26 & -07:43:17 & $160.9 \pm  2.6$ & H & M5.0V & M3.0 &  $10.57 \pm 0.01$ & $6.71 \pm 0.03$ & $6.09 \pm 0.03$ & $5.84 \pm 0.02$ & 3/8/8/8 \\
Gl588 & 15:32:13 & -41:16:36 & $168.7 \pm  1.3$ & H & M2.5V & M2.5 &  $9.31 \pm 0.02$ & $5.65 \pm 0.02$ & $5.03 \pm 0.02$ & $4.76 \pm 0.02$ & 1/8/8/8 \\
Gl618A & 16:20:04 & -37:31:41 & $119.8 \pm  2.5$ & H & M3 & M3.0 &  $10.59 \pm 0.02$ & $6.79 \pm 0.02$ & $6.22 \pm 0.02$ & $5.95 \pm 0.02$ & 1/8/8/8 \\
Gl628 & 16:30:18 & -12:39:47 & $233.0 \pm  1.6$ & H & M3V & M3.5 &  $10.07 \pm 0.02$ & $5.95 \pm 0.02$ & $5.37 \pm 0.04$ & $5.08 \pm 0.02$ & 1/8/8/8 \\
GJ1214 & 17:15:19 & +04:57:50 & $68.7 \pm  0.6$ & A13 & M4.5 & M4.0 &  $14.64 \pm 0.03$ & $9.75 \pm 0.02$ & $9.09 \pm 0.02$ & $8.78 \pm 0.02$ & 7/8/8/8 \\
Gl667C & 17:18:58 & -34:59:42 & $146.3 \pm  9.0$ & H & M1.5V & M2.0 &  $10.27 \pm 0.04$ & $6.85 \pm 0.02$ & $6.32 \pm 0.04$ & $6.04 \pm 0.02$ & 2/8/8/8 \\
Gl674 & 17:28:40 & -46:53:42 & $220.2 \pm  1.4$ & H & M3V & M2.5 &  $9.41 \pm 0.02$ & $5.71 \pm 0.02$ & $5.15 \pm 0.03$ & $4.86 \pm 0.02$ & 1/8/8/8 \\
GJ676A & 17:30:11 & -51:38:13 & $60.8 \pm  1.6$ & H & M0V & M0.0 &  $9.59 \pm 0.02$ & $6.71 \pm 0.02$ & $6.08 \pm 0.02$ & $5.83 \pm 0.03$ & 1/8/8/8 \\
Gl678.1A & 17:30:22 & +05:32:53 & $100.2 \pm  1.1$ & H & M1V & M1.0 &  $9.33 \pm 0.01$ & $6.24 \pm 0.02$ & $5.65 \pm 0.04$ & $5.42 \pm 0.03$ & 3/8/8/8 \\
Gl680 & 17:35:13 & -48:40:53 & $102.8 \pm  2.8$ & H & M3V & M2.0 &  $10.13 \pm 0.02$ & $6.67 \pm 0.02$ & $6.08 \pm 0.03$ & $5.83 \pm 0.02$ & 1/8/8/8 \\
Gl682 & 17:37:03 & -44:19:11 & $196.9 \pm  2.1$ & H & M3.5 & M3.5 &  $10.95 \pm 0.02$ & $6.54 \pm 0.02$ & $5.92 \pm 0.04$ & $5.61 \pm 0.02$ & 1/8/8/8 \\
Gl686 & 17:37:53 & +18:35:30 & $123.0 \pm  1.6$ & H & M1.0V & M1.5 &  $9.58 \pm 0.02$ & $6.36 \pm 0.02$ & $5.79 \pm 0.02$ & $5.57 \pm 0.02$ & 1/8/8/8 \\
Gl693 & 17:46:35 & -57:19:11 & $171.5 \pm  2.3$ & H & M2.0 & M3.0 &  $10.78 \pm 0.02$ & $6.86 \pm 0.02$ & $6.30 \pm 0.04$ & $6.02 \pm 0.02$ & 1/8/8/8 \\
Gl699 & 17:57:49 & +04:41:36 & $549.0 \pm  1.6$ & H & M4.0V & M3.5 &  $9.51 \pm 0.02$ & $5.24 \pm 0.02$ & $4.83 \pm 0.03$ & $4.52 \pm 0.02$ & 1/8/8/8 \\
Gl701 & 18:05:07 & -03:01:53 & $128.9 \pm  1.4$ & H & M2.0V & M1.0 &  $9.36 \pm 0.02$ & $6.16 \pm 0.02$ & $5.57 \pm 0.04$ & $5.31 \pm 0.02$ & 1/8/8/8 \\
Gl752A & 19:16:55 & +05:10:05 & $170.4 \pm  1.0$ & H & M3V B & M2.0 &  $9.12 \pm 0.02$ & $5.58 \pm 0.03$ & $4.93 \pm 0.03$ & $4.67 \pm 0.02$ & 1/8/8/8 \\
Gl832 & 21:33:34 & -49:00:36 & $201.9 \pm  1.0$ & H & M1.5 & M1.5 &  $8.67 \pm 0.02$ & $5.36 \pm 0.02$ & $4.69 \pm 0.02$ & $4.47 \pm 0.02$ & 1/1/1/1 \\
Gl846 & 22:02:10 & +01:24:00 & $97.6 \pm  1.5$ & H & M0 & M0.5 &  $9.15 \pm 0.02$ & $6.20 \pm 0.02$ & $5.56 \pm 0.05$ & $5.32 \pm 0.02$ & 1/8/8/8 \\
Gl849 & 22:09:40 & -04:38:30 & $109.9 \pm  2.1$ & H & M3.5V & M3.0 &  $10.37 \pm 0.02$ & $6.51 \pm 0.02$ & $5.90 \pm 0.04$ & $5.59 \pm 0.02$ & 1/8/8/8 \\
Gl876 & 22:53:17 & -14:15:48 & $213.3 \pm  2.1$ & H & M5.0V & M3.5 &  $10.18 \pm 0.02$ & $5.93 \pm 0.02$ & $5.35 \pm 0.05$ & $5.01 \pm 0.02$ & 1/8/8/8 \\
Gl877 & 22:55:46 & -75:27:36 & $116.1 \pm  1.2$ & H & M3V & M2.5 &  $10.38 \pm 0.02$ & $6.62 \pm 0.02$ & $6.08 \pm 0.03$ & $5.81 \pm 0.02$ & 1/8/8/8 \\
Gl880 & 22:56:35 & +16:33:12 & $146.1 \pm  1.0$ & H & M2.0V & M1.5 &  $8.64 \pm 0.02$ & $5.36 \pm 0.02$ & $4.75 \pm 0.05$ & $4.52 \pm 0.02$ & 1/8/9/8 \\
Gl887 & 23:05:52 & -35:51:12 & $303.9 \pm  0.9$ & H & M2V & M1.0 &  $7.35 \pm 0.01$ & $4.17 \pm 0.05$ & $3.61 \pm 0.05$ & $3.36 \pm 0.03$ & 3/9/9/9 \\
Gl908 & 23:49:13 & +02:24:06 & $167.3 \pm  1.2$ & H & M2V & M1.0 &  $8.98 \pm 0.01$ & $5.83 \pm 0.02$ & $5.28 \pm 0.03$ & $5.04 \pm 0.02$ & 3/8/8/8 \\
LTT9759 & 23:53:50 & -75:37:53 & $100.1 \pm  1.1$ & H & Ma & M2.5 &  $10.02 \pm 0.02$ & $6.45 \pm 0.02$ & $5.78 \pm 0.02$ & $5.55 \pm 0.03$ & 1/8/8/8 \\
\\
\hline
\hline
\end{tabular}
}
{
\\
\raggedright{
references: H -- \citep{van-Leeuwen-2007}; H06 -- \citet{Henry-2006}; A12 -- \citet{Anglada-Escude-2012}; R10 -- \citet{Riedel-2010}; A13 -- \citet{Anglada-Escude-2013}; 1-- \citet{Koen-2010}; 2 -- \citet{Henden-2009,Henden-2012}; 3 -- \citet{Perryman-1997}; 4 -- \citet{Weis-1993}; 5 -- \citet{Laing-1989}; 6 -- \citet{Fabricius-2002}; 7 -- \citet{Dawson-1992}; 8 -- \citet{Skrutskie-2006}; 9 -- \citet{Leggett-1992}; S -- Simbad; $\dagger$ -- \citet{Hawley-1997} \\
}
}
\end{table*}




From the 110 stars of our sample we first selected 69 stars with spectra having a signal to noise higher than 100. The final spectrum of each star was constructed from median normalized individual observations. The S/N of the individual spectra were added in quadrature. Our final sample is determined by an activity cut, as detailed in Sect. \ref{sec:sample}.

\subsection{Method}
\label{sec:method}

\begin{figure}[h]
\begin{center}
\includegraphics[scale=0.45]{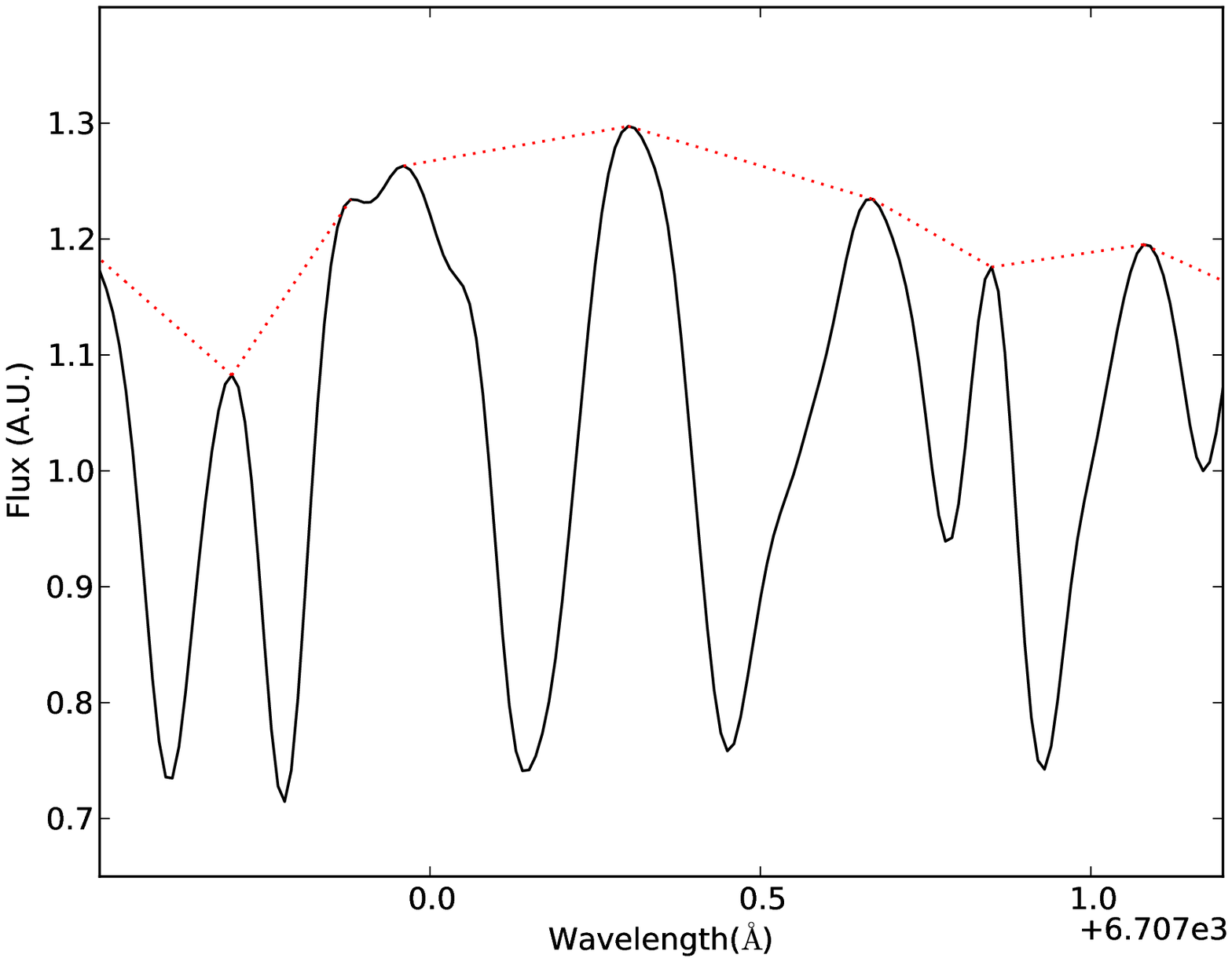}
\end{center}
\caption{Small region of the Gl 205 spectra illustrating pseudo equivalent width line measurement. The red dotted line represents the `peak-to-peak' flux.}
\label{fig:spec}
\end{figure}

From our final sample we measured pseudo EWs of lines and features (blended lines) from the spectra in the region between 530 and 690 nm, but excluded the features from the regions between 588-590.5, 656.1-656.4, and 686-690 nm due to the location of the activity sensitive Na doublet and H$\alpha$ lines, and the heavy presence of telluric lines respectively. We define the pseudo equivalent widths as 

\begin{equation}
W = \sum{\frac{F_{pp}-F_{\lambda}}{F_{pp}}\Delta\lambda},
\label{ew}
\end{equation}
where $F_{pp}$ is the value of the flux between the peaks of the line/feature at each integration step and $F_{\lambda}$ the flux of the line/feature. The measurements of the EWs are illustrated in Fig. \ref{fig:spec}, where the `peak-to-peak' flux corresponds to the red dotted lines and the flux of the star is shown as a black line. 

The very high S/N spectrum of the star Gl205 was used as a reference to establish the line/feature regions that were going to be measured in all spectra. We rejected all lines/features with a EW lower than 8 $m\AA$ to ensure that all lines in stars with lower [Fe/H] or/and $T_{eff}$ can be properly measured.  Lines with steep slopes are usually joined with adjacent lines, and measured as one feature. At the end of the line selection we obtained 4104 lines/features. An automatic search of the maximum values of $\pm0.02 \AA$ at the extremes of each line/feature is made to make sure that the `peak-to-peak' regions of all lines/features the spectra are effectively covered.

\begin{figure}[h]
\begin{center}
\includegraphics[scale=0.45]{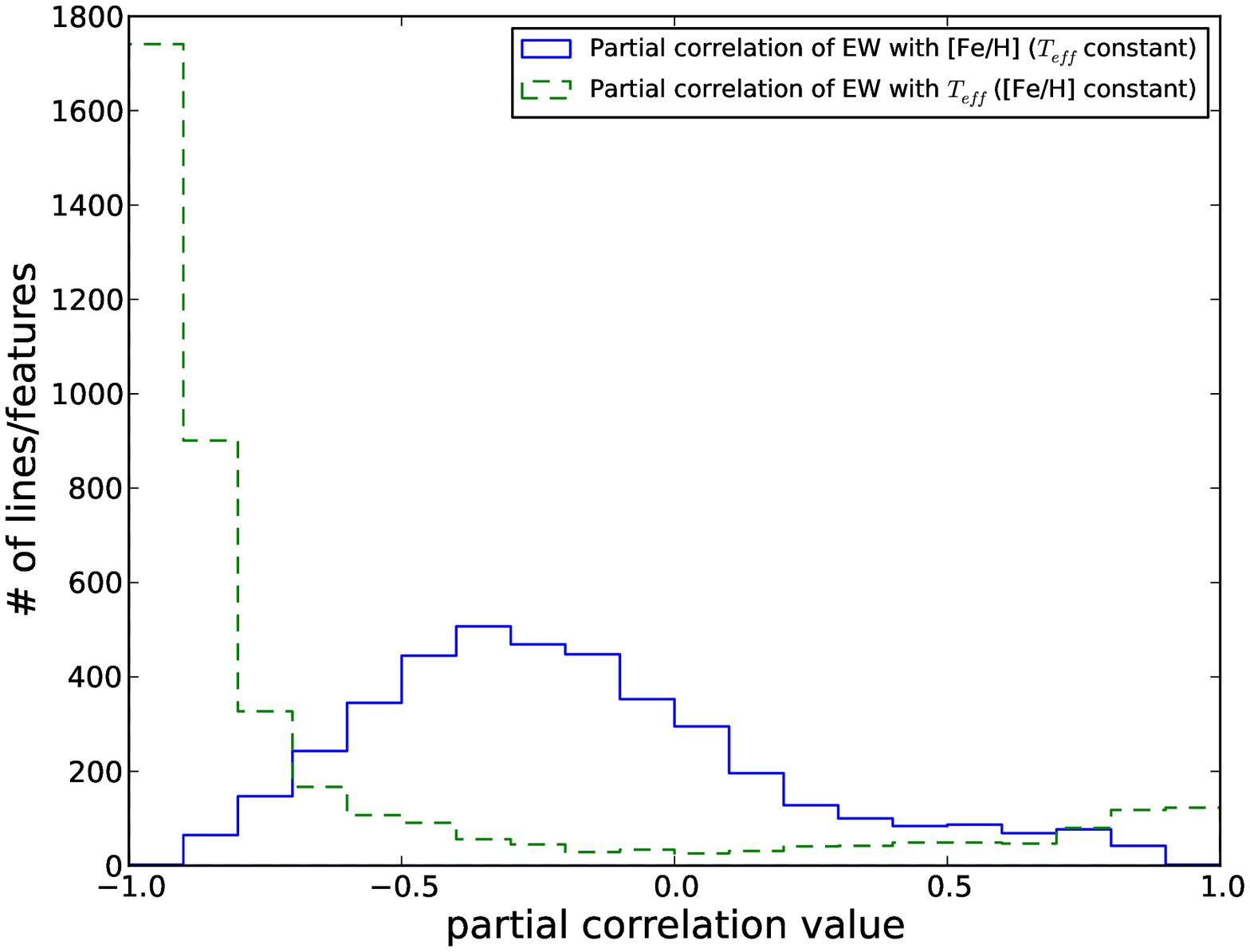}
\end{center}
\caption{Histograms of the partial correlations of [Fe/H] (solid blue histogram) and $T_{eff}$ (dashed green histogram).}
\label{fig:pcorr}
\end{figure}

The next step consisted in the investigation of the correlation between the measured EWs and the reference values for [Fe/H] and $T_{eff}$. Fig. \ref{fig:pcorr} shows the histograms of the partial correlation coefficient values of the EWs with the value of the metallicity and effective temperature (solid blue and dashed green lines respectively). The partial correlation coefficient is defined as the correlation coefficient of one parameter keeping the other fixed. 
We observe, in Fig. \ref{fig:pcorr} that a significant amount of lines have good correlation values with the parameters.

Then we did a least squares linear fit of the EWs with the metallicity and effective temperature. The reference values were calculated with the calibration of \citet{Neves-2012}, for [Fe/H], and with the three ($V-J$, $V-H$, and $V-K_{S}$) photometric calibrations of \citet{Casagrande-2008}, for $T_{eff}$, where we took the average value. From each line/feature $i$ of every star $m$ we calculate a EW value. Then we have

\begin{equation}
W_{i,m} = \alpha_{i}[Fe/H]_{m}^{T} + \beta_{i}T_{eff,m}^{T} + \gamma_{i}, 
\label{eq:fit}
\end{equation}

where $W$ is the matrix containing the EWs, and both $[Fe/H]^{T}$, and $T_{eff}^{T}$ are the transpose vectors of the parameter values. The $\alpha$ and the $\beta$ are the coefficients related to metallicity and effective temperature, respectively, while $\gamma$ is an independent coefficient. The error associated to each parameter $p$ is calculated as 
\begin{equation}
\label{eq:rss}
\epsilon_{p} = \sqrt{RSS.J},
\end{equation}
where RSS is the residual sum of squares, expressed as

\begin{equation}
RSS = \frac{\sum{(x_{i,model}-x_{i})^{2}}}{n_{obs}-n_{coef}},
\end{equation}
and $J$ is the diagonal of the estimate of the jacobian matrix around the solution. 
The $x_{i,model}$, $x_{i}$, $n_{obs}$, and $n_{coef}$ from Eq. \ref{eq:rss} are, respectively, the predicted value of the data, $x_{i}$, by the regression model, the data values, the number of data points, and the number of coefficients. We assume that both metallicity and effective temperature are independent and do not correlate with each other. This assumption was tested by perturbing each parameter in turn by introducing an positive or negative offset and then calculating both parameters. There was no difference in the obtained values of the unperturbed parameter. We also tried to use the full covariance matrix to calculate the uncertainties but in the end we got a worse result for the dispersion. Therefore, we decided to use only the diagonal values of the covariance matrix. The total error of the coefficients associated to each line $i$ can then be written as

\begin{equation} 
\epsilon_{i} = \sqrt{\epsilon_{\alpha}^{2}+\epsilon_{\beta}^{2}+\epsilon_{\gamma}^{2}}.
\end{equation}

The aim of our technique is to increase the precision of both [Fe/H] and $T_{eff}$ determinations. To do that we need to obtain the values of the metallicity and effective temperature via a weighted least squares refit, that is obtained after a left multiplication of $(C^{T}C)^{-1}C^{T}$ on both terms of Eq. \ref{eq:fit}, where $C$ is the calibration matrix or the coefficient matrix, that can be written as
\begin{equation}
\label{eq:matrix}
C = \left[\begin{array}{ccc} \alpha_{1,1} & \beta_{1,2} & \gamma_{1,3} \\ \alpha_{2,1} & \beta_{2,2} & \gamma_{2,3} \\... & ... & ...\\ \alpha_{I,1} & \beta_{I,2} & \gamma_{I,3} \end{array}\right],
\end{equation}
and $C^{T}$ is the transpose of $C$. The refit is then expressed, for each star $m$, as

\begin{equation}
\label{eq:refit}
\left[\begin{array}{c} [Fe/H] \\T_{eff} \\ \Gamma \end{array}\right] =  (C^{T}C)^{-1}C^{T}W,
\end{equation}
where $\Gamma$ is the parameter related to the independent  $\gamma$ coefficients.

In order to correct the offset of our method we added an extra parameter, while adding a corresponding dimension in Eq. \ref{eq:matrix}, that corresponds to an array of ones. The updated matrix $C$ has now dimension $I \times 4$ instead of $I \times 3$, where $I$ has the value of the number of lines.


Finally we introduce a weight to Eq. \ref{eq:refit}, using a \textit{Levenberg-Marquardt} \citep[][]{Press-1992} algorithm. We can write the elements of the normalised weight $E$ as



\begin{equation}
E_{i} = \frac{1/\epsilon_{i}^{2}}{\sum{1/\epsilon_{i}^{2}}}.
\label{eq:weight}
\end{equation}

Other methods were tested, such as choosing lines/features with the best correlations or partial correlations with the parameters. However, the weighted least squares approach performed best at minimising the uncertainties of both metallicity and temperature. 

\subsection{\textit{A posteriori} sample selection}
\label{sec:sample}

\begin{figure}[]
\centering
\includegraphics[scale=0.45]{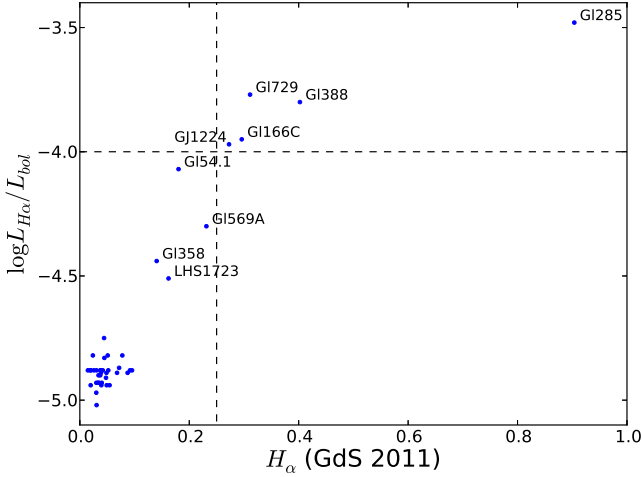}
\caption{Normalized H$\alpha$ luminosity, taken from \citet{Reiners-2012} versus the H$_{\alpha}$ index of \citet{Gomes_da_Silva-2011} for the stars in common with our sample.The black dashed lines depict the limits that we have established for the sample selection.}
\label{fig:activity}
\end{figure}

At this stage we observed that some stars appeared as outliers in the plots of the pseudo EWs versus the reference [Fe/H] and $T_{eff}$ for many lines. We suspected that this behaviour was due to activity or rotation and did a \textit{a posteriori} study of the impact of the activity with our technique. To this end, we used the normalized H$\alpha$ luminosity, $\log{L_{H_{\alpha}}/L_{bol}}$, from \citet{Reiners-2012}, for the stars in common with our full sample, as well as the median of individual measurements of the $H_{\alpha}$ index defined by \citet{Gomes_da_Silva-2011}, kindly provided by the author. Table \ref{table:full} lists both activity indicators, in columns 5 and 6, for the stars in common with our sample. Fig. \ref{fig:activity} depicts the relation between both indices, where we observe the inactive stars, in the bottom left corner of the diagram, a linear trend between the indices for increasingly active stars, and a very active star, Gl285, in the top right corner of diagram, where the  $\log{L_{H_{\alpha}}/L_{bol}}$ indicator seems to have saturated. The dashed black lines show the limits above which the stars were excluded from the final sample, as described in the following paragraph. 

\begin{figure*}[]
\begin{center}
\subfigure[]{\includegraphics[scale=0.44]{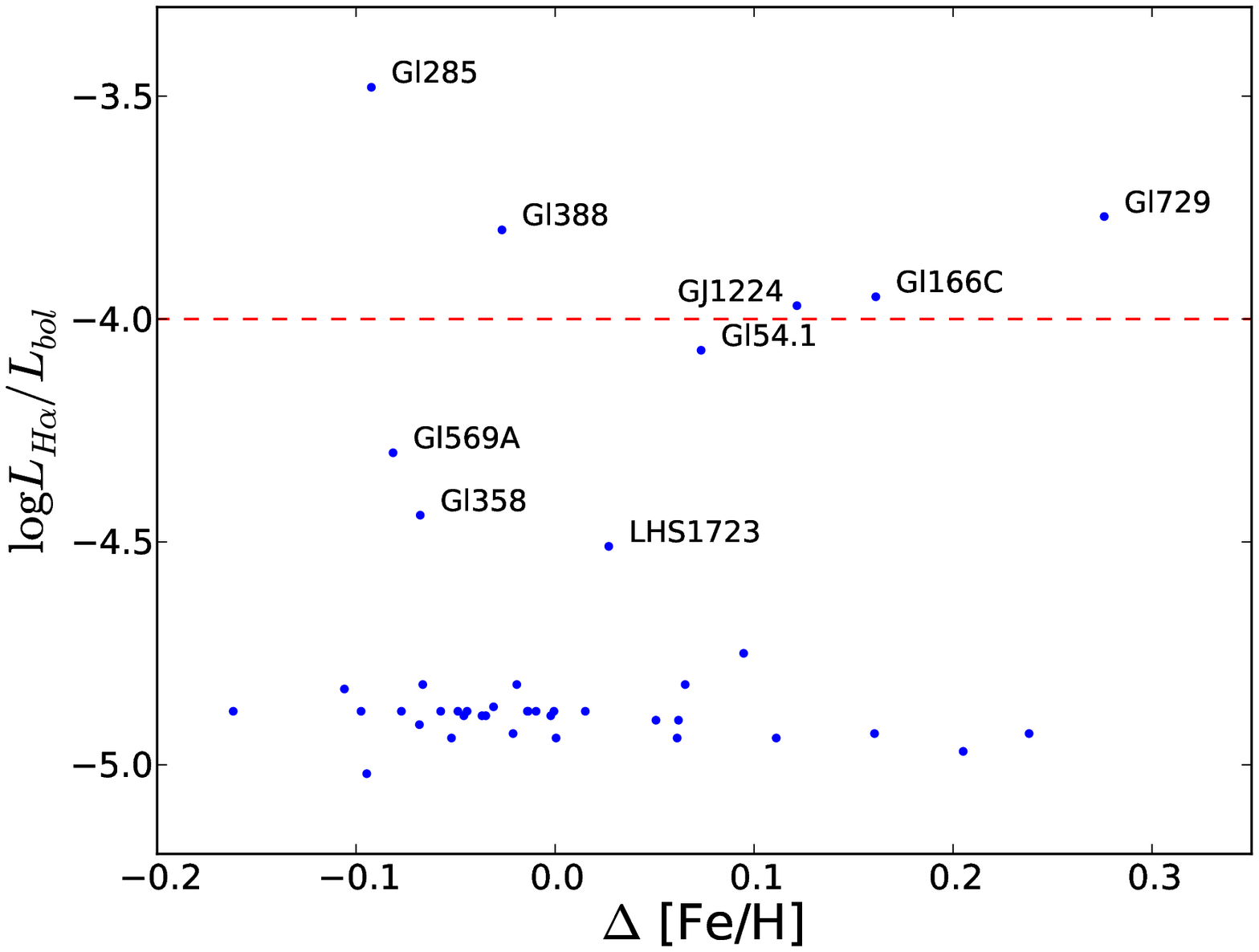}}
\subfigure[]{\includegraphics[scale=0.44]{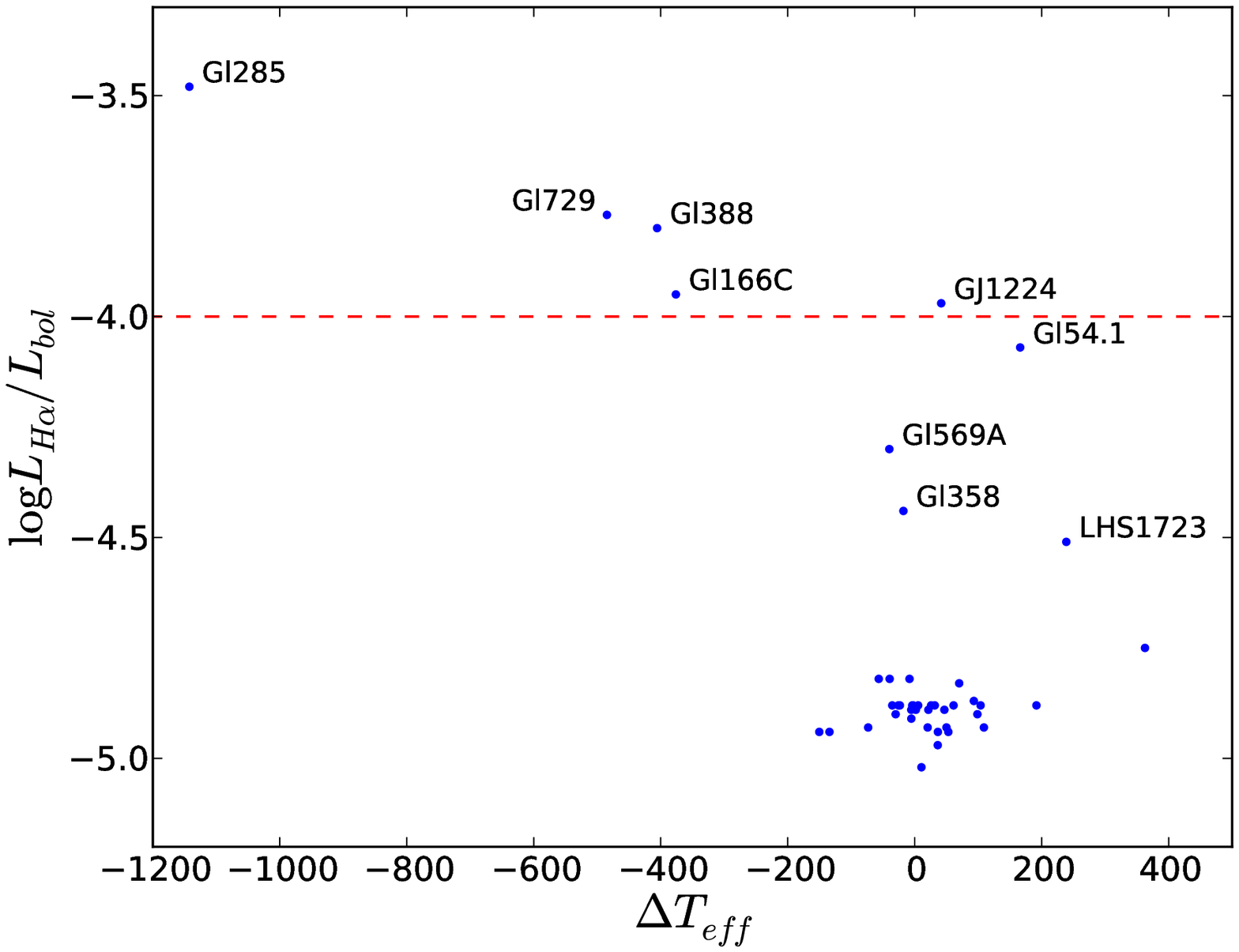}}
\subfigure[]{\includegraphics[scale=0.44]{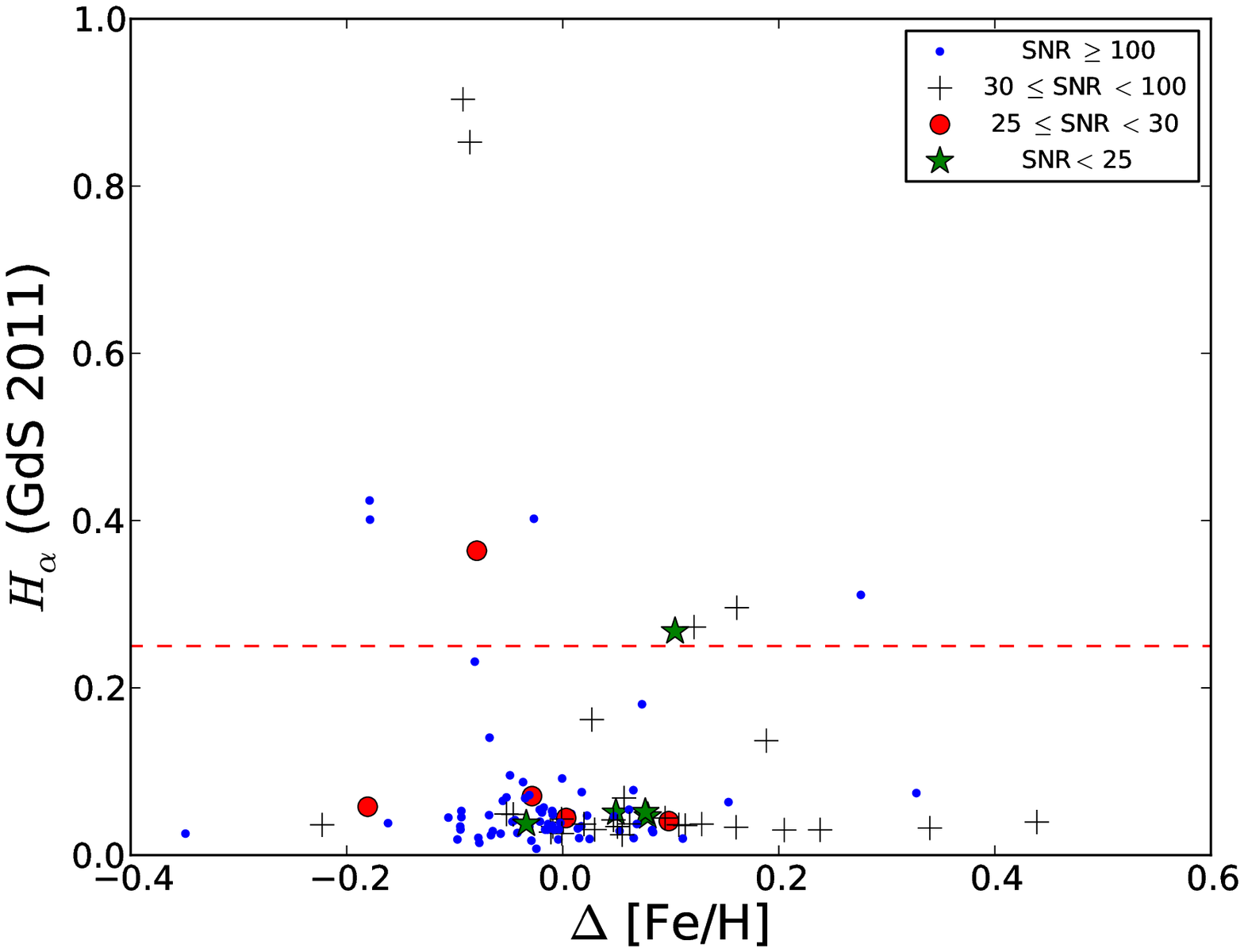}}
\subfigure[]{\includegraphics[scale=0.44]{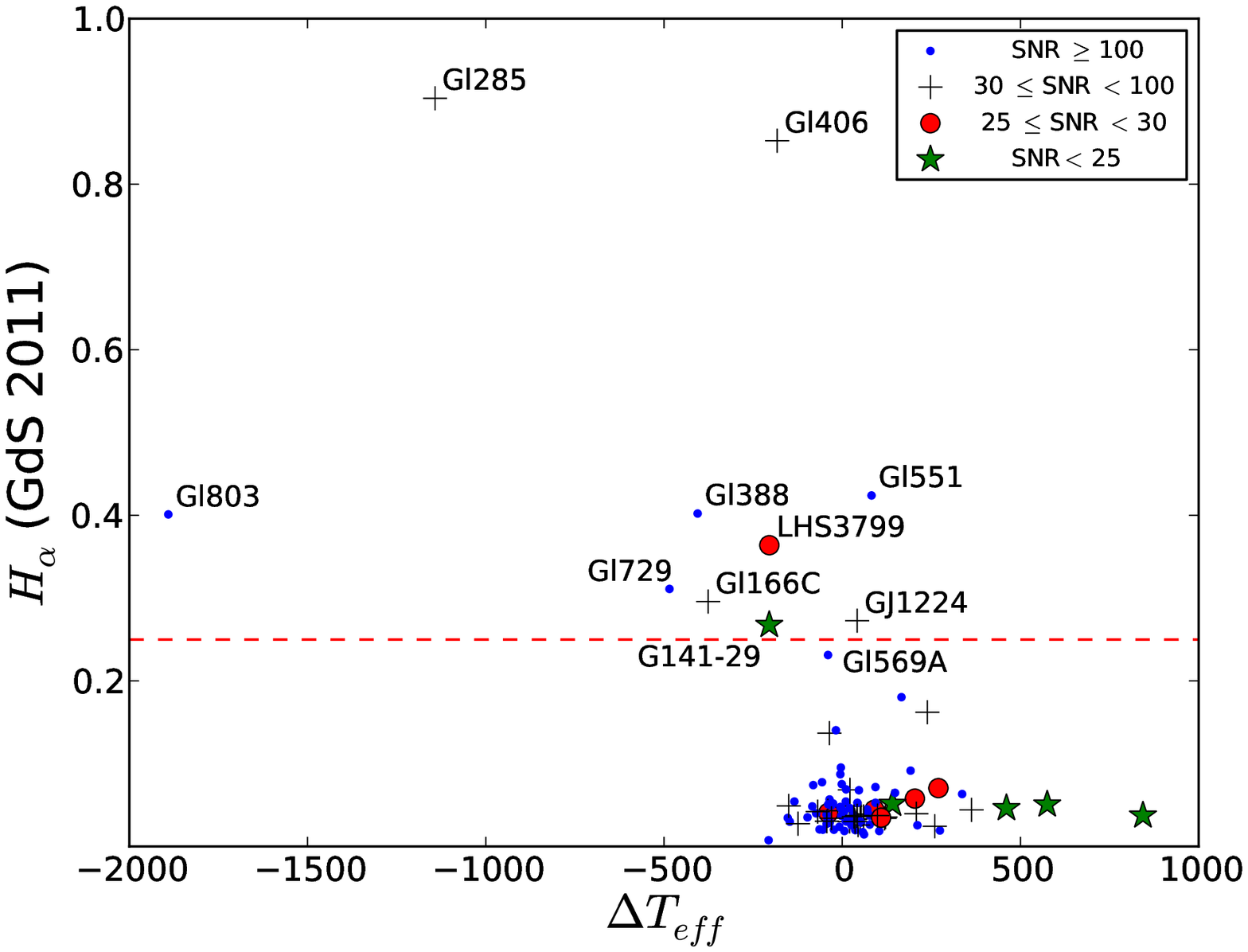}}
\end{center}
\caption{ a) Normalized H$\alpha$ luminosity taken from \citet{Reiners-2012} as a function of $\Delta$[Fe/H]; b) Normalized H$\alpha$ luminosity taken from \citet{Reiners-2012} as a function of $\Delta T_{eff}$; c) H${\alpha}$ index defined by \citet{Gomes_da_Silva-2011} versus $\Delta$[Fe/H]; d) H${\alpha}$ index defined by \citet{Gomes_da_Silva-2011} versus $\Delta T_{eff}$. The stars in common were taken from our full sample. The red dashed lines mark the limits above which the stars were excluded from the final sample.}
\label{fig:act_test}
\end{figure*}

Fig. \ref{fig:act_test} displays the normalized H$\alpha$ luminosity and the $H_{\alpha}$ index defined by \citet{Gomes_da_Silva-2011}  as a function of the difference between the parameters obtained with our method and the initial parameters. The legend in panels c) and d) depict the stellar spectra with S/N $\geq$ 100 (blue dots), S/N between 30 and 100 (black crosses), S/N between 30 and 25 (red circles), and S/N lower than 25 (green stars). We observe no clear correlation of the activity indices or S/N with [Fe/H]. Regarding $T_{eff}$ however, we can see that there is a clear trend towards lower temperatures with both activity indices. To take this trend into account we decided to perform an activity cut, excluding all stars with $\log{L_{H_{\alpha}}/L_{bol}} \geq -4.0 $ and H$\alpha \geq 0.25$ from our final sample. We also note a trend of H$\alpha$ with $T_{eff}$ towards higher temperatures for stars with S/N $<$ 25 (see bottom right corner of Fig. \ref{fig:act_test} d). The trends of our method with S/N are studied in detail in Sect. \ref{sec:testcal}.

\begin{figure}[]
\begin{center}
\subfigure[]{\includegraphics[scale=0.5]{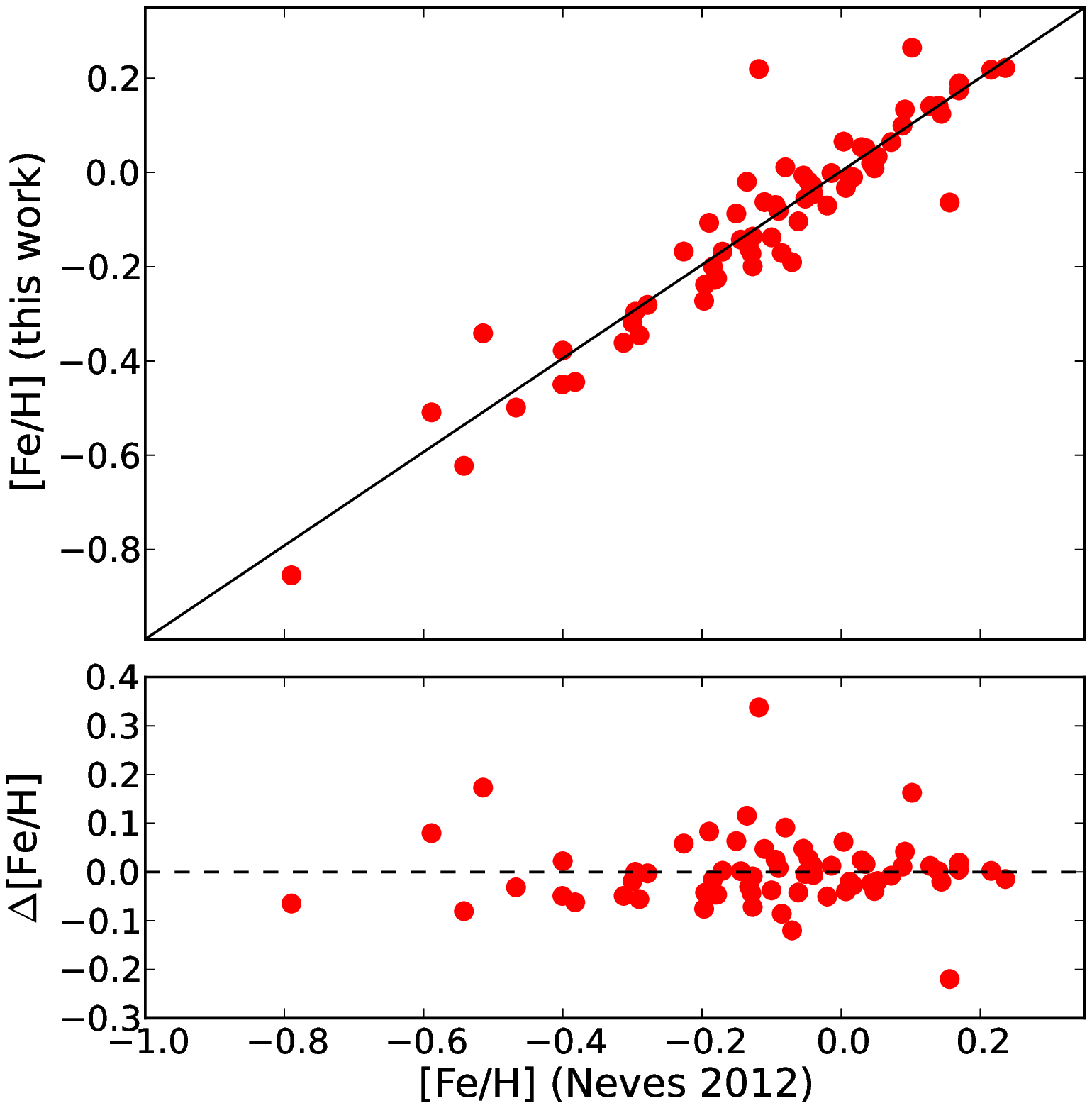}}
\subfigure[]{\includegraphics[scale=0.5]{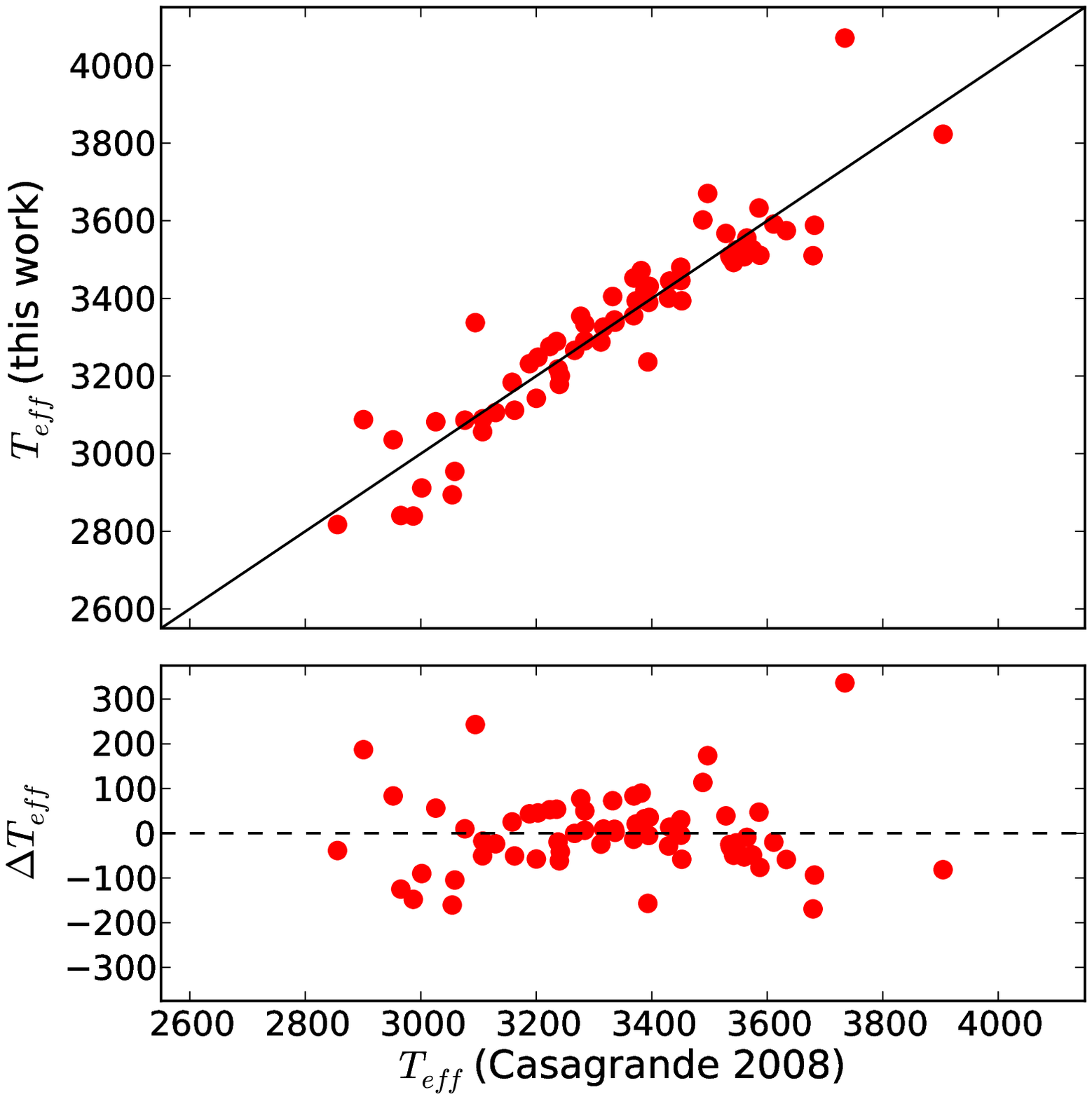}}
\end{center}
\caption{(a) [Fe/H] comparison between this work and the photometric calibration of \citet{Neves-2012}; (b) $T_{eff}$ comparison between this work and the photometric calibration of \citet{Casagrande-2008}.}
\label{fig:fehfeh}
\end{figure}

In the end, with a final sample of 65 stars, we obtain a dispersion of 0.08 dex for the metallicity and 91K for the effective temperature, as shown in Fig. \ref{fig:fehfeh}. The technique is valid between -0.85 to 0.26 dex for [Fe/H], 2800 to 4100 K for $T_{eff}$, and between M0.0 to M5.0. The dispersion around the calibration is quantified by the root mean square error (RMSE), and defined as

\begin{equation}
RMSE = \sqrt{\frac{\sum{(x_{i}-x_{ref})^{2}}}{n_{obs}-n_{coef}}},
\end{equation}
where $x_{i}$ is the estimated quantity, $x_{ref}$ the reference value for the same quantity, $n_{obs}$ the number of calibrators and $n_{coef}$ the number of parameters used in the method (four in this case). 

The calculated parameters as well as the reference determinations for [Fe/H] and $T_{eff}$ are listed in Table \ref{table:cal}. Columns 1 and 3 contain the values for the reference calibrations, while columns 2 and 4 show the values obtained with our technique. We emphasise here that we only get an improvement on the precision. The accuracy of the method as well as its systematics are tied to the original determinations of the parameters. 

\begin{table}[]
\tiny
\caption{ Our sample table with the reference and calibrated metallicity and effective temperature. Sorted by right ascension.}
\label{table:cal}
\scalebox{0.98}{
\begin{tabular}{ l r r r r}
\hline
\hline
star & [Fe/H]$_{N12}$ & [Fe/H]$_{NEW}$ & $T_{eff C08}$ & $T_{eff NEW}$ \\
       &         [dex]          &      [dex]              &       [K]                &  [K] \\
\hline
Gl1 & -0.40 & -0.45 & 3528 & 3567 \\
Gl54.1 & -0.40 & -0.38 & 2901 & 3088 \\
Gl87 & -0.30 & -0.32 & 3565 & 3555 \\
Gl105B & -0.14 & -0.02 & 3054 & 2894 \\
HIP12961 & -0.12 & 0.22 & 3904 & 3823 \\
LP771-95A & -0.51 & -0.34 & 3393 & 3236 \\
GJ163 & 0.00 & 0.07 & 3223 & 3276 \\
Gl176 & 0.02 & -0.01 & 3369 & 3355 \\
GJ179 & 0.14 & 0.12 & 3076 & 3086 \\
Gl191 & -0.79 & -0.85 & 3679 & 3510 \\
Gl205 & 0.17 & 0.19 & 3497 & 3670 \\
Gl213 & -0.19 & -0.11 & 3026 & 3082 \\
Gl229 & -0.04 & -0.03 & 3586 & 3633 \\
HIP31293 & -0.04 & -0.05 & 3312 & 3288 \\
HIP31292 & -0.11 & -0.06 & 3158 & 3184 \\
Gl250B & -0.09 & -0.08 & 3369 & 3453 \\
Gl273 & -0.05 & -0.01 & 3107 & 3090 \\
Gl300 & 0.09 & 0.13 & 2965 & 2841 \\
GJ2066 & -0.09 & -0.17 & 3388 & 3421 \\
GJ317 & 0.22 & 0.22 & 3130 & 3106 \\
Gl341 & -0.14 & -0.14 & 3633 & 3575 \\
GJ1125 & -0.15 & -0.09 & 3162 & 3112 \\
Gl357 & -0.30 & -0.30 & 3335 & 3344 \\
Gl358 & 0.01 & -0.01 & 3240 & 3178 \\
Gl367 & -0.09 & -0.07 & 3452 & 3394 \\
Gl382 & 0.04 & 0.02 & 3429 & 3401 \\
Gl393 & -0.13 & -0.20 & 3396 & 3431 \\
GJ3634 & -0.02 & -0.07 & 3332 & 3405 \\
Gl413.1 & -0.06 & -0.10 & 3373 & 3394 \\
Gl433 & -0.13 & -0.17 & 3450 & 3480 \\
Gl436 & 0.01 & -0.03 & 3277 & 3354 \\
Gl438 & -0.31 & -0.36 & 3536 & 3505 \\
Gl447 & -0.23 & -0.17 & 2952 & 3036 \\
Gl465 & -0.54 & -0.62 & 3382 & 3472 \\
Gl479 & 0.05 & 0.01 & 3238 & 3218 \\
Gl514 & -0.13 & -0.16 & 3574 & 3526 \\
Gl526 & -0.18 & -0.22 & 3545 & 3515 \\
Gl536 & -0.13 & -0.14 & 3546 & 3525 \\
Gl555 & 0.13 & 0.14 & 2987 & 2839 \\
Gl569A & 0.16 & -0.06 & 3235 & 3289 \\
Gl581 & -0.18 & -0.20 & 3203 & 3248 \\
Gl588 & 0.07 & 0.06 & 3284 & 3291 \\
Gl618A & -0.05 & -0.06 & 3242 & 3200 \\
Gl628 & -0.05 & -0.02 & 3107 & 3057 \\
GJ1214 & 0.03 & 0.05 & 2856 & 2817 \\
Gl667C & -0.47 & -0.50 & 3431 & 3445 \\
Gl674 & -0.18 & -0.23 & 3284 & 3334 \\
GJ676A & 0.10 & 0.26 & 3734 & 4071 \\
Gl678.1A & -0.10 & -0.14 & 3611 & 3591 \\
Gl680 & -0.07 & -0.19 & 3395 & 3390 \\
Gl682 & 0.09 & 0.10 & 3002 & 2912 \\
Gl686 & -0.29 & -0.35 & 3542 & 3493 \\
Gl693 & -0.28 & -0.28 & 3188 & 3232 \\
Gl699 & -0.59 & -0.51 & 3094 & 3338 \\
Gl701 & -0.20 & -0.27 & 3535 & 3510 \\
Gl752A & 0.04 & 0.05 & 3336 & 3339 \\
Gl832 & -0.17 & -0.17 & 3450 & 3446 \\
Gl846 & -0.08 & 0.01 & 3682 & 3588 \\
Gl849 & 0.24 & 0.22 & 3200 & 3143 \\
Gl876 & 0.14 & 0.14 & 3059 & 2954 \\
Gl877 & -0.01 & -0.00 & 3266 & 3266 \\
Gl880 & 0.05 & 0.03 & 3488 & 3602 \\
Gl887 & -0.20 & -0.24 & 3560 & 3507 \\
Gl908 & -0.38 & -0.44 & 3587 & 3511 \\
LTT9759 & 0.17 & 0.17 & 3316 & 3326 \\
\hline
\end{tabular}}
\end{table}


\subsection{Estimation of the uncertainties}
\label{sec:uncertain}
To validate our method and have a better understanding of the uncertainties of our measurements we performed a bootstrap resampling and calculated the root mean square error of validation (RMSE$_{V}$). 

\begin{figure}[]
\begin{center}
\subfigure[]{\includegraphics[scale=0.44]{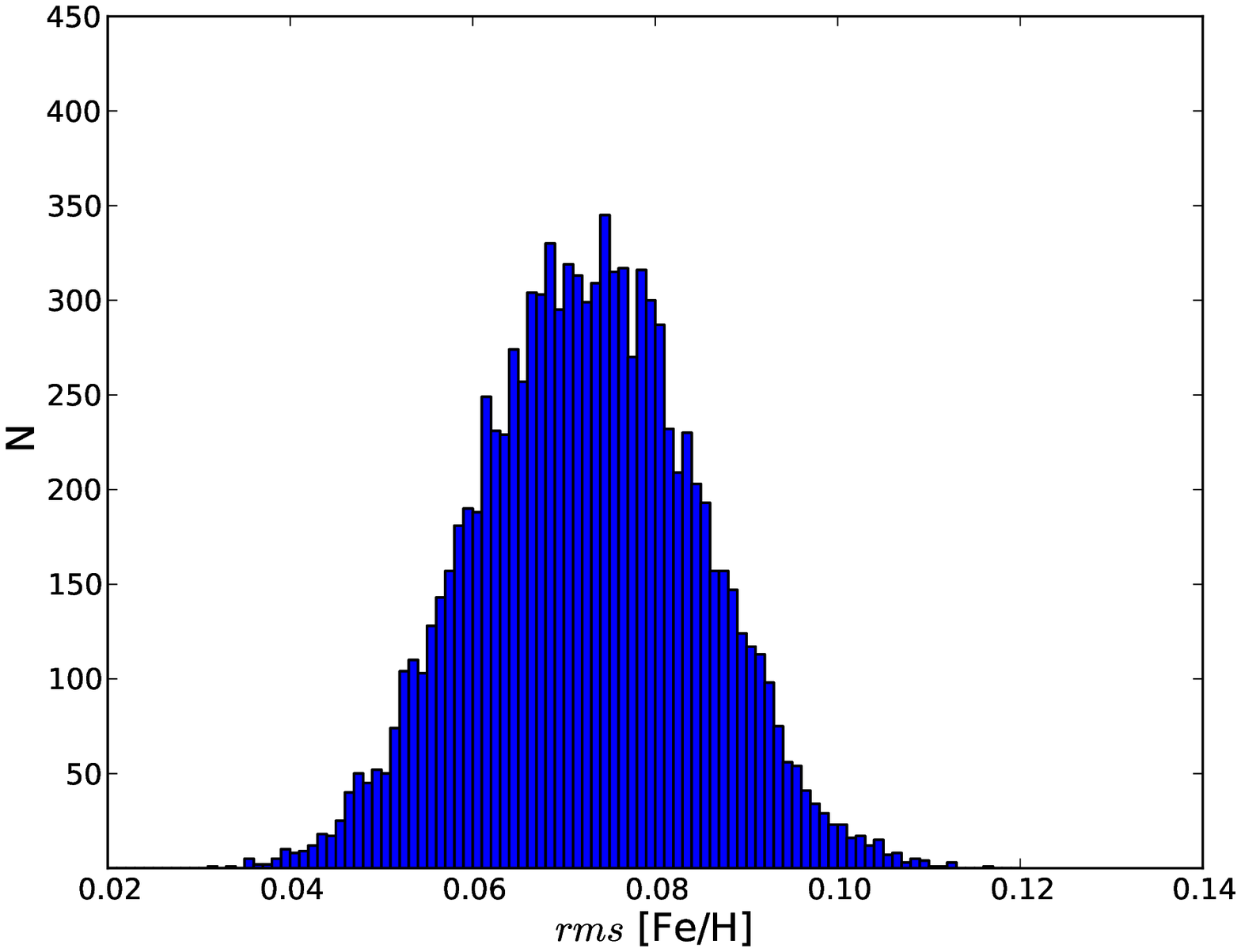}}
\subfigure[]{\includegraphics[scale=0.44]{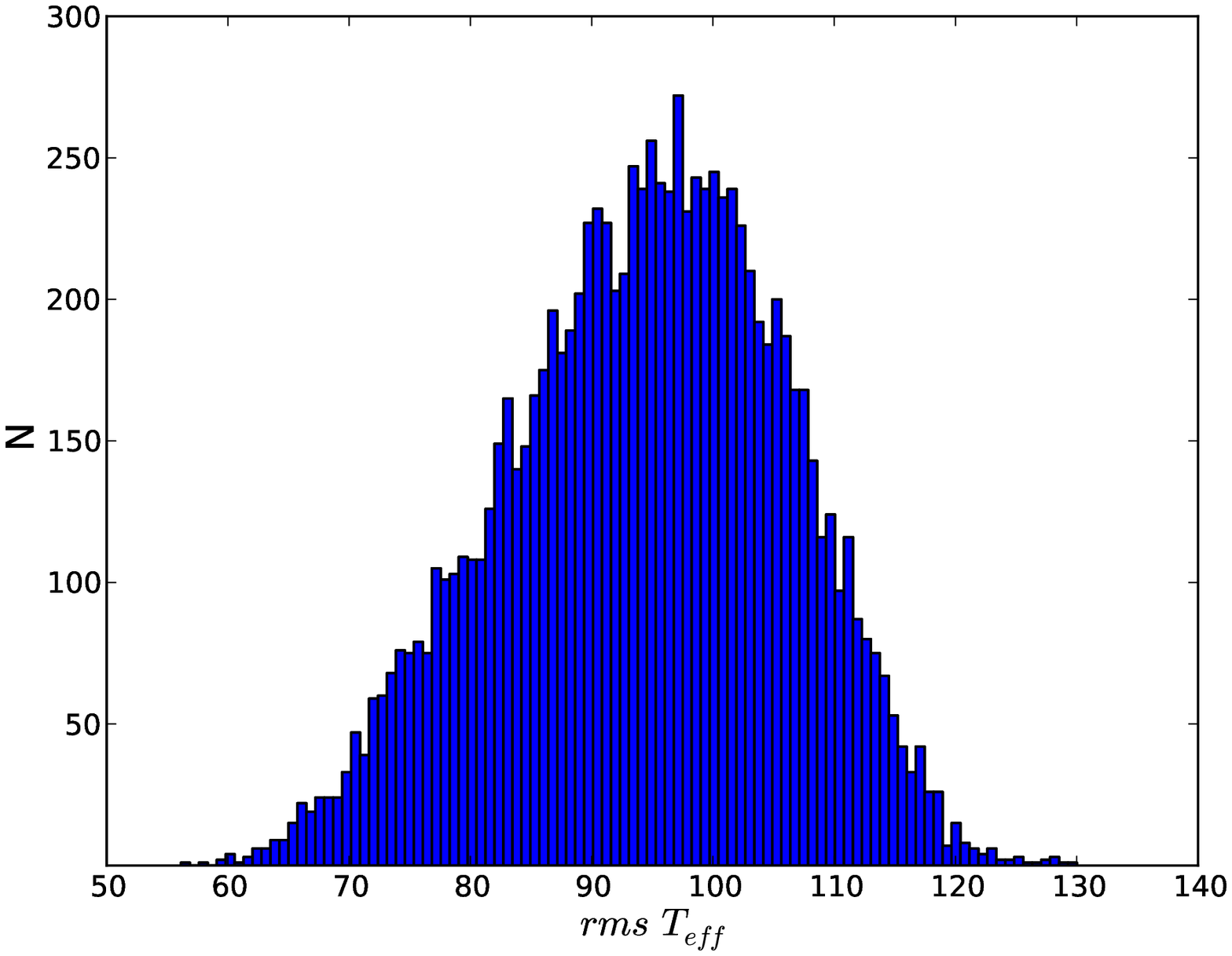}}
\end{center}
\caption{Histogram of the dispersion given by bootstrap for [Fe/H] (a) and $T_{eff}$ (b). The N is the number of trials.}
\label{fig:strap}
\end{figure}

The bootstrap method we implemented tests how the $rms$ of the method changes when using slightly different `bootstrapped' samples. To have a statistical significative number we first created 10.000 virtual samples by randomly drawing with repetition, for each virtual sample, a number of stars equal to the size of our sample. The random drawing followed a random uniform distribution. Then we calculated the $rms$ for each trial and measured the $1\sigma$ gaussian equivalent interval between 15.9\% and 84.1\% from the resulting distribution, following the procedure of e.g. \citet{Burgasser-2003,Neves-2013}. The distributions of the $rms$ for both parameters are depicted in Fig. \ref{fig:strap}. The final result shows a variation of the $rms$ of the [Fe/H] and $T_{eff}$ by  $\pm 0.01$ dex and $\pm 13$ K respectively.

The calculation of the RMSE$_{V}$ is a predicted residual sum of squares (PRESS) procedure \citep{Weisberg-2005} and follows the description in the Appendix of \citet{Rojas-Ayala-2012}. In short, we try to obtain the original value of the metallicity and temperature of each star $i$ of the technique leaving that star out when calculating the parameters. Then, we calculate the residuals, or the difference between the original and obtained value for each star and add them up in quadrature. The PRESS statistic is then defined as 
\begin{equation}
PRESS = \sum{(y_{i}-y_{ref}})^{2},
\end{equation}
where $y_{i}$ is the estimated value of the parameter and $y_{ref}$ is the reference value of the measured quantity. From here we can calculate the root mean squared error of validation,

\begin{equation}
RMSE_{V} = \sqrt{\frac{PRESS}{n_{V}}},
\end{equation}
where $n_{V}$ is the number of calibrators. The RMSE$_{V}$ value can then be used to obtain confidence intervals. We obtain a RMSE$_{V}$ value of 0.12 dex and $\sim$ 293 K for the [Fe/H] and $T_{eff}$ respectively and will use these values as $1\sigma$ confidence intervals, assuming a normal cumulative distribution function. Table \ref{table:errors} summarises our results. 

\begin{table}[h!]
\centering
\caption[]{Uncertainty estimators for [Fe/H] and $T_{eff}$.}
\label{table:errors}
\begin{tabular}{l c c}
\hline
\hline
Estimator & [Fe/H] & $T_{eff}$ \\ 
                &  [dex]  &  [K] \\
\hline
RMSE & 0.08 & 91\\
Bootstrap & 0.08$\pm$0.01 & 91$\pm$13 \\
RMSE$_{V}$ & 0.12 & 293 \\
\hline
\hline
\end{tabular}
\end{table}

We observe that the uncertainties calculated with the different techniques are consistent with each other. The uncertainty for $T_{eff}$ is large but is in line with the expected uncertainties. We also perturbed our sample by introducing an offset in [Fe/H] or $T_{eff}$, as explained in Sect. \ref{sec:method} but found that it does not affect the measurement of the parameters. In the end we assume our final uncertainty to be the maximum uncertainty of the RMSE given by the bootstrap, that translates into 0.09 dex for [Fe/H] and 110 K for $T_{eff}$.   

\subsection{Testing our technique as a function of resolution and S/N}
\label{sec:testcal}

\begin{figure}[]
\begin{center}
\includegraphics[scale=0.40]{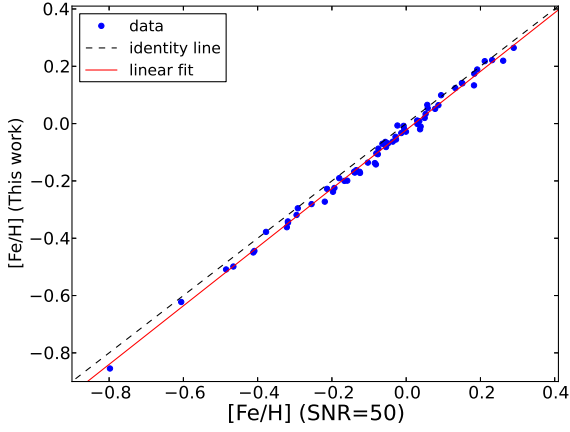}
\end{center}
\caption{[Fe/H] of our procedure versus the [Fe/H] obtained with S/N=50.The identity line is depicted in dashed black. The solid red line shows the linear fit.}
\label{fig:fehfehsnr}
\end{figure}

To further test our method, we calculated the dispersion of the parameters as a function of the resolution and S/N. The first step consisted in the study of the behaviour of the procedure as a function of S/N only. We injected random Gaussian noise in the spectra of the sample to obtain spectra with S/N @ 5500 \AA~ between 100 and 10. Then, we calculated linear fits between the [Fe/H] and $T_{eff}$ obtained with the lower S/N spectra and the values of our method, for the full S/N range, as shown in Fig. \ref{fig:fehfehsnr} for [Fe/H] and S/N=50. The dashed black line marks the identity line while the red line represents a linear fit to the data.

\begin{figure*}[]
\centering
\subfigure[]{\includegraphics[scale=0.27]{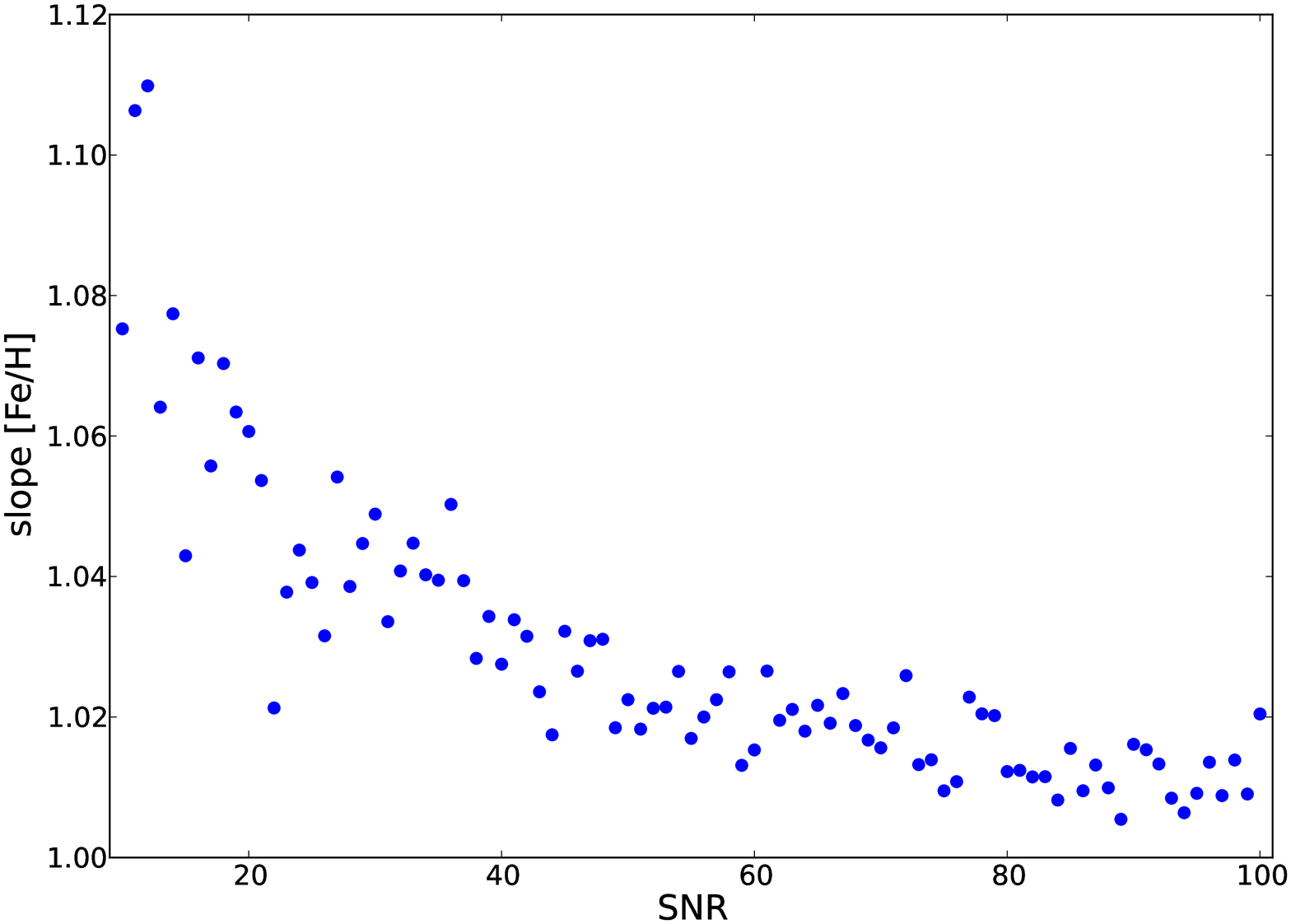}}
\subfigure[]{\includegraphics[scale=0.27]{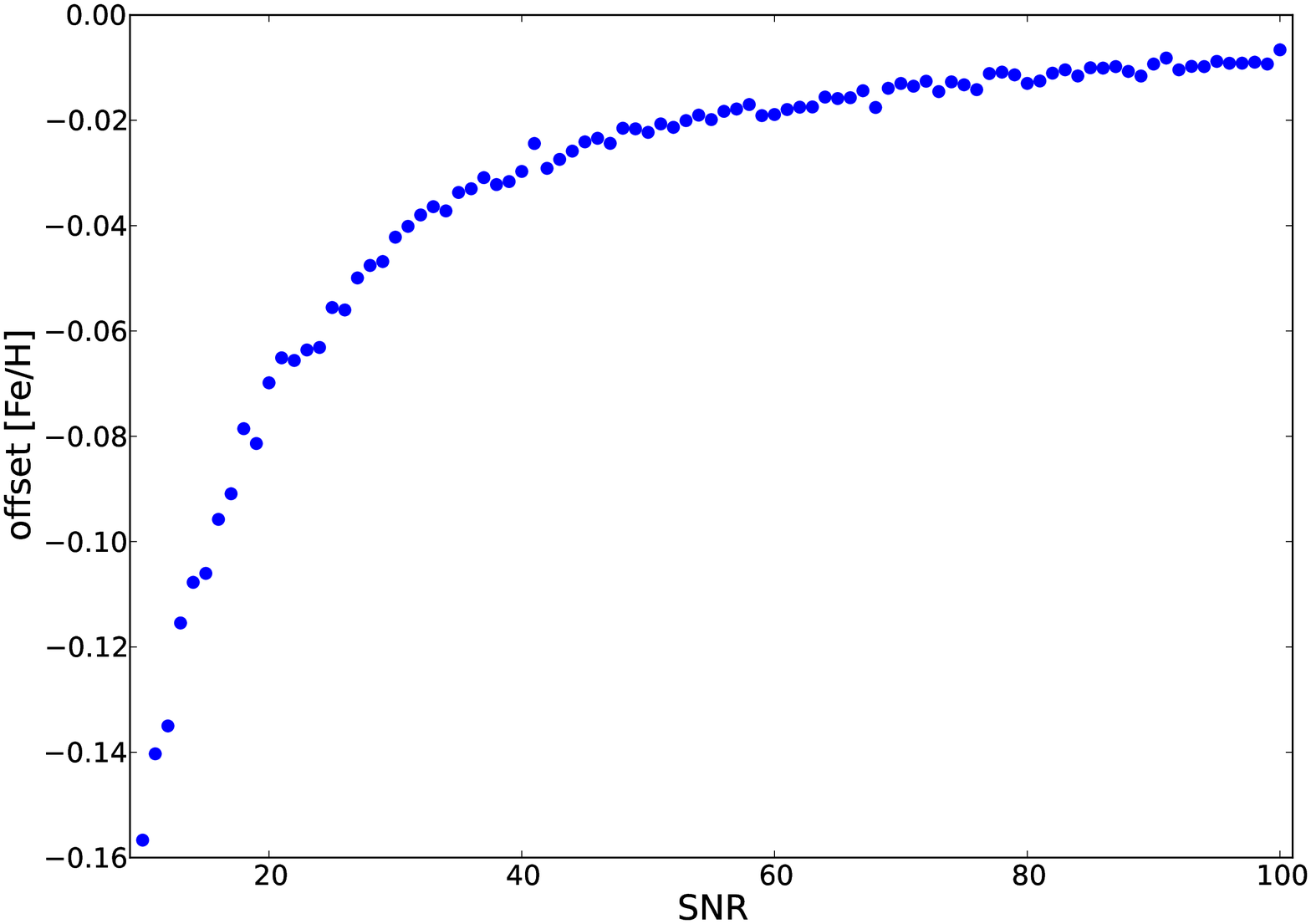}}
\subfigure[]{\includegraphics[scale=0.27]{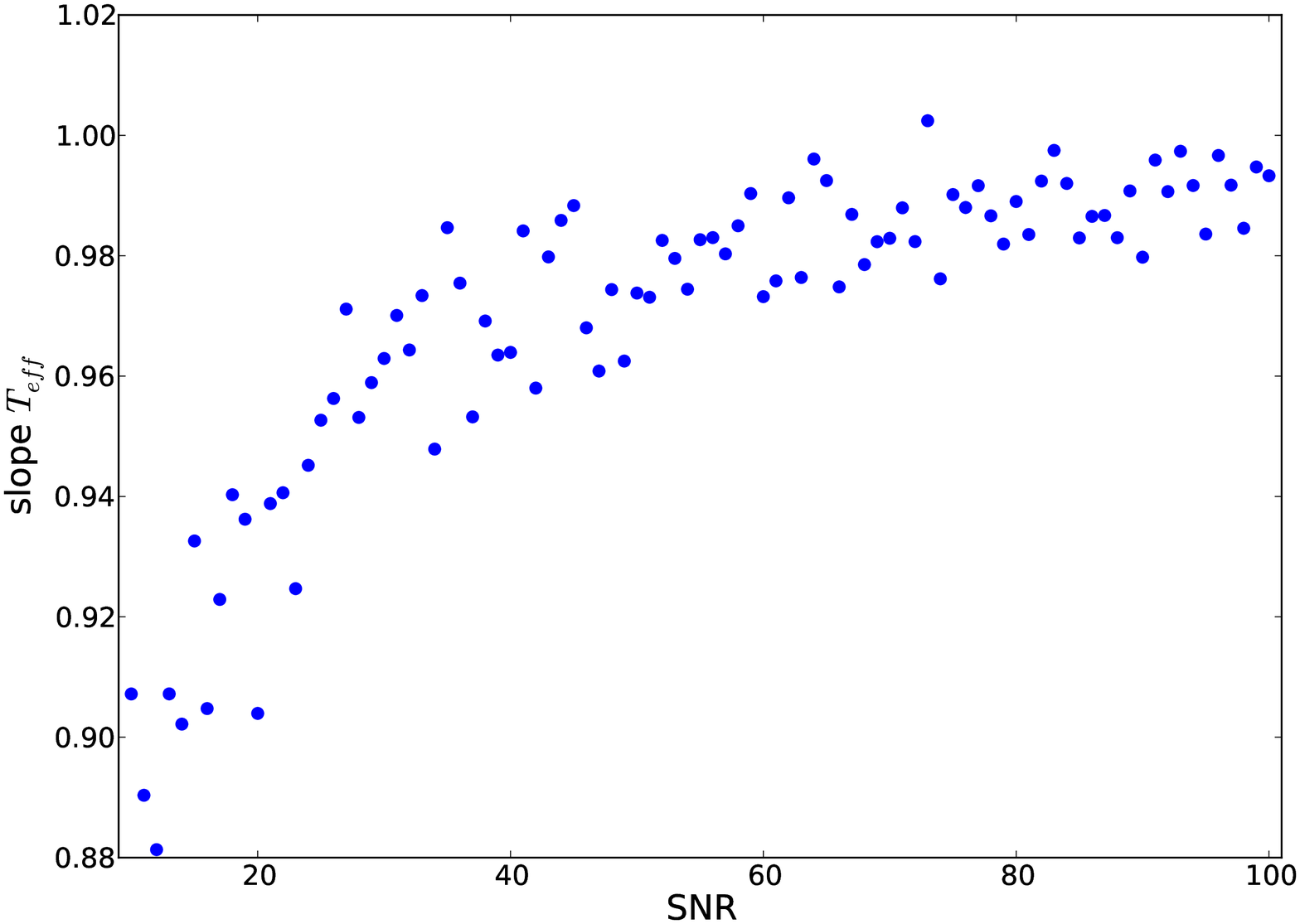}}
\subfigure[]{\includegraphics[scale=0.27]{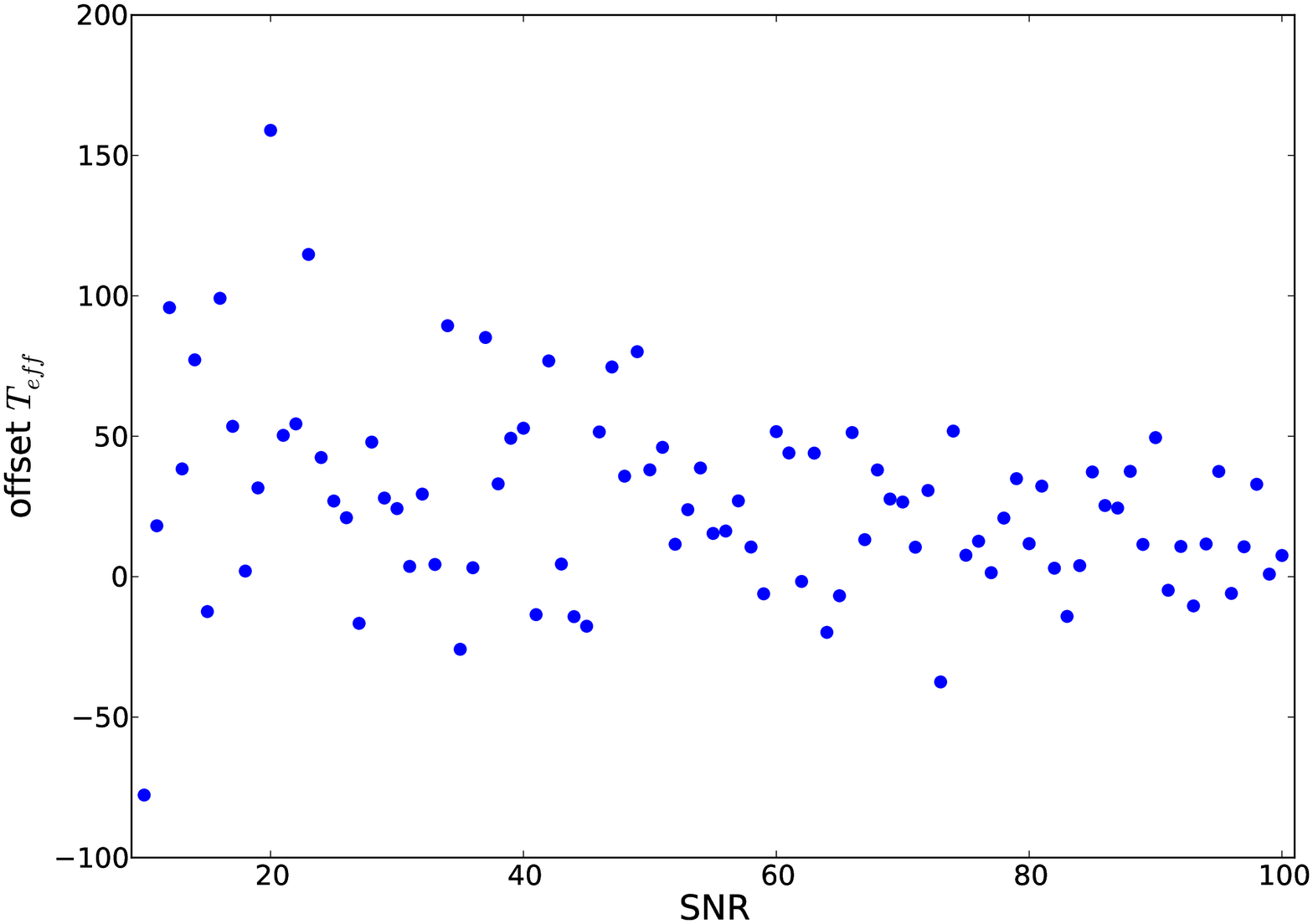}}
\caption{ Slope and offset of [Fe/H] and $T_{eff}$ of our work as a function of S/N for star Gl479.}
\label{fig:fit}
\end{figure*}

\begin{figure*}[]
\centering
\subfigure[]{\includegraphics[scale=0.27]{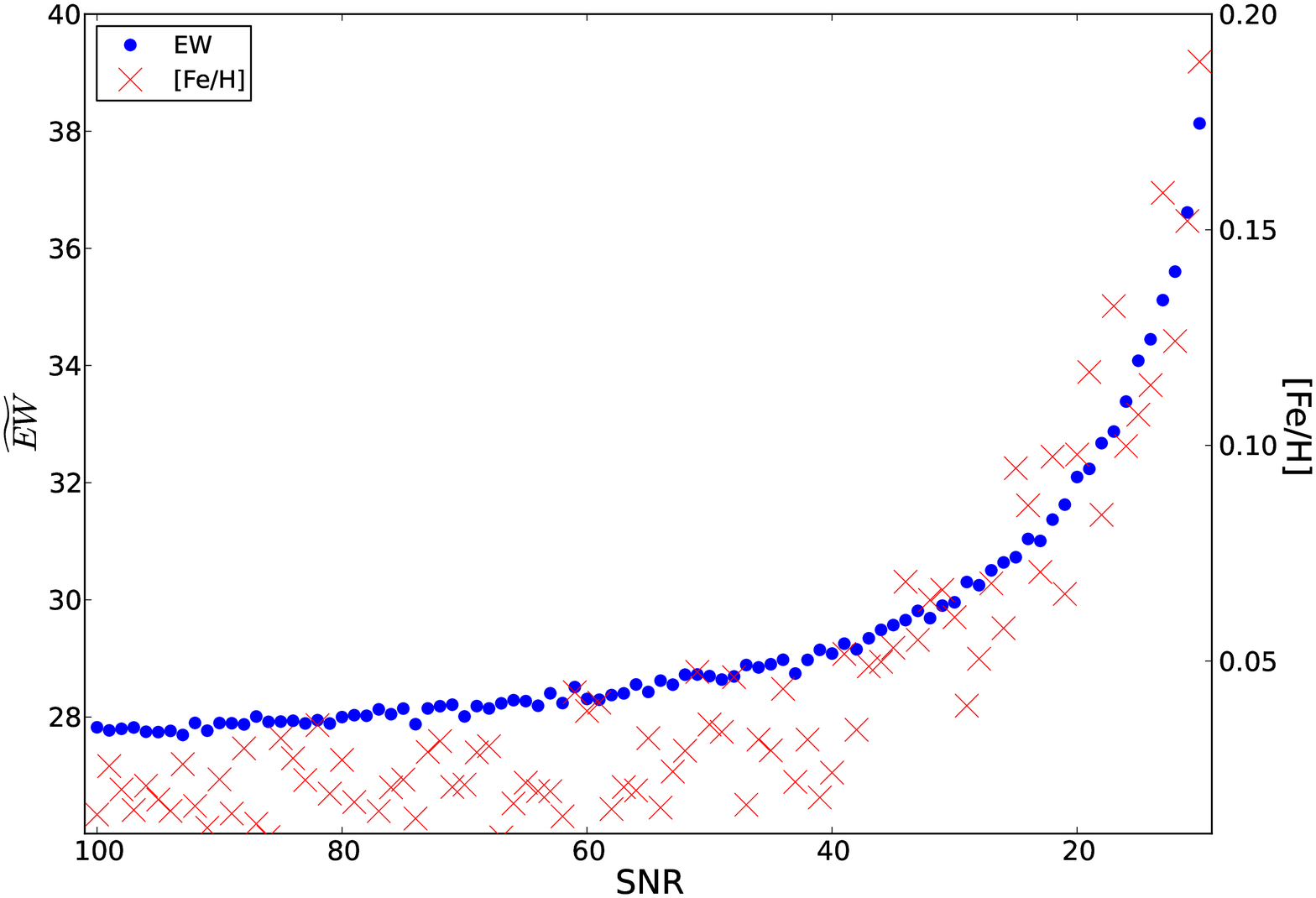}}
\subfigure[]{\includegraphics[scale=0.27]{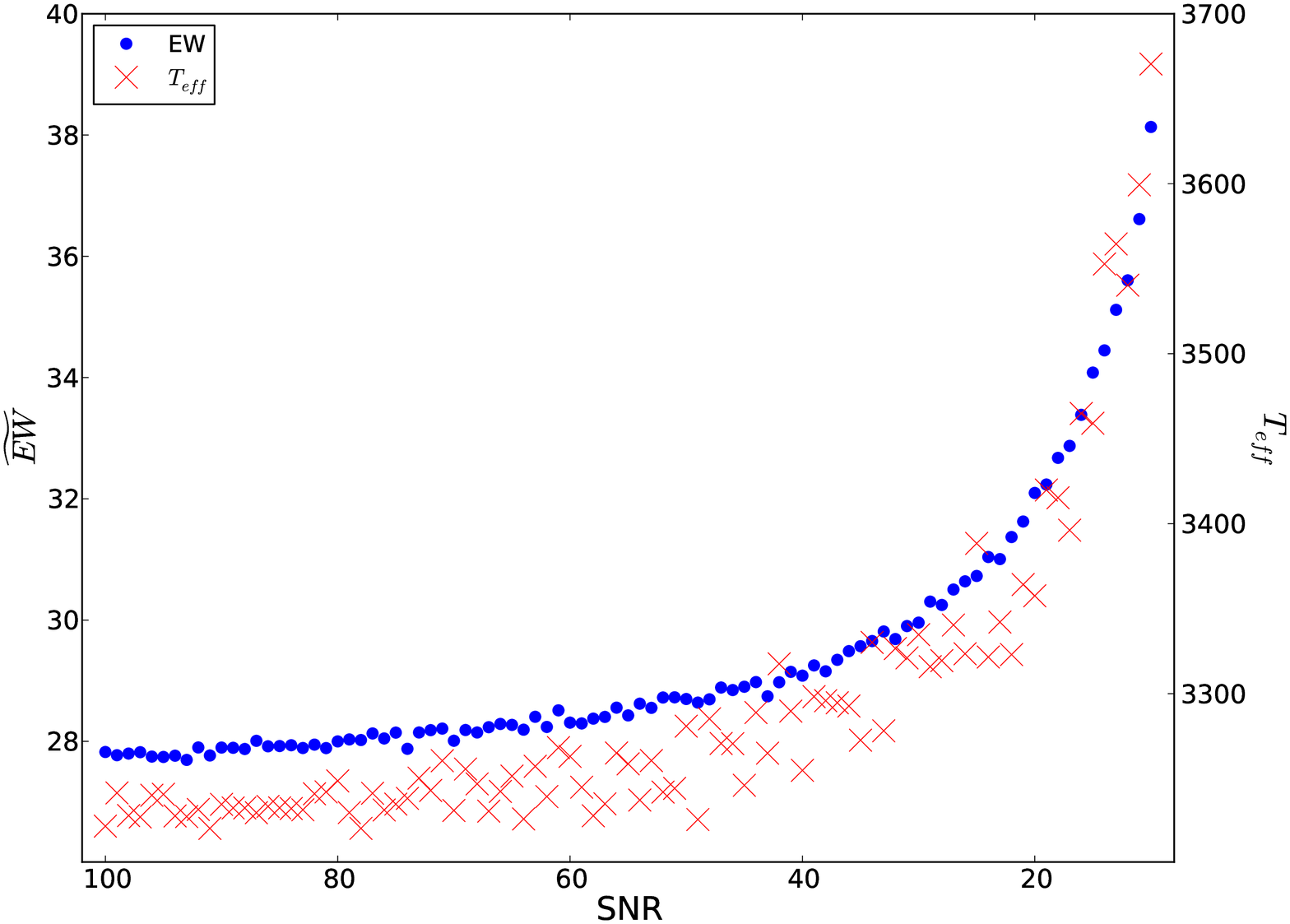}}
\caption{Median of the pseudo EWs and [Fe/H] (a) or $T_{eff}$ (b) as a function of S/N for star Gl479.}
\label{fig:ewsnr}
\end{figure*}

Fig. \ref{fig:fit} depicts the slope and offset of the linear fits for [Fe/H] and $T_{eff}$, as a function of S/N. We observe a deviation from the identity line as the S/N decreases as expected, except in d), where the offset of $T_{eff}$ is reasonably constant, on average, with S/N. From here we investigated the cause of this trend and found that the measured EWs, roughly quantified as the median of all EWs, follow similar trends with S/N, as shown in Fig. \ref{fig:ewsnr} the for the star Gl479 ([Fe/H] = 0.01 dex, $T_{eff}$ = 3218 K) as example. The observed trends are similar in other stars, with different metallicities and effective temperatures.


Using this information, we corrected the values of both parameters for all stars of our sample with S/N $<$ 100, where [Fe/H] (Corrected) = $a$[Fe/H] (Degraded) + $b$, and $T_{eff}$ (Corrected) = $aT_{eff}$ (Degraded) + $b$. The $a$ is the slope and $b$ is the offset. The uncertainties associated with each correction were estimated by calculating the dispersion of the residuals. However, we observed that these uncertainties are very small compared to the the final uncertainty of our method, as shown in Sect. \ref{sec:uncertain}. Therefore, we assumed the uncertainty of our method for all stars, except for the active ones, according to the two estimated activity thresholds calculated in Sect. \ref{sec:sample}. In this case, we used the initial values of the parameters, calculated with the calibrations of \citet{Neves-2012} and \citet{Casagrande-2008}. At this point, we also decided not to use our new values for stars with S/N $\leq$ 25, because the correction was not good enough to obtain a reasonable value for the objects depicted as green stars in Fig. \ref{fig:act_test}. Moreover, we note that the stars LHS 1513 and Gl 803 give an estimated [Fe/H] value of $-1.51$ and $0.46$ dex respectively, calculated with the calibration of \citet{Neves-2012}, our reference scale. Both values are outside the range of this method and will not be used. Table \ref{table:full} shows the corrected results, along with the previous values of [Fe/H] from \citet{Neves-2013}. Column 1 describes the star designation and column 2 and 3, the right ascension and declination of the star, respectively. Column 4 depicts the S/N, column 5 the normalized $H\alpha$ index taken from \citet{Reiners-2012}, and column 6 the $H\alpha$ index described by \citet{Gomes_da_Silva-2011}. Column 7 shows the metallicity calculated with the coefficient matrix used in our previous work, \citet{Neves-2013}, column 8 the [Fe/H] obtained in our work, along with its associated error in column 9. The $T_{eff}$ values of our stars, and its uncertainties are described in columns 10 and 11.    

\addtocounter{table}{1}

In the second step of this test we modified both resolution and S/N of our sample spectra. The resolution of the HARPS spectra was degraded by convolving the spectra with a normalised Gaussian curve, simulating the instrumental profile, with $FWHM = \lambda/R$, where $\lambda$ is the wavelength and $R$ the resolution we intend to obtain. 
From here, the standard deviation of the Gaussian is calculated with the well known formula 

\begin{equation}
\label{eq:fwhm}
\sigma = \frac{FWHM}{2\sqrt{2\log{2}}}. 
\end{equation}

The $\sigma$ was adjusted to the HARPS resolution ($R\sim115.000$) and to the original S/N @ $5500 \AA$  of each spectrum. The final value for $\sigma$ is then



\begin{equation}
\sigma' = \sqrt{\sigma^2-\sigma_{HARPS}^2}.
\end{equation}

\begin{figure*}[]
\begin{center}
\includegraphics[scale=0.9]{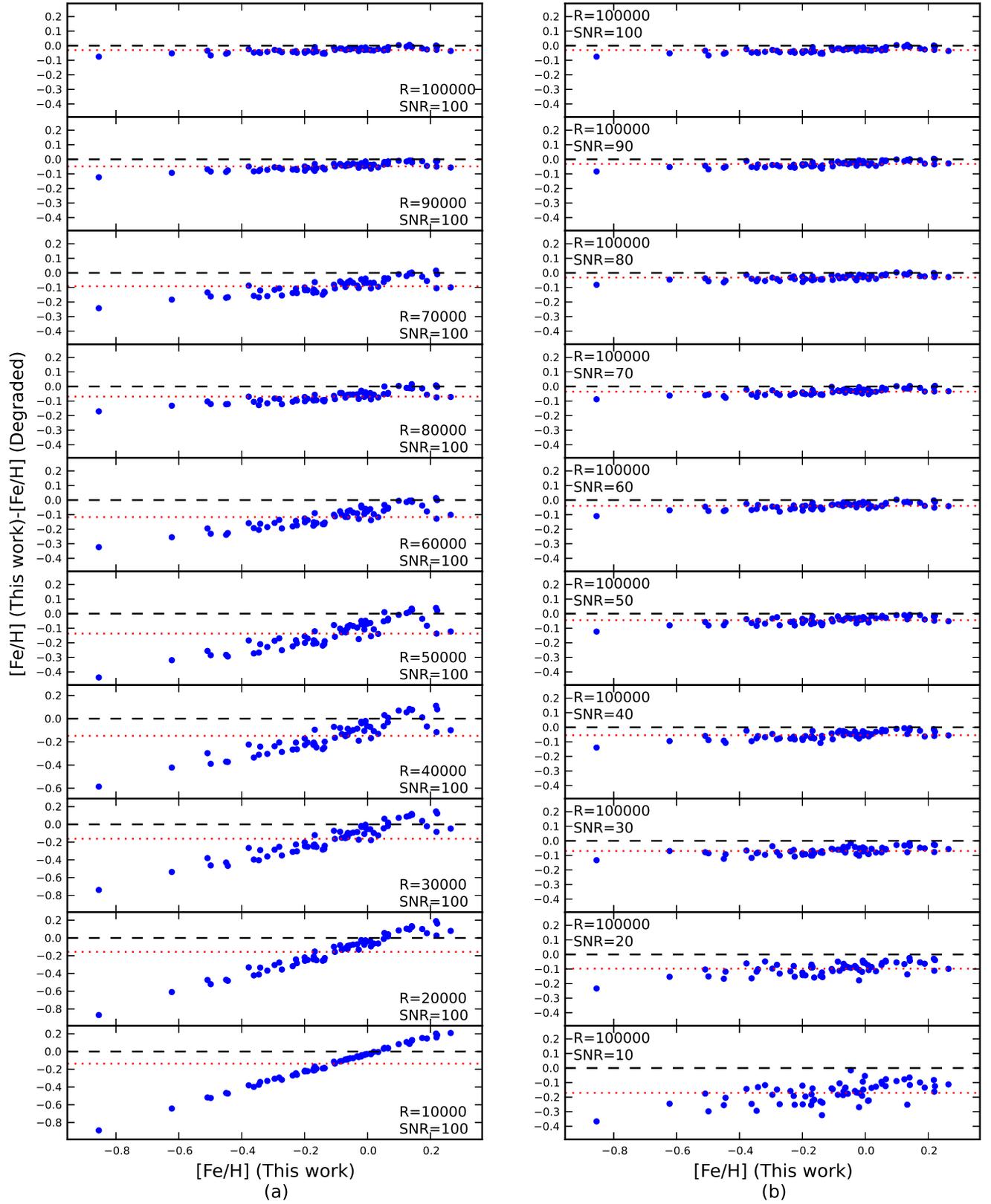}
\end{center}
\caption{Difference of the [Fe/H] for this work and the [Fe/H] calculated with different resolution/signal-to-noise combinations as a function of the resolution and S/N.}
\label{fig:resnr-feh}
\end{figure*}

\begin{figure*}[h!]
\begin{center}
\includegraphics[scale=0.9]{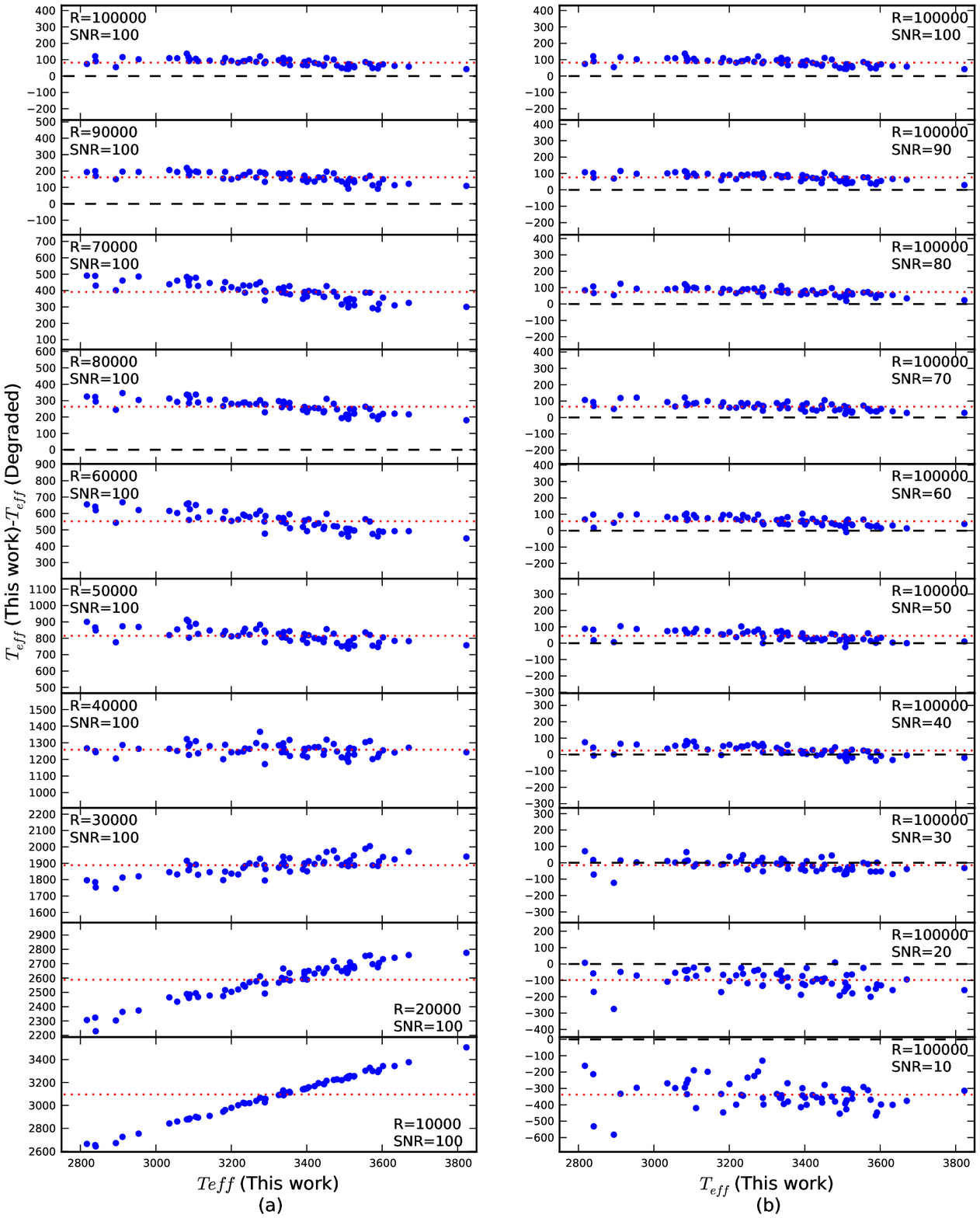}
\end{center}
\caption{Difference of the $T_{eff}$ for this work and the $T_{eff}$ calculated with different resolution/signal-to-noise combinations as a function of the resolution and S/N.}
\label{fig:resnr-teff}
\end{figure*}

Figures \ref{fig:resnr-feh} and \ref{fig:resnr-teff} show the difference of our parameters against the degraded values, as a function of resolution (while the S/N is kept constant at 100) and as a function of S/N (while the resolution is kept constant at 100.000) for [Fe/H] and $T_{eff}$ respectively. The red dotted line depicts the offset of the residuals. 




From the two Figures we observe the existence of linear trends for different resolutions and S/N. To correct these trends we performed a linear fit for each Resolution/signal-to-noise combination with the functional form [Fe/H] (This work)$ = a$[Fe/H] (Degraded) + $b$ for metallicity, and 
$T_{eff}$ (This work)$ = aT_{eff}$ (Degraded) + $b$ for effective temperature. The values for each combination are shown in Table \ref{table:linlin}.

\begin{table*}[]
\caption{Linear fit coefficients $a$ and $b$ from the relation between the values of our parameters and the values calculated for different combinations of resolution and S/N.}
\label{table:linlin}
\begin{center}
\scalebox{0.77}{
\subtable[{[Fe/H]}]{
\begin{tabular}{l | r r | r r | r r | r r | r r}
\hline
\hline
S/N & \multicolumn{2}{c|}{100} & \multicolumn{2}{c|}{90} & \multicolumn{2}{c|}{80} & \multicolumn{2}{c|}{70}  & \multicolumn{2}{c}{60} \\
Resolution & a & b & a & b & a & b & a & b & a & b \\
\hline
100000 & 1.0573 & -0.0257 & 1.0645 & -0.0269 & 1.0572 & -0.0275 & 1.0721 & -0.0304 & 1.0686 & -0.0353 \\
95000 & 1.0793 & -0.0304 & 1.0786 & -0.0343 & 1.0849 & -0.0358 & 1.0766 & -0.0401 & 1.0740 & -0.0455 \\
90000 & 1.0947 & -0.0428 & 1.1111 & -0.0430 & 1.0973 & -0.0472 & 1.1024 & -0.0488 & 1.1029 & -0.0521 \\
85000 & 1.1164 & -0.0529 & 1.1345 & -0.0511 & 1.1313 & -0.0557 & 1.1451 & -0.0592 & 1.1209 & -0.0639 \\
80000 & 1.1522 & -0.0630 & 1.1527 & -0.0652 & 1.1545 & -0.0670 & 1.1595 & -0.0694 & 1.1587 & -0.0749 \\
75000 & 1.1819 & -0.0751 & 1.1859 & -0.0790 & 1.1774 & -0.0801 & 1.1891 & -0.0838 & 1.1967 & -0.0877 \\
70000 & 1.2184 & -0.0877 & 1.2271 & -0.0905 & 1.2309 & -0.0920 & 1.2354 & -0.0939 & 1.2441 & -0.1018 \\
65000 & 1.2614 & -0.0997 & 1.2781 & -0.1045 & 1.2739 & -0.1075 & 1.2850 & -0.1094 & 1.2849 & -0.1149 \\
60000 & 1.3350 & -0.1191 & 1.3306 & -0.1174 & 1.3361 & -0.1206 & 1.3408 & -0.1255 & 1.3437 & -0.1305 \\
55000 & 1.3949 & -0.1372 & 1.4058 & -0.1370 & 1.4039 & -0.1417 & 1.3962 & -0.1439 & 1.4440 & -0.1518 \\
50000 & 1.4859 & -0.1499 & 1.5020 & -0.1559 & 1.5317 & -0.1590 & 1.5211 & -0.1631 & 1.5289 & -0.1666 \\
45000 & 1.5768 & -0.1623 & 1.5924 & -0.1670 & 1.5550 & -0.1700 & 1.5998 & -0.1726 & 1.5686 & -0.1811 \\
40000 & 1.6667 & -0.1739 & 1.6608 & -0.1837 & 1.6628 & -0.1912 & 1.6434 & -0.1903 & 1.6910 & -0.1976 \\
35000 & 1.6830 & -0.1907 & 1.7071 & -0.1930 & 1.7183 & -0.1984 & 1.6553 & -0.2061 & 1.6662 & -0.2076 \\
30000 & 1.7031 & -0.1998 & 1.6868 & -0.2016 & 1.6715 & -0.2080 & 1.6561 & -0.2102 & 1.6526 & -0.2109 \\
25000 & 1.7480 & -0.2035 & 1.7205 & -0.2016 & 1.7607 & -0.2103 & 1.7139 & -0.2077 & 1.7232 & -0.2126 \\
20000 & 1.8697 & -0.1965 & 1.8721 & -0.1996 & 1.9415 & -0.2054 & 1.7788 & -0.1984 & 1.7328 & -0.1986 \\
15000 & 2.2339 & -0.1950 & 2.1304 & -0.1933 & 2.3825 & -0.2044 & 2.4904 & -0.2125 & 2.3611 & -0.2164 \\
10000 & 3.6228 & -0.2062 & 2.6636 & -0.1899 & 2.8711 & -0.1999 & 2.6645 & -0.2022 & 0.1404 & -0.1154 \\
\hline
S/N & \multicolumn{2}{c|}{50} & \multicolumn{2}{c|}{40} & \multicolumn{2}{c|}{30} & \multicolumn{2}{c|}{20} & \multicolumn{2}{c}{10} \\
Resolution & a & b & a & b & a & b & a & b & a & b  \\
\hline
100000 & 1.0749 & -0.0400 & 1.0863 & -0.0496 & 1.0652 & -0.0669 & 1.0962 & -0.0962 & 1.1196 & -0.1781 \\
95000 & 1.1026 & -0.0480 & 1.0960 & -0.0566 & 1.0912 & -0.0775 & 1.1092 & -0.1012 & 1.0946 & -0.1912 \\
90000 & 1.1050 & -0.0574 & 1.1229 & -0.0683 & 1.1226 & -0.0842 & 1.1506 & -0.1184 & 1.1492 & -0.2043 \\
85000 & 1.1322 & -0.0698 & 1.1658 & -0.0794 & 1.1645 & -0.0921 & 1.1767 & -0.1251 & 1.1727 & -0.2200 \\
80000 & 1.1730 & -0.0826 & 1.1852 & -0.0877 & 1.2021 & -0.1082 & 1.2164 & -0.1443 & 1.1864 & -0.2345 \\
75000 & 1.2013 & -0.0947 & 1.2168 & -0.1039 & 1.2283 & -0.1229 & 1.2608 & -0.1553 & 1.1913 & -0.2321 \\
70000 & 1.2298 & -0.1103 & 1.2395 & -0.1177 & 1.2697 & -0.1359 & 1.2593 & -0.1605 & 1.2609 & -0.2620 \\
65000 & 1.3065 & -0.1210 & 1.3072 & -0.1315 & 1.3125 & -0.1482 & 1.3743 & -0.1880 & 1.2488 & -0.2867 \\
60000 & 1.3658 & -0.1417 & 1.3886 & -0.1524 & 1.3698 & -0.1694 & 1.3925 & -0.1991 & 1.2093 & -0.2847 \\
55000 & 1.4498 & -0.1584 & 1.4282 & -0.1685 & 1.4266 & -0.1878 & 1.4316 & -0.2206 & 1.4105 & -0.3137 \\
50000 & 1.5178 & -0.1755 & 1.5140 & -0.1952 & 1.5464 & -0.2043 & 1.5296 & -0.2523 & 1.2569 & -0.3076 \\
45000 & 1.6032 & -0.1942 & 1.5460 & -0.2028 & 1.6107 & -0.2200 & 1.5216 & -0.2526 & 1.2952 & -0.3159 \\
40000 & 1.6677 & -0.2050 & 1.5879 & -0.2159 & 1.5324 & -0.2321 & 1.4905 & -0.2579 & 1.2183 & -0.3194 \\
35000 & 1.6524 & -0.2123 & 1.6276 & -0.2232 & 1.6483 & -0.2486 & 1.4189 & -0.2579 & 1.0835 & -0.2904 \\
30000 & 1.7535 & -0.2277 & 1.6359 & -0.2315 & 1.5524 & -0.2371 & 1.4348 & -0.2617 & 1.0072 & -0.2814 \\
25000 & 1.7370 & -0.2223 & 1.5786 & -0.2180 & 1.6060 & -0.2398 & 1.5160 & -0.2570 & 0.9624 & -0.2582 \\
20000 & 1.7250 & -0.2059 & 1.8548 & -0.2279 & 1.6821 & -0.2321 & 1.1003 & -0.2154 & 0.8352 & -0.2465 \\
15000 & 1.6573 & -0.1903 & 1.7403 & -0.2065 & 1.9868 & -0.2428 & 1.0153 & -0.2100 & 0.3012 & -0.1615 \\
10000 & 1.5151 & -0.1772 & 1.3777 & -0.1797 & 0.8703 & -0.1678 & 1.2768 & -0.2351 & 0.5612 & -0.2087 \\
\hline
\end{tabular}}}
\scalebox{0.77}{
\subtable[$T_{eff}$]{
\begin{tabular}{l | r r | r r | r r | r r | r r}
\hline
\hline
S/N & \multicolumn{2}{c|}{100} & \multicolumn{2}{c|}{90} & \multicolumn{2}{c|}{80} & \multicolumn{2}{c|}{70}  & \multicolumn{2}{c}{60} \\
Resolution & a & b & a & b & a & b & a & b & a & b \\
\hline
100000 & 0.9391 & 279 & 0.9395 & 273 & 0.9394 & 271 & 0.9306 & 293 & 0.9294 & 288 \\
95000 & 0.9305 & 342 & 0.9156 & 387 & 0.9182 & 377 & 0.9098 & 396 & 0.9164 & 364 \\
90000 & 0.9137 & 435 & 0.8986 & 482 & 0.9167 & 416 & 0.9087 & 435 & 0.9028 & 447 \\
85000 & 0.8802 & 582 & 0.8873 & 558 & 0.8811 & 575 & 0.8729 & 590 & 0.8879 & 538 \\
80000 & 0.8726 & 654 & 0.8705 & 658 & 0.8646 & 672 & 0.8676 & 655 & 0.8524 & 694 \\
75000 & 0.8524 & 767 & 0.8502 & 769 & 0.8497 & 767 & 0.8444 & 778 & 0.8511 & 750 \\
70000 & 0.8339 & 880 & 0.8360 & 872 & 0.8478 & 829 & 0.8305 & 876 & 0.8418 & 830 \\
65000 & 0.8286 & 957 & 0.8290 & 952 & 0.8277 & 953 & 0.8396 & 912 & 0.8354 & 917 \\
60000 & 0.8298 & 1025 & 0.8240 & 1039 & 0.8457 & 977 & 0.8389 & 989 & 0.8576 & 932 \\
55000 & 0.8553 & 1050 & 0.8583 & 1040 & 0.8743 & 994 & 0.8712 & 997 & 0.8711 & 993 \\
50000 & 0.8864 & 1101 & 0.8870 & 1096 & 0.8892 & 1088 & 0.8896 & 1083 & 0.9061 & 1036 \\
45000 & 0.9083 & 1227 & 0.9243 & 1184 & 0.9139 & 1204 & 0.9318 & 1159 & 0.9243 & 1169 \\
40000 & 0.9733 & 1313 & 0.9820 & 1294 & 0.9832 & 1284 & 0.9873 & 1274 & 1.0037 & 1227 \\
35000 & 1.0585 & 1451 & 1.0587 & 1445 & 1.0673 & 1424 & 1.0664 & 1414 & 1.0656 & 1408 \\
30000 & 1.2080 & 1587 & 1.2091 & 1573 & 1.2198 & 1554 & 1.2030 & 1565 & 1.2386 & 1498 \\
25000 & 1.4745 & 1717 & 1.4831 & 1699 & 1.4847 & 1683 & 1.5107 & 1641 & 1.5034 & 1624 \\
20000 & 1.9977 & 1842 & 2.0116 & 1817 & 1.9598 & 1834 & 1.9840 & 1793 & 1.9362 & 1792 \\
15000 & 3.1029 & 1946 & 3.1103 & 1914 & 2.9692 & 1936 & 3.0119 & 1867 & 3.0566 & 1775 \\
10000 & 6.8985 & 1681 & 6.2175 & 1775 & 6.2539 & 1685 & 6.2833 & 1548 & 5.6847 & 1562 \\
\hline
S/N & \multicolumn{2}{c|}{50} & \multicolumn{2}{c|}{40} & \multicolumn{2}{c|}{30} & \multicolumn{2}{c|}{20} & \multicolumn{2}{c}{10} \\
Resolution & a & b & a & b & a & b & a & b & a & b  \\
\hline
100000 & 0.9249 & 291 & 0.9223 & 281 & 0.9363 & 198 & 0.9001 & 245 & 0.8592 & 179 \\
95000 & 0.9148 & 358 & 0.9194 & 327 & 0.9000 & 352 & 0.8961 & 290 & 0.8793 & 130 \\
90000 & 0.9083 & 419 & 0.8891 & 459 & 0.8993 & 390 & 0.8898 & 353 & 0.8804 & 161 \\
85000 & 0.8733 & 572 & 0.8741 & 547 & 0.8894 & 462 & 0.8595 & 492 & 0.8308 & 361 \\
80000 & 0.8665 & 640 & 0.8566 & 647 & 0.8626 & 597 & 0.8649 & 512 & 0.8964 & 178 \\
75000 & 0.8492 & 742 & 0.8506 & 714 & 0.8534 & 674 & 0.8623 & 571 & 0.8284 & 465 \\
70000 & 0.8494 & 796 & 0.8407 & 806 & 0.8479 & 750 & 0.8816 & 572 & 0.8621 & 402 \\
65000 & 0.8438 & 878 & 0.8257 & 915 & 0.8466 & 816 & 0.8567 & 713 & 0.8433 & 519 \\
60000 & 0.8477 & 948 & 0.8705 & 863 & 0.8817 & 797 & 0.8801 & 725 & 0.9280 & 332 \\
55000 & 0.8968 & 912 & 0.8730 & 954 & 0.8778 & 913 & 0.8882 & 817 & 0.8757 & 608 \\
50000 & 0.9027 & 1032 & 0.9235 & 962 & 0.9400 & 889 & 0.9670 & 730 & 0.9068 & 629 \\
45000 & 0.9462 & 1109 & 0.9496 & 1078 & 0.9855 & 956 & 1.0141 & 791 & 1.0176 & 467 \\
40000 & 1.0046 & 1211 & 1.0191 & 1157 & 1.0381 & 1074 & 1.0690 & 901 & 0.9749 & 794 \\
35000 & 1.1054 & 1317 & 1.1249 & 1252 & 1.1198 & 1204 & 1.1365 & 1059 & 1.0122 & 939 \\
30000 & 1.2608 & 1441 & 1.2615 & 1401 & 1.2422 & 1353 & 1.2352 & 1216 & 1.2106 & 780 \\
25000 & 1.5634 & 1525 & 1.5462 & 1491 & 1.4461 & 1513 & 1.4680 & 1287 & 1.1921 & 1163 \\
20000 & 2.0226 & 1665 & 2.0712 & 1542 & 1.9740 & 1475 & 1.8071 & 1364 & 1.3502 & 1271 \\
15000 & 3.0308 & 1698 & 2.8844 & 1623 & 2.7979 & 1454 & 2.3031 & 1400 & 1.4130 & 1486 \\
10000 & 4.8757 & 1643 & 4.4861 & 1557 & 3.1564 & 1817 & 1.9481 & 2046 & 0.3888 & 2893 \\
\hline
\end{tabular}
}}
\end{center}
\end{table*}

From here we calculated the dispersion of the difference between the corrected parameter values and the ones obtained from our original determinations, and added this dispersion with the one from our method in quadrature. Table \ref{table:snres} shows the results. The horizontal header of both tables correspond to the S/N of the spectra, between 100 and 10,  while the vertical header depicts their resolution, from 100.000 to 10.000. The row with the resolution number is the value of the dispersion of our technique using the corresponding resolution/signal-to-noise combination. This table should be used as a guideline for the uncertainties of the parameters when using spectra other than HARPS.

\begin{table*}[]
\caption{Dispersion of the residuals of the parameters as a function of the resolution and S/N. }
\label{table:snres}
\begin{center}
\subtable[{[Fe/H]}]{
\begin{tabular}{l r r r r r r r r r r}

\hline
\hline
S/N & 100 & 90 & 80 & 70 & 60 & 50 & 40 & 30 & 20 & 10 \\
Resolution & & &    &    &    &    &    &    &    &    \\
\hline
100000 & 0.081 & 0.081 & 0.081 & 0.081 & 0.082 & 0.082 & 0.083 & 0.083 & 0.089 & 0.105 \\
95000 & 0.082 & 0.082 & 0.082 & 0.082 & 0.082 & 0.083 & 0.083 & 0.086 & 0.091 & 0.107 \\
90000 & 0.082 & 0.082 & 0.082 & 0.082 & 0.083 & 0.083 & 0.084 & 0.087 & 0.089 & 0.103 \\
85000 & 0.083 & 0.083 & 0.083 & 0.083 & 0.085 & 0.085 & 0.085 & 0.085 & 0.094 & 0.116 \\
80000 & 0.084 & 0.084 & 0.084 & 0.085 & 0.085 & 0.085 & 0.087 & 0.089 & 0.095 & 0.113 \\
75000 & 0.085 & 0.085 & 0.086 & 0.086 & 0.086 & 0.087 & 0.089 & 0.092 & 0.097 & 0.123 \\
70000 & 0.088 & 0.087 & 0.088 & 0.088 & 0.088 & 0.091 & 0.091 & 0.094 & 0.100 & 0.122 \\
65000 & 0.090 & 0.090 & 0.091 & 0.092 & 0.094 & 0.095 & 0.096 & 0.100 & 0.107 & 0.125 \\
60000 & 0.092 & 0.096 & 0.093 & 0.095 & 0.097 & 0.099 & 0.098 & 0.102 & 0.109 & 0.144 \\
55000 & 0.098 & 0.099 & 0.101 & 0.102 & 0.102 & 0.101 & 0.109 & 0.113 & 0.126 & 0.144 \\
50000 & 0.108 & 0.109 & 0.109 & 0.113 & 0.115 & 0.118 & 0.120 & 0.124 & 0.132 & 0.161 \\
45000 & 0.125 & 0.126 & 0.129 & 0.128 & 0.135 & 0.131 & 0.138 & 0.138 & 0.148 & 0.174 \\
40000 & 0.146 & 0.147 & 0.148 & 0.151 & 0.149 & 0.152 & 0.156 & 0.161 & 0.174 & 0.185 \\
35000 & 0.168 & 0.169 & 0.170 & 0.172 & 0.174 & 0.175 & 0.174 & 0.176 & 0.187 & 0.203 \\
30000 & 0.189 & 0.189 & 0.190 & 0.191 & 0.192 & 0.189 & 0.194 & 0.199 & 0.202 & 0.212 \\
25000 & 0.205 & 0.207 & 0.205 & 0.208 & 0.207 & 0.207 & 0.211 & 0.212 & 0.209 & 0.219 \\
20000 & 0.220 & 0.217 & 0.218 & 0.220 & 0.219 & 0.221 & 0.219 & 0.219 & 0.228 & 0.225 \\
15000 & 0.227 & 0.228 & 0.226 & 0.224 & 0.226 & 0.231 & 0.229 & 0.227 & 0.231 & 0.235 \\
10000 & 0.231 & 0.233 & 0.233 & 0.233 & 0.236 & 0.234 & 0.234 & 0.235 & 0.231 & 0.233 \\
\hline
\end{tabular}}

\subtable[$T_{eff}$]{
\begin{tabular}{l r r r r r r r r r r }
\hline
\hline
S/N & 100 & 90 & 80 & 70 & 60 & 50 & 40 & 30 & 20 & 10 \\
Resolution & & &    &    &    &    &    &    &    &    \\
\hline
100000 & 92 & 92 & 92 & 92 & 93 & 93 & 94 & 96 & 103 & 118 \\
95000 & 92 & 92 & 92 & 93 & 93 & 93 & 95 & 97 & 103 & 116 \\
90000 & 93 & 93 & 94 & 93 & 94 & 95 & 95 & 98 & 99 & 116 \\
85000 & 93 & 94 & 93 & 94 & 94 & 94 & 95 & 95 & 102 & 115 \\
80000 & 94 & 94 & 94 & 95 & 94 & 95 & 96 & 99 & 102 & 117 \\
75000 & 93 & 94 & 94 & 95 & 94 & 96 & 96 & 97 & 101 & 120 \\
70000 & 94 & 95 & 95 & 95 & 95 & 96 & 97 & 99 & 103 & 116 \\
65000 & 95 & 95 & 94 & 95 & 95 & 96 & 97 & 97 & 102 & 117 \\
60000 & 95 & 95 & 96 & 96 & 95 & 97 & 97 & 100 & 102 & 115 \\
55000 & 95 & 95 & 96 & 95 & 95 & 97 & 98 & 101 & 107 & 127 \\
50000 & 95 & 95 & 96 & 96 & 96 & 98 & 99 & 103 & 104 & 123 \\
45000 & 97 & 96 & 97 & 97 & 97 & 96 & 99 & 101 & 108 & 129 \\
40000 & 98 & 98 & 97 & 99 & 99 & 99 & 100 & 106 & 113 & 128 \\
35000 & 99 & 100 & 100 & 99 & 99 & 100 & 103 & 106 & 114 & 148 \\
30000 & 101 & 101 & 103 & 102 & 103 & 101 & 106 & 111 & 123 & 147 \\
25000 & 104 & 104 & 108 & 105 & 106 & 108 & 113 & 115 & 129 & 161 \\
20000 & 112 & 110 & 111 & 118 & 115 & 116 & 116 & 125 & 143 & 179 \\
15000 & 116 & 120 & 126 & 125 & 130 & 129 & 146 & 154 & 179 & 216 \\
10000 & 135 & 133 & 155 & 146 & 167 & 177 & 185 & 214 & 235 & 259 \\
\hline
\end{tabular}}
\end{center}
\end{table*}

\begin{figure*}[]
\centering
\subfigure[]{\includegraphics[scale=0.29]{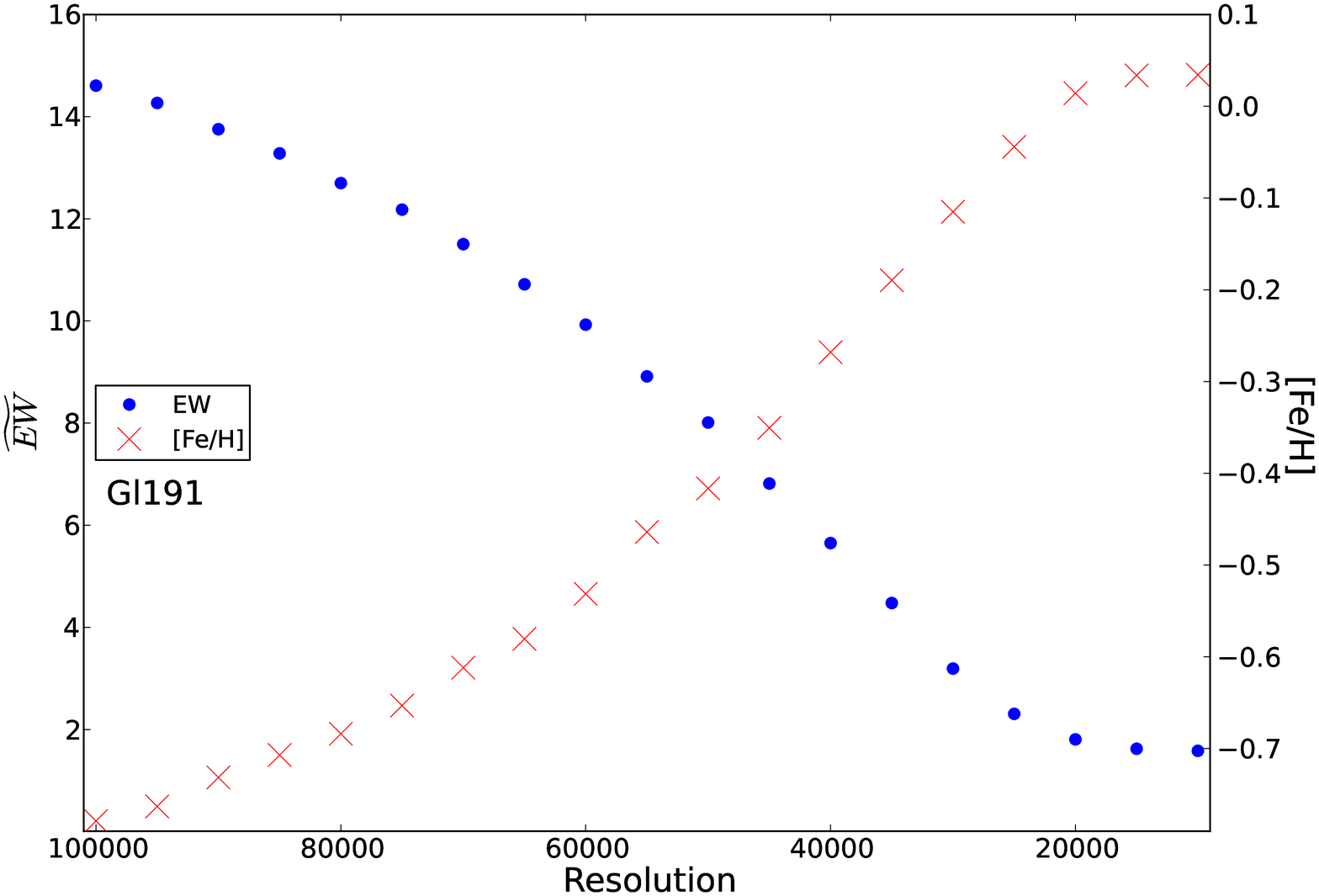}}
\subfigure[]{\includegraphics[scale=0.29]{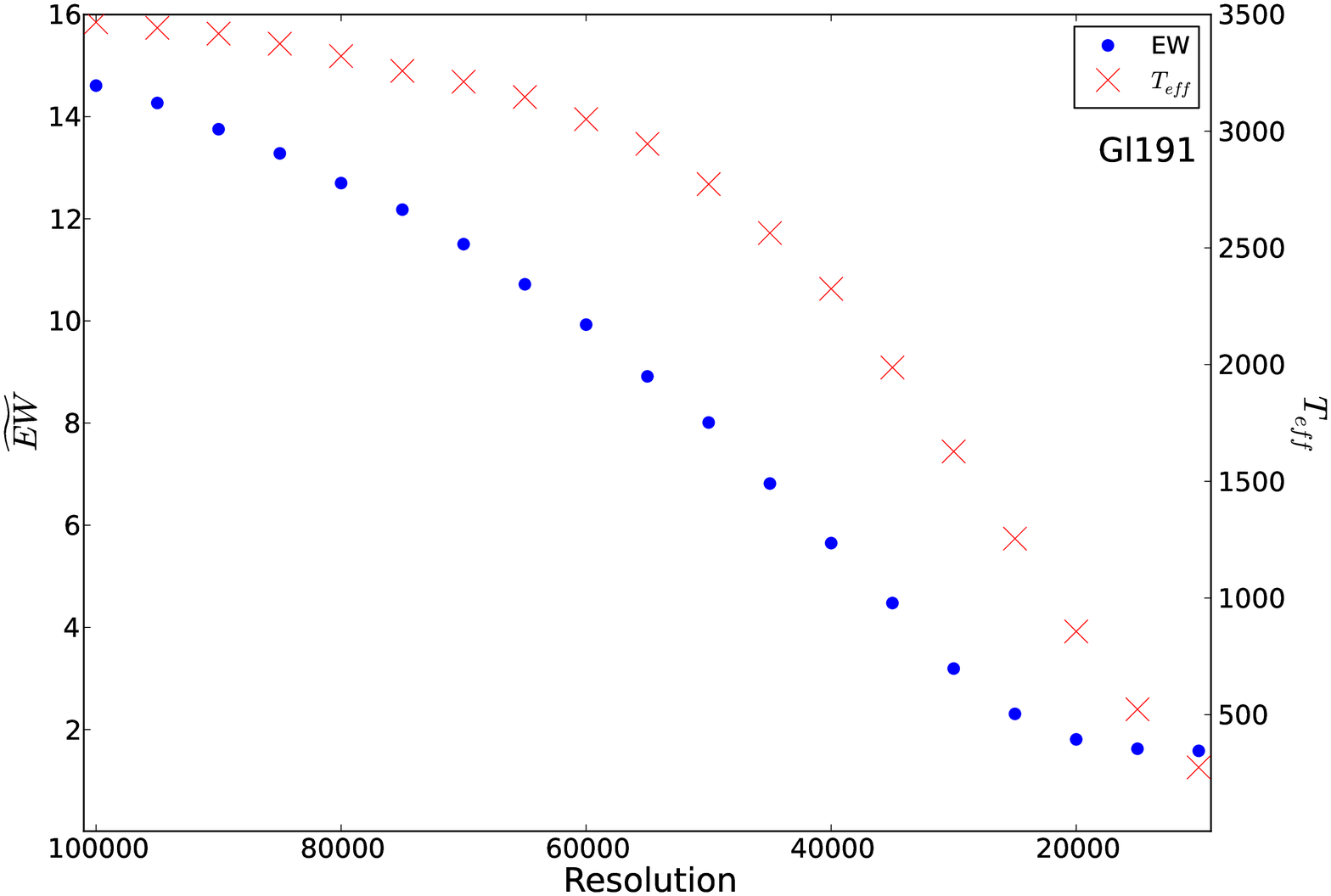}}
\subfigure[]{\includegraphics[scale=0.29]{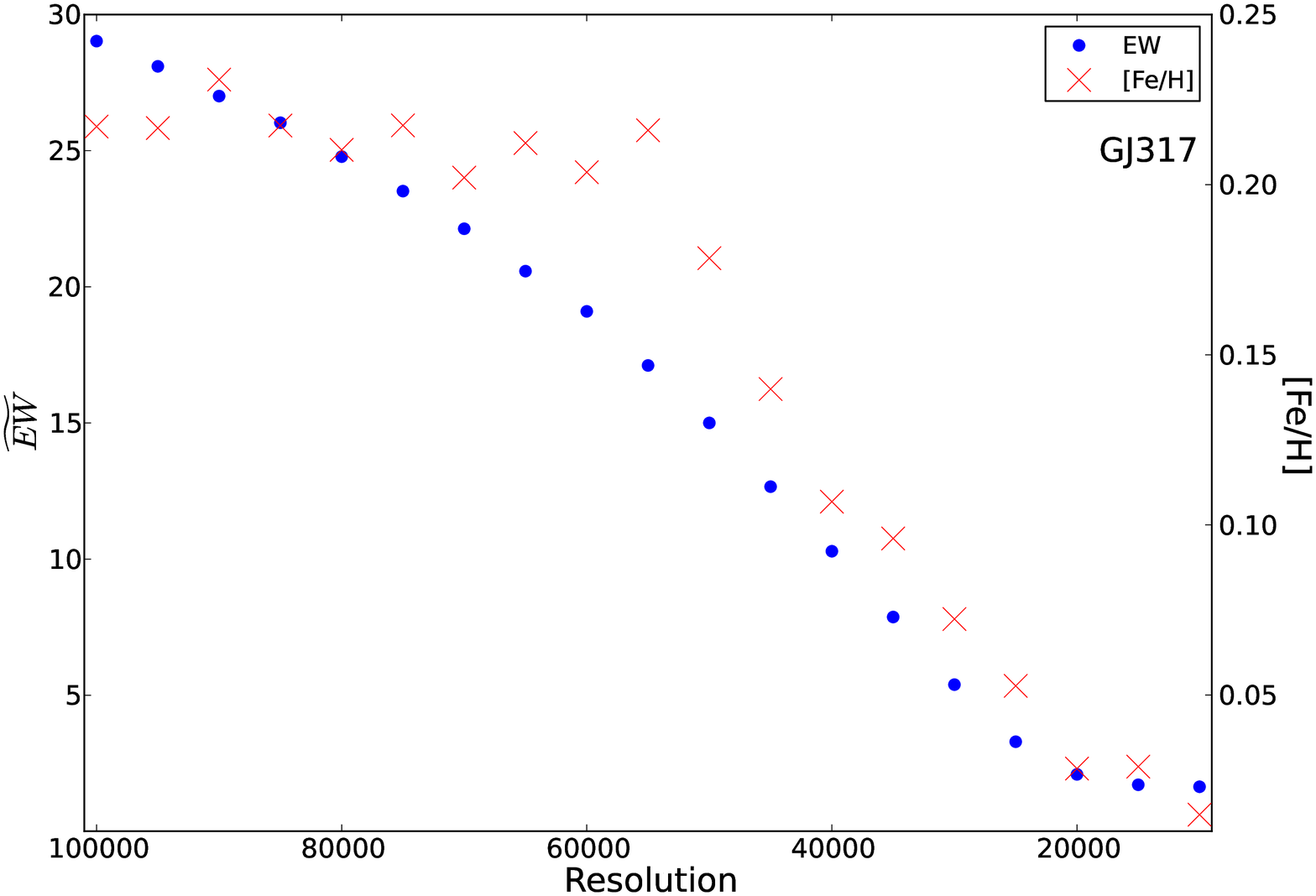}}
\subfigure[]{\includegraphics[scale=0.29]{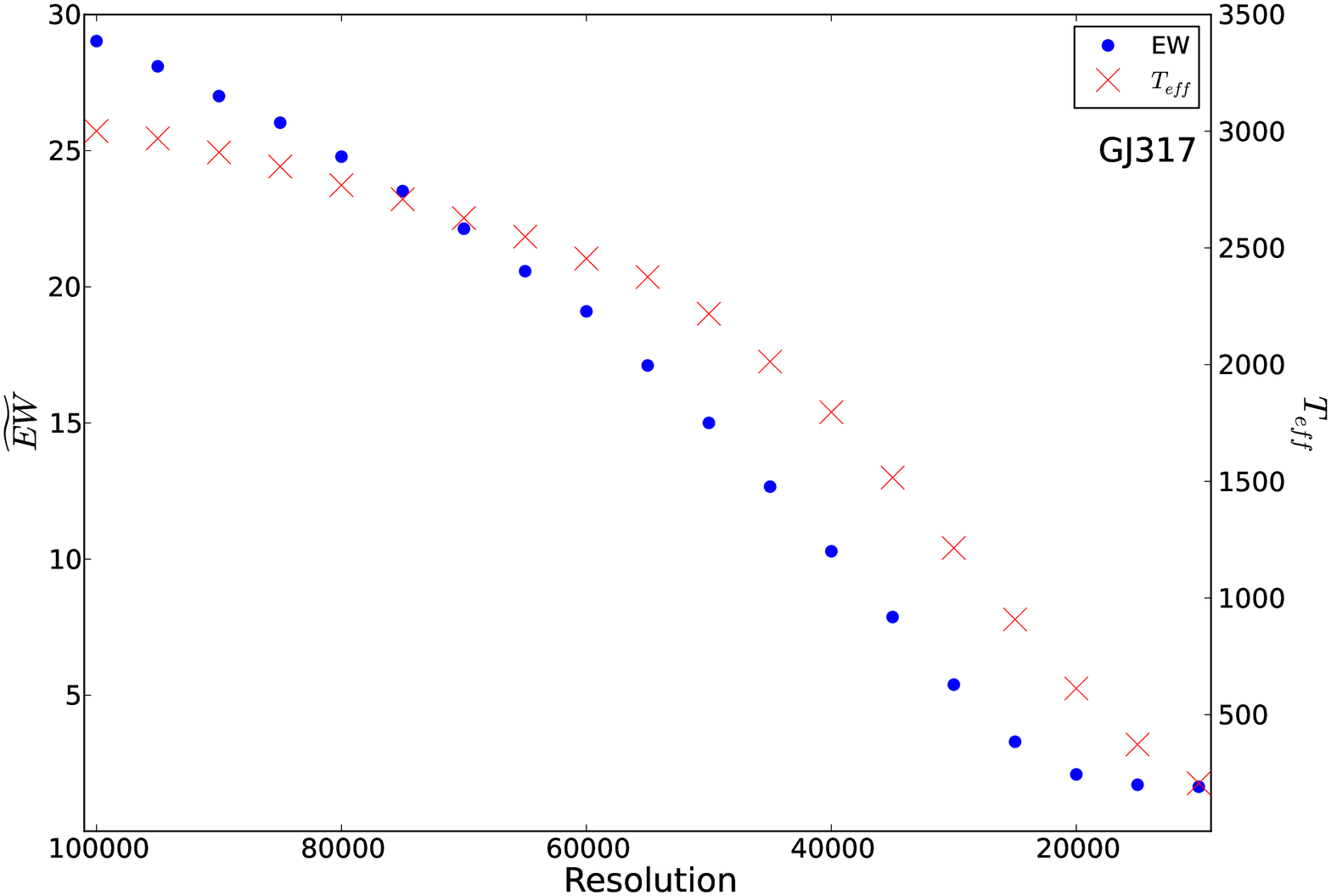}}
\caption{Measured median of the pseudo EWs, [Fe/H] and $T_{eff}$ as a function of resolution for the stars Gl191 (a and b) and GJ317 (c and d). The blue dots depict the median of the pseudo EWs while the red crosses show the metallicity or the effective temperature.}
\label{fig:ewres}
\end{figure*}

From Table \ref{table:snres} and Figs. \ref{fig:resnr-feh} and \ref{fig:resnr-teff} we observe that, as the resolution degrades, the dispersion and offset of the residuals increase. In the case of [Fe/H], the dispersion value holds well for a resolution higher and equal to 40.000 respectively. From 35.000 and lower resolutions we observe that the uncertainties of the residuals are similar or greater than the original dispersion (0.17 dex), meaning that the method is not useful any more, providing we have precise parallaxes and visual magnitudes. 
Regarding $T_{eff}$, we consider that the method is valid for the same resolution and S/N intervals as in [Fe/H]. We also limit the use of the technique for spectra with S/N greater than 25, as we cannot properly correct the parameters, as we have previously seen in this Section. 
From here we investigated the nature of these correlations by plotting the median of all EWs with the parameters, as a function of resolution (with S/N=100), for a metal-poor star Gl191 ([Fe/H] = -0.85 dex, $T_{eff}$ = 3510 K) and a metal-rich star, GJ 317 ([Fe/H] = 0.22 dex, $T_{eff}$ = 3106 K), as it was previously done for S/N, and shown in Fig. \ref{fig:ewsnr}. Fig. \ref{fig:ewres} pictures the results. The blue dots depict the median of the pseudo EWs while the red crosses show the metallicity or the effective temperature.

We observe that the EW and the parameters follow similar trends as the resolution degrades, as expected, except in Fig. \ref{fig:ewres} a) where metallicity has an opposite trend to the EW. We do not know the cause of the different trend.

The increasing dispersions and offsets with the resolution observed in Fig. \ref{fig:resnr-feh} and \ref{fig:resnr-teff} should originate from the nature of our `peak-to-peak' technique because it does not consider the continuum. As the resolution gets worse, more and more flux from the `peak-to-peak' region is lost to the line wings. However, as the resolution decreases there is an increase of line blending that makes the measurement of the correct flux of the each line/feature increasingly difficult. Moreover, we also observe from Fig. \ref{fig:pcorr} that we have similar numbers of correlation and anti-correlations of the lines with [Fe/H], and the overall effect of the weighted lines may change as the resolution degrades. The mix of any of these effects may be the reason behind we get different trends for [Fe/H] from Gl191 (Fig. \ref{fig:ewres} a) and GJ436 (Fig. \ref{fig:ewres} c).


In order to validate our estimation of the uncertainties we used a sample of spectra of stars in common taken from the SOPHIE spectrograph \citep{Bouchy-2006} archive\footnote{\url{http://atlas.obs-hp.fr/sophie/}}. We downloaded individual observations taken with the `HR' ($R\sim75.000$) and `HE' ($R\sim40.000$) modes, with the reference fiber exposed to the sky rather than to the Thorium-Argon calibration lamp, to avoid potential contamination. The `HR' sample is comprised of 12 stars in common while the `HE' sample contains only 5 stars.

First, we summed the individual spectra of each star, after correcting the radial velocity of each spectrum, and calculated their pseudo EWs. The S/N of the summed spectra for each star in common range from 34 to more than 600. Then, we calculated the metallicity and effective temperature and applied the appropriate corrections shown in Table \ref{table:linlin}, following the resolution and S/N of each spectra. The final values of the parameters are displayed in Table \ref{table:sophie}. Column 2 lists the resolution, column 3 the S/N, and column 4 to 7 the parameters of our method and the ones obtained with the SOPHIE spectra, respectively. 

\begin{table*}[]
\centering
\caption[]{[Fe/H] and $T_{eff}$ calculated from a sample in common with SOPHIE observations.}
\label{table:sophie}
\begin{center}
\begin{tabular}{l r r r r r r}
\hline
\hline
Star & Resolution & S/N & [Fe/H] (This work) & [Fe/H] (SOPHIE) & $T_{eff}$ (this work) & $T_{eff}$ (SOPHIE) \\
     &            &     &  [dex] & [dex]       &  [K] & [K]       \\
\hline
Gl699 & 75.000 & 102 & -0.51 & -0.39 & 3338 & 3292 \\
Gl686 & 75.000 & 557 & -0.35 & -0.28 & 3493 & 3529 \\
Gl87 & 75.000 & 147 & -0.32 & -0.21 & 3555 & 3472 \\
GJ1125 & 75.000 & 64 & -0.09 & 0.07 & 3112 & 3049 \\
Gl273 & 75.000 & 62 & -0.01 & -0.02 & 3090 & 3057 \\
GJ2066 & 75.000 & 86 & -0.17 & -0.07 & 3421 & 3249 \\
Gl393 & 75.000 & 91 & -0.20 & -0.10 & 3431 & 3326 \\
Gl514 & 75.000 & 636 & -0.16 & -0.11 & 3526 & 3501 \\
Gl846 & 75.000 & 175 & 0.01 & 0.04 & 3588 & 3551 \\
Gl176 & 75.000 & 79 & -0.01 & 0.06 & 3355 & 3196 \\
GJ179 & 75.000 & 34 & 0.12 & 0.06 & 3086 & 2878 \\
Gl436 & 75.000 & 150 & -0.03 & 0.04 & 3354 & 3304 \\
Gl447 & 40.000 & 56 & -0.17 & -0.15 & 3036 & 3202 \\
Gl213 & 40.000 & 85 & -0.11 & -0.15 & 3082 & 3246 \\
GJ1125 & 40.000 & 62 & -0.09 & -0.03 & 3112 & 3340 \\
Gl846 & 40.000 & 79 & 0.01 & 0.27 & 3588 & 3762 \\
GJ179 & 40.000 & 141 & 0.12 & 0.13 & 3086 & 3188 \\
\hline
\end{tabular}
\end{center}
\end{table*}

We obtain a $rms$ for the [Fe/H] of 0.088 and 0.123 dex for the `HR' and the `HE' samples respectively. Both dispersions are very close the expected dispersions for R=75.000 and R=40.000, as depicted in Table \ref{table:snres}. For $T_{eff}$, we obtain a dispersion of 104 and 172 K for the `HR' and `HE' samples respectively. Both values are above the expected values, but in the case of the 'HR' sample, the difference is only a few Kelvin. Regarding the 'HE' sample, this difference is larger ($\sim$ 75 K), but the dispersion value is close to the one considered for the photometric calibration ($\sim$ 150 K). Moreover, we note that the sample size is very small (N=5). These results give us confidence and validate our uncertainty estimation method.

\section{Comparison with the literature}
\label{sec:comp}
A comparison with other studies in the literature was performed. This comparison allow us to evaluate the accuracy of our method and the possible systematics that it may suffer. Table \ref{table:compfull} \addtocounter{table}{1} shows our spectroscopic results compared to the ones found in the literature, for the stars in common. Column 1 depicts the star designations and column 2 informs if the star belongs or not to the selected sample that we used to calibrate our method. Columns 3 to 13 and columns 14 to 21 describe our [Fe/H] and $T_{eff}$ against the metallicity found in the literature, respectively. We note here that we only used the spectroscopic derived values for comparison. All active stars were excluded from this exercise.

We show in Table \ref{table:comp} the results of the comparison for the sample used in our technique and the full sample. The results for [Fe/H] are separated by photometric and spectroscopic techniques. The first column depicts the name of the method along with its reference. Column two and three describe the dispersion and offset. The last column reports the number of stars in common with our sample. We do not display the calibration of \citet{Mann-2013a} in Table \ref{table:comp} because we only have two stars in common with them. However, we include the 4 measurements from the V- and K-band calibrations in the row `All [Fe/H] values'. We note here that we only compared stars in common for which we could calculate precise values of [Fe/H] and $T_{eff}$ with our methodology.

\begin{table}[]
\centering
\caption[]{Dispersion and offset of [Fe/H] and $T_{eff}$ from the residuals of the sample of the method and full sample against other studies. The last column shows the number of stars in common. }
\label{table:comp}
\subtable[Selected sample]{
\begin{tabular}{l r r r}
\hline
\hline
Photometric [Fe/H] calibrations  &  $rms$  & offset  &  N \\
\citet{Bonfils-2005}  &  0.11 & -0.06 & 65 \\
\citet{Schlaufman-2010}  &  0.11 & 0.02 & 65 \\
\citet{Johnson-2012}  &  0.19 & 0.05 & 65 \\
\hline
Spectroscopic [Fe/H] determinations  &  $rms$  &  offset  &  N \\
All [Fe/H] values &  0.11 & 0.05 & 55 \\
\citet{Woolf-2005} & 0.09 & -0.02 & 5 \\
\citet{Rojas-Ayala-2012}  &  0.12 & 0.03 & 19 \\
\citet{Onehag-2012}  &  0.08 & 0.05 & 8 \\
\citet{Terrien-2012}  &  0.07 & 0.06 & 7 \\
\citet{Mann-2013b}  &  0.16 & 0.11 & 7 \\
\citet{Newton-2013}  &  0.11 & 0.07 & 5 \\
\hline
$T_{eff}$ determinations & $rms$ & offset & N \\
\citet{Woolf-2005} & 122 & 116 & 5 \\
\citet{Rojas-Ayala-2012}  &  299 & 246 & 19 \\
\citet{Onehag-2012}  &  160 & 64 & 8 \\
\citet{Mann-2013b}  &  167 & 133 & 7 \\
\citet{Boyajian-2012}  &  157 & 129 & 49 \\
\citet{Rajpurohit-2013a}  &  132 & 100 & 8 \\
\hline
\end{tabular}}
\subtable[Full Sample]{
\begin{tabular}{l r r r}
\hline
\hline
Spectroscopic [Fe/H] determinations  &  $rms$  &  offset  &  N \\
\citet{Rojas-Ayala-2012}  &  0.13 & 0.06 & 25 \\
\citet{Newton-2013}  &  0.15 & 0.1 & 13 \\
\hline
$T_{eff}$ values & $rms$ & offset & N \\
\citet{Rojas-Ayala-2012}  &  304 & 222 & 25 \\
\citet{Boyajian-2012}  &  149 & 111 & 55 \\
\citet{Rajpurohit-2013a}  &  181 & 133 & 12 \\
\hline
\end{tabular}}
\end{table}

\begin{figure*}[]
\begin{center}
\subfigure[]{\includegraphics[scale=0.5]{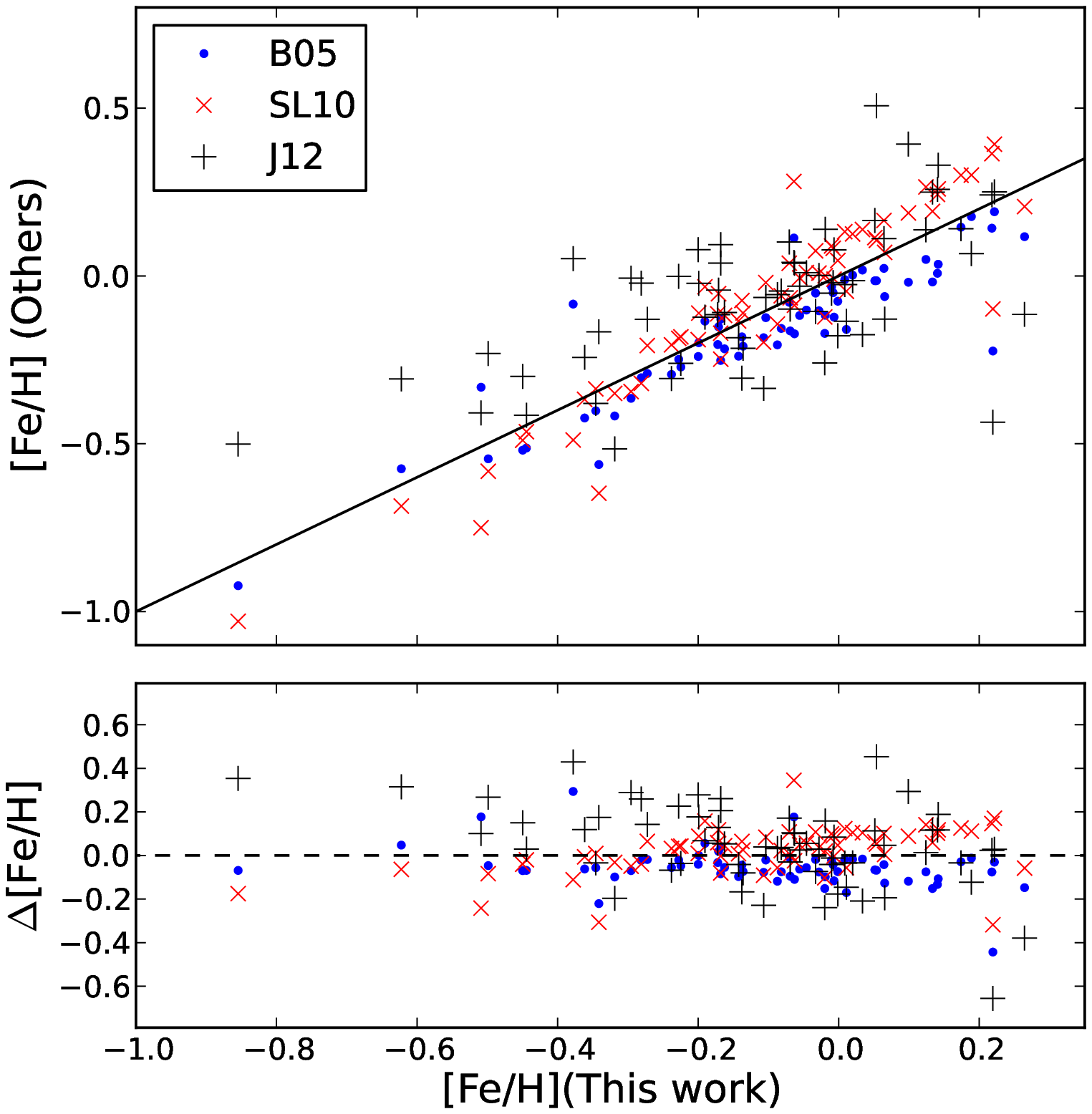}}
\subfigure[]{\includegraphics[scale=0.5]{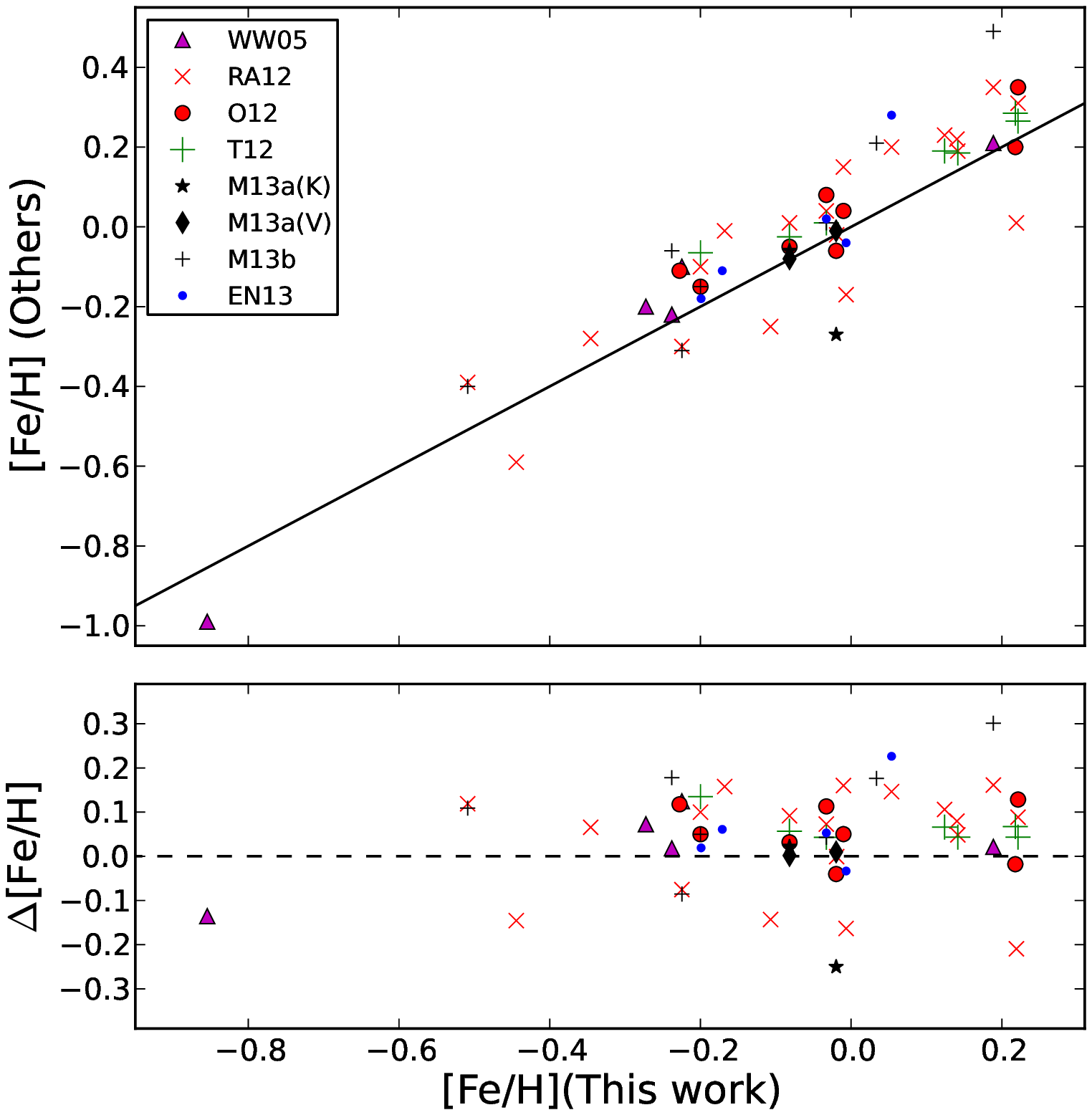}}
\end{center}
\caption{\textit{Upper panel:} [Fe/H]-[Fe/H] plots comparing the values of this work, from the selected sample, against others in the literature. The solid black line of both (a) and (b) depict the identity line; \textit{Lower panel:} Comparison plot residuals. The dashed black line marks the zero point of our technique. The plot (a) shows the results of our work versus three photometric calibrations taken from the literature, while plot (b) depicts the comparison between our results against other spectroscopic measurements.}
\label{fig:compfeh}
\end{figure*}

The photometric [Fe/H] was calculated with the relations of \citet{Bonfils-2005}, \citet{Schlaufman-2010} and \citet{Johnson-2012}, while the spectroscopic [Fe/H] was taken directly from the works of \citet{Woolf-2005}, \citet{Rojas-Ayala-2012}, \citet{Onehag-2012}, \citet{Terrien-2012}, \citet{Newton-2013}, and \citet{Mann-2013b}, except in the case of \citet{Mann-2013a}, where the values of the visible (their Eq. 8) and K-band (their Eq. 16) calibrations were provided directly by the author. We note that we used the average of the H- and K-band spectroscopic relations of \citet{Terrien-2012}. Also, we restricted the calculation of the photometric calibrations to the stars from the sample selection, for which we have precise photometries and parallaxes, to insure the best possible results in the comparison exercise.

Figure \ref{fig:compfeh} portrays two [Fe/H]-[Fe/H] plots, with data taken from the selected sample. The left plot (a) shows the comparison of our sample with the works based on photometric scales. The blue dots, red crosses and black plus signs indicate the results of \citet{Bonfils-2005}, \citet{Schlaufman-2010}, and \citet{Johnson-2012} respectively. The right plot (b) depicts the comparison of our selected sample with other spectroscopic methods. The purple triangles, red crosses, red circles, green plus signs, black stars, black diamonds, black plus signs, and blue dots correspond to the measurements of \citet[][]{Woolf-2005}, \citet{Rojas-Ayala-2012}, \citet{Onehag-2012}, \citet{Terrien-2012}, \citet{Mann-2013a}, \citet{Mann-2013b}, and \citet{Newton-2013} respectively. The (V) and (K) in \citet{Mann-2013a} correspond to measurements performed with a V- and K-band calibration respectively. The solid black line in the upper panel of both plots defines an identity line. The lower panels show the residuals. The dashed black line marks the zero-point of the calibration.

For metallicity we observe a general agreement between our results and the ones from the literature. We note here that the calibration of \citet{Schlaufman-2010} is very similar to our reference calibration, from \citet{Neves-2012}, and this is the reason why we obtain a value of dispersion smaller than the one of the original calibration (0.11 vs 0.17 dex). The dispersion of the oldest photometric calibration \citep{Bonfils-2005} is surprisingly low (0.11 dex), considering that the original dispersion for this calibration is 0.20 dex. However, the \citet{Bonfils-2005} is also similar to \citet{Schlaufman-2010} and \citet{Neves-2012} which may explain part of the low dispersion. Regarding the \citet{Johnson-2012} calibration, we obtain a $rms$ of 0.19 dex, higher that their reported value of 0.15 dex. The dispersion of the spectroscopic determinations are within the expected values ($\sim$0.11 dex), considering the uncertainties of each method, except in the case of \citet{Mann-2013b}, where we obtain a dispersion of 0.16 dex, and in two stars in common with \citet{Woolf-2005}, where the [Fe/H] difference for Gl191 and Gl526 is higher that the uncertainties reported here and in their work (0.14 and -0.12 dex, respectively). The offset of each calibration is smaller than the dispersion value of our calibration, aside from \citet{Mann-2013b} (0.11 dex). We should note, however, that we only have seven stars in common with \citet{Mann-2013b} and one of these stars, Gl205, has a [Fe/H] difference of 0.30 dex. When we consider the full sample we detect a slight increase of dispersion for \citet{Rojas-Ayala-2012}, and a considerable increase in both dispersion and offset for \citet{Newton-2013}. This increase in the \citet{Newton-2013} dispersion is due to the addition of several stars in common with high [Fe/H], where the two calibrations show most disagreement.  

\begin{figure}[]
\begin{center}
\includegraphics[scale=0.5]{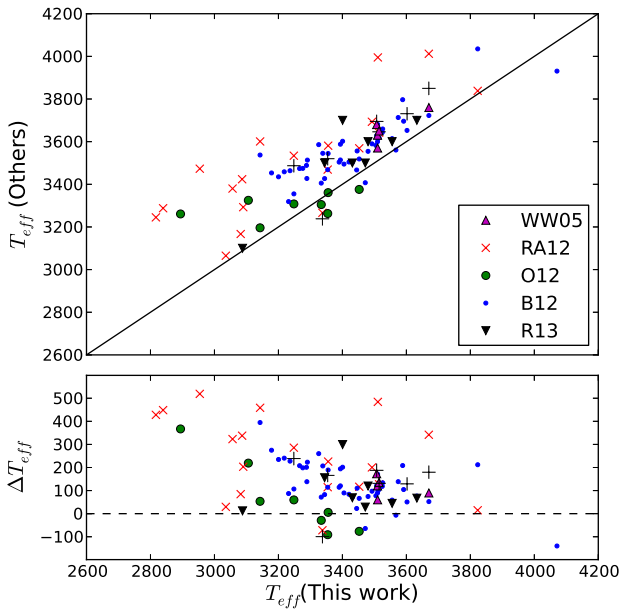}
\end{center}
\caption{\textit{Upper panel:} $T_{eff}$-$T_{eff}$ plot comparing the values of this work, taken from the selected sample, against others in the literature. The solid black line depicts the identity line; \textit{Lower panel:} Comparison plot residuals. The dashed black line marks the zero point of our calibration.}
\label{fig:compteff}
\end{figure}

Regarding the effective temperature, the photometric temperature scale of \citet{Boyajian-2012} was calculated using the average value of the three colour-metallicity $T_{eff}$ relations ($V-J$, $V-H$, and $V-K_{S}$) from their Table 9, and imposing a cutoff of $V-K < 4.5$ for the three scales, according to their limits.  
The $T_{eff}$ values of \citet{Woolf-2005}, \citet{Rojas-Ayala-2012}, \citet{Onehag-2012}, \citet{Mann-2013a}, and \citet{Rajpurohit-2013a} were taken directly from their works. Figure \ref{fig:compteff} describes the comparison between our $T_{eff}$ results and those of the other authors. The purple (pointing up) triangles, red crosses, green circles, blue dots, and black (pointing down) triangles correspond to the measurements of \citet{Woolf-2005}, \citet{Rojas-Ayala-2012}, \citet{Onehag-2012}, \citet{Mann-2013a}, and \citet{Rajpurohit-2013a} respectively. The solid black line in the upper panel defines an identity line. The lower panels show the residuals. 
The photometric [Fe/H] measurements as well as the $T_{eff}$ determinations using the calibration of \citet{Boyajian-2012} were calculated with the data from Table \ref{table:params}.

From Figure \ref{fig:compteff} and Table \ref{table:comp} we observe a good agreement with the results from \citet{Woolf-2005} where we obtain a low dispersion and offset. However, we only have 5 stars in common with them, and they occupy a very narrow region of the T${_eff}$ range, around 3500 K. Our results also match well the BT-SETTL based work of \citet{Rajpurohit-2013a}. However, when we look at the results of the full sample we observe a significant increase in the $rms$ of \citet{Rajpurohit-2013a}, and we also witness a considerable dispersion with \citet{Rojas-Ayala-2012} in both samples. The \citet{Onehag-2012} and \citet{Boyajian-2012} determinations tend to converge with ours as the $T_{eff}$ increases. 
We also note a systematic underestimation of our values of temperature in general that increases below 3200K. The \citet{Onehag-2012} determinations have the smallest offset, but this result is expected since they use the same reference $T_{eff}$ calibration as we do. When we consider the full sample we observe that the $rms$ and offset do not change considerably, except in the case of \citet{Rajpurohit-2013a}.
Finally, we calculated linear fits  for the different $T_{eff}$ methods, where $T_{eff}$ (This work) = $aT_{eff}$ (Others) + $b$. The only exception concerns \citet{Woolf-2005}, because the 5 stars we have in common only cover a very narrow range in the effective temperature region. The coefficients $a$, $b$ and respective uncertainties are reported in Table \ref{table:fit_teff}.

\begin{table}[]
\centering
\caption[]{Linear fit coefficients for each $T_{eff}$ method}
\label{table:fit_teff}
\begin{tabular}{l r r}
\hline
\hline
$T_{eff}$ method & $a$ & $b$ \\
\citet{Rojas-Ayala-2012}  &  0.840$\pm$0.158 & 315$\pm$555 \\
\citet{Onehag-2012} &  1.663$\pm$1.060 & -2250$\pm$3499 \\
\citet{Mann-2013b}  &  0.643$\pm$0.184 & 1149$\pm$663 \\
\citet{Boyajian-2012} &  1.079$\pm$0.098 & -409$\pm$349 \\
\citet{Rajpurohit-2013a}  &  0.745$\pm$0.171 & 798$\pm$603 \\
\hline
\end{tabular}
\end{table}



\section{Discussion}
\label{sec:discussion}

In this paper we present a new high-precision technique to calculate metallicities and effective temperatures for M dwarfs. Within the activity and S/N limits of our method, we achieve a $rms$ of 0.08 dex for metallicity and 91 K for effective temperature. Alternatively we obtain a RMSE$_{V}$ value of 0.12 dex for [Fe/H] and 293 K for $T_{eff}$. A bootstrap resampling was also conducted, showing a variation of the $rms$ of [Fe/H] and $T_{eff}$ of the order of $\pm$ 0.01 dex and $\pm$ 13 K respectively. Our technique is available for download at \url{http://www.astro.up.pt/resources/mcal}. The procedure to use our method is detailed in this webpage as well as in the Annex. A test of the behaviour of the technique as a function of the resolution and S/N was also performed. We estimate that our method behaves properly down to R = 40.000 and S/N = 25, after correcting the observed trends. We also validated our results against a sample of stars in common with SOPHIE high resolution spectra.


To have a measure of the accuracy of our method, we tested it against several studies from the literature. Most studies agree well with our [Fe/H] determinations, and the offset is almost always below the precision of the method. For $T_{eff}$ however, the same agreement could not be met. Despite reaching a good agreement with the results of \citet{Woolf-2005}, and \citet{Rajpurohit-2013a}, that use synthetic spectra from the latest BT-SETTL models, the dispersion as well as the systematics between our determinations and the other works is considerable and beyond the calibration errors. Further studies are needed to investigate the nature of these systematics. 




\begin{acknowledgements}
We would like to thank Barbara Rojas-Ayala for useful discussions. We would also like to thank Andrew Mann and Jo\~ao Gomes da Silva for kindly providing several metallicity values and the $H\alpha$ determinations, respectively. We acknowledge the support by the European Research Council/European Community under the FP7 through Starting Grant agreement number 239953. The financial support from the "Programme National de Plan\'etologie'' (PNP) of CNRS/INSU, France, is also gratefully acknowledged. VN acknowledges the support from Funda\c{c}\~ao para a Ci\^encia e a Tecnologia (FCT) of the fellowship SFRH/BD/60688/2009. NCS also acknowledges the support in the form of a Investigador FCT contract funded by Funda\c{c}\~ao para a Ci\^encia e a Tecnologia (FCT) /MCTES (Portugal) and POPH/FSE (EC). XB, TF and XD would like to acknowledge the support of the French Agence Nationale de la Recherche (ANR), under program ANR-12-BS05-0012 Exo-Atmos. This research has made use of the SIMBAD database, operated at CDS, Strasbourg, France. This publication makes use of data products from the Two Micron All Sky Survey, which is a joint project of the University of Massachusetts and the Infrared Processing and Analysis Center/California Institute of Technology, funded by the National Aeronautics and Space Administration and the National Science Foundation.

\end{acknowledgements}

\onecolumn
\longtab{4}{
\begin{longtable}{l r r r r r r r r r r}
\caption{ Full HARPS M dwarf GTO sample. Sorted by right ascension.} \\
\hline
\hline
star & $\alpha$ (2000) & $\delta$ (2000) & S/N & $\log{L_{H_{\alpha}}/L_{bol}}$ & $H\alpha(GdS)$ & [Fe/H]$_{N13}$ & [Fe/H] & $\sigma$[Fe/H] & $T_{eff}$ & $\sigma T_{eff}$ \\
     & [hour] & [deg] &    &        &        & [dex] & [dex] & [dex] & [K] & [K] \\
\hline
\endfirsthead
\caption{continued.}\\
\hline
\hline
star & $\alpha$ (2000) & $\delta$ (2000) & S/N & $\log{L_{H_{\alpha}}/L_{bol}}$ & $H\alpha(GdS)$ & [Fe/H]$_{N13}$ & [Fe/H] & $\sigma$[Fe/H] & $T_{eff}$ & $\sigma T_{eff}$ \\
     & [hour] & [deg] &    &        &        & [dex] & [dex] & [dex] & [K] & [K] \\
\hline
\endhead
\hline
\multicolumn{11}{l}{$^1$ Active star or star with S/N $\leq$ 25. The [Fe/H] and $T_{eff}$ were calculated using the original photometric calibrations.}\\
\multicolumn{11}{l}{$^2$ LHS 1513 is a metal poor star outside the calibration region of \citet{Neves-2012}.} \\
\multicolumn{11}{l}{$^3$ Gl 803 is a young, metal rich star outside the calibration region of \citet{Neves-2012}.}
\endfoot
Gl1 & 00:05:25 & -37:21:23 & 870 & -4.88 & 0.01 & -0.45 & -0.45 & 0.09 & 3567 &  110 \\
GJ1002$^1$ & 00:06:44 & -07:32:23 & 21 &    - & 0.05 & -0.19 & -0.27 & 0.20 & 2718 &  150 \\
Gl12 & 00:15:49 & +13:33:17 & 46 &    - & 0.03 & -0.34 & -0.29 & 0.09 & 3239 &  110 \\
LHS1134 & 00:43:26 & -41:17:36 & 35 &    - & 0.03 & -0.10 & -0.13 & 0.09 & 2950 &  110 \\
Gl54.1 & 01:12:31 & -17:00:00 & 107 & -4.07 & 0.18 & -0.40 & -0.38 & 0.09 & 3088 &  110 \\
L707-74 & 01:23:18 & -12:56:23 & 33 &    - & 0.03 & -0.35 & -0.38 & 0.09 & 3353 &  110 \\
Gl87 & 02:12:21 & +03:34:30 & 281 & -4.88 & 0.02 & -0.31 & -0.32 & 0.09 & 3555 &  110 \\
Gl105B & 02:36:16 & +06:52:12 & 143 &    - & 0.02 & -0.02 & -0.02 & 0.09 & 2894 &  110 \\
CD-44-836A & 02:45:11 & -43:44:30 & 68 &    - & 0.13 & -0.08 & -0.07 & 0.09 & 3032 &  110 \\
HIP12961 & 02:46:43 & -23:05:12 & 384 &    - & 0.07 & 0.22 & 0.22 & 0.09 & 3823 &  110 \\
LHS1481 & 02:58:10 & -12:53:06 & 65 &    - & 0.02 & -0.72 & -0.76 & 0.09 & 3510 &  110 \\
LP771-95A & 03:01:51 & -16:35:36 & 109 &    - & 0.02 & -0.34 & -0.34 & 0.09 & 3236 &  110 \\
LHS1513$^{2}$ & 03:11:36 & -38:47:17 & 26 & -4.93 & 0.03 & -0.11 & - & - &  3197 &  110 \\
GJ1057 & 03:13:23 & +04:46:30 & 28 &    - & 0.05 & 0.10 & -0.10 & 0.09 & 2916 &  110 \\
Gl145 & 03:32:56 & -44:42:06 & 93 & -4.89 & 0.04 & -0.28 & -0.28 & 0.09 & 3270 &  110 \\
GJ1061 & 03:36:00 & -44:30:48 & 28 &    - & 0.07 & -0.08 & -0.09 & 0.09 & 2882 &  110 \\
GJ1065 & 03:50:44 & -06:05:42 & 38 & -4.93 & 0.03 & -0.22 & -0.23 & 0.09 & 3062 &  110 \\
GJ163 & 04:09:16 & -53:22:25 & 312 &    - & 0.02 & 0.05 & 0.07 & 0.09 & 3276 &  110 \\
GJ1068$^1$ & 04:10:28 & -53:36:06 & 21 &    - & 0.04 & -0.30 & -0.43 & 0.20 & 2887 &  150 \\
Gl166C$^1$ & 04:15:22 & -07:39:23 & 82 & -3.95 & 0.29 & 0.08 & -0.12 & 0.20 & 3018 &  150 \\
Gl176 & 04:42:56 & +18:57:29 & 576 &    - & 0.06 & -0.01 & -0.01 & 0.09 & 3355 &  110 \\
GJ179 & 04:52:06 & +06:28:36 & 999 &    - & 0.05 & 0.11 & 0.12 & 0.09 & 3086 &  110 \\
LHS1723 & 05:01:57 & -06:56:47 & 62 & -4.51 & 0.16 & -0.25 & -0.24 & 0.09 & 3167 &  110 \\
LHS1731 & 05:03:20 & -17:22:23 & 79 & -4.9 & 0.03 & -0.26 & -0.19 & 0.09 & 3273 &  110 \\
Gl191 & 05:11:40 & -45:01:06 & 788 &    - & 0.00 & -0.88 & -0.85 & 0.09 & 3510 &  110 \\
Gl203 & 05:28:00 & +09:38:36 & 68 & -4.93 & 0.03 & -0.25 & -0.22 & 0.09 & 3138 &  110 \\
Gl205 & 05:31:27 & -03:40:42 & 1430 & -4.88 & 0.09 & 0.22 & 0.19 & 0.09 & 3670 &  110 \\
Gl213 & 05:42:09 & +12:29:23 & 105 & -4.94 & 0.01 & -0.11 & -0.11 & 0.09 & 3082 &  110 \\
Gl229 & 06:10:34 & -21:51:53 & 727 & -4.87 & 0.07 & -0.01 & -0.03 & 0.09 & 3633 &  110 \\
HIP31293 & 06:33:43 & -75:37:47 & 147 &    - & 0.05 & -0.04 & -0.05 & 0.09 & 3288 &  110 \\
HIP31292 & 06:33:47 & -75:37:30 & 125 &    - & 0.04 & -0.10 & -0.06 & 0.09 & 3184 &  110 \\
G108-21 & 06:42:11 & +03:34:53 & 56 &    - & 0.03 & -0.01 & -0.02 & 0.09 & 3186 &  110 \\
Gl250B & 06:52:18 & -05:11:24 & 167 &    - & 0.05 & -0.10 & -0.08 & 0.09 & 3453 &  110 \\
Gl273 & 07:27:24 & +05:13:30 & 644 &    - & 0.02 & -0.01 & -0.01 & 0.09 & 3090 &  110 \\
LHS1935 & 07:38:41 & -21:13:30 & 77 & -4.9 & 0.03 & -0.24 & -0.22 & 0.09 & 3181 &  110 \\
Gl285$^1$ & 07:44:40 & +03:33:06 & 85 & -3.48 & 0.90 & 0.18 & 0.27 & 0.20 & 2946 &  150 \\
Gl299 & 08:11:57 & +08:46:23 & 69 & -4.75 & 0.04 & -0.50 & -0.53 & 0.09 & 3373 &  110 \\
Gl300 & 08:12:41 & -21:33:12 & 189 & -4.94 & 0.05 & 0.14 & 0.13 & 0.09 & 2841 &  110 \\
GJ2066 & 08:16:08 & +01:18:11 & 196 &    - & 0.03 & -0.18 & -0.17 & 0.09 & 3421 &  110 \\
GJ317 & 08:40:59 & -23:27:23 & 131 &    - & 0.07 & 0.21 & 0.22 & 0.09 & 3106 &  110 \\
GJ1123 & 09:17:05 & -77:49:17 & 28 &    - & 0.04 & 0.20 & 0.15 & 0.09 & 2779 &  110 \\
Gl341 & 09:21:38 & -60:16:53 & 471 &    - & 0.05 & -0.13 & -0.14 & 0.09 & 3575 &  110 \\
GJ1125 & 09:30:44 & +00:19:18 & 110 &    - & 0.02 & -0.11 & -0.09 & 0.09 & 3112 &  110 \\
Gl357 & 09:36:02 & -21:39:42 & 258 &    - & 0.01 & -0.34 & -0.30 & 0.09 & 3344 &  110 \\
Gl358 & 09:39:47 & -41:04:00 & 295 & -4.44 & 0.14 & -0.01 & -0.01 & 0.09 & 3178 &  110 \\
Gl367 & 09:44:30 & -45:46:36 & 352 & -4.88 & 0.04 & -0.07 & -0.07 & 0.09 & 3394 &  110 \\
GJ1129 & 09:44:48 & -18:12:48 & 44 & -4.94 & 0.03 & 0.07 & 0.05 & 0.09 & 3017 &  110 \\
Gl382 & 10:12:17 & -03:44:47 & 532 & -4.89 & 0.08 & 0.04 & 0.02 & 0.09 & 3401 &  110 \\
Gl388$^1$ & 10:19:36 & +19:52:12 & 589 & -3.8 & 0.40 & 0.07 & 0.12 & 0.20 & 3171 &  150 \\
Gl393 & 10:28:55 & +00:50:23 & 464 & -4.83 & 0.04 & -0.22 & -0.20 & 0.09 & 3431 &  110 \\
LHS288$^1$ & 10:44:32 & -61:11:35 & 19 &    - & 0.03 & -0.60 & -0.55 & 0.20 & 2760 &  150 \\
Gl402 & 10:50:52 & +06:48:30 & 70 &    - & 0.04 & 0.06 & 0.03 & 0.09 & 2943 &  110 \\
Gl406$^1$ & 10:56:29 & +07:00:54 & 36 &    - & 0.85 & 0.18 & 0.19 & 0.20 & 2523 &  150 \\
GJ3634 & 10:58:35 & -31:08:38 & 204 &    - & 0.03 & -0.10 & -0.07 & 0.09 & 3405 &  110 \\
Gl413.1 & 11:09:31 & -24:36:00 & 279 & -4.88 & 0.03 & -0.12 & -0.10 & 0.09 & 3394 &  110 \\
Gl433 & 11:35:27 & -32:32:23 & 599 &    - & 0.03 & -0.17 & -0.17 & 0.09 & 3480 &  110 \\
Gl436 & 11:42:11 & +26:42:23 & 573 &    - & 0.02 & -0.06 & -0.03 & 0.09 & 3354 &  110 \\
Gl438 & 11:43:20 & -51:50:23 & 254 & -4.82 & 0.02 & -0.39 & -0.36 & 0.09 & 3505 &  110 \\
Gl447 & 11:47:44 & +00:48:16 & 133 &    - & 0.03 & -0.18 & -0.17 & 0.09 & 3036 &  110 \\
Gl465 & 12:24:53 & -18:14:30 & 191 & -4.88 & 0.01 & -0.66 & -0.62 & 0.09 & 3472 &  110 \\
Gl479 & 12:37:53 & -52:00:06 & 468 & -4.88 & 0.09 & 0.02 & 0.01 & 0.09 & 3218 &  110 \\
LHS337 & 12:38:50 & -38:22:53 & 60 & -4.97 & 0.03 & -0.25 & -0.27 & 0.09 & 3007 &  110 \\
Gl480.1 & 12:40:46 & -43:34:00 & 77 &    - & 0.03 & -0.48 & -0.48 & 0.09 & 3211 &  110 \\
Gl486 & 12:47:57 & +09:45:12 & 70 &    - & 0.02 & 0.06 & 0.03 & 0.09 & 2941 &  110 \\
Gl514 & 13:30:00 & +10:22:36 & 433 & -4.91 & 0.04 & -0.16 & -0.16 & 0.09 & 3526 &  110 \\
Gl526 & 13:45:44 & +14:53:30 & 729 & -5.02 & 0.03 & -0.20 & -0.22 & 0.09 & 3515 &  110 \\
Gl536 & 14:01:03 & -02:39:18 & 390 & -4.88 & 0.05 & -0.12 & -0.14 & 0.09 & 3525 &  110 \\
Gl551$^1$ & 14:29:43 & -62:40:47 & 291 &    - & 0.42 & -0.00 & 0.16 & 0.20 & 2654 &  150 \\
Gl555 & 14:34:17 & -12:31:06 & 107 &    - & 0.03 & 0.17 & 0.14 & 0.09 & 2839 &  110 \\
Gl569A & 14:54:29 & +16:06:04 & 182 & -4.3 & 0.23 & -0.08 & -0.06 & 0.09 & 3289 &  110 \\
Gl581 & 15:19:26 & -07:43:17 & 773 &    - & 0.01 & -0.21 & -0.20 & 0.09 & 3248 &  110 \\
Gl588 & 15:32:13 & -41:16:36 & 602 & -4.89 & 0.03 & 0.07 & 0.06 & 0.09 & 3291 &  110 \\
Gl618A & 16:20:04 & -37:31:41 & 255 & -4.88 & 0.03 & -0.08 & -0.06 & 0.09 & 3200 &  110 \\
Gl628 & 16:30:18 & -12:39:47 & 451 &    - & 0.03 & -0.02 & -0.02 & 0.09 & 3057 &  110 \\
Gl643 & 16:55:25 & -08:19:23 & 83 &    - & 0.02 & -0.28 & -0.26 & 0.09 & 3102 &  110 \\
GJ1214 & 17:15:19 & +04:57:50 & 999 &    - & 0.04 & 0.06 & 0.05 & 0.09 & 2817 &  110 \\
Gl667C & 17:18:58 & -34:59:42 & 1025 & -4.88 & 0.02 & -0.53 & -0.50 & 0.09 & 3445 &  110 \\
Gl674 & 17:28:40 & -46:53:42 & 686 & -4.89 & 0.06 & -0.25 & -0.23 & 0.09 & 3334 &  110 \\
GJ676A & 17:30:11 & -51:38:13 & 432 &    - & 0.06 & 0.25 & 0.26 & 0.09 & 4071 &  110 \\
Gl678.1A & 17:30:22 & +05:32:53 & 387 & -4.82 & 0.05 & -0.11 & -0.14 & 0.09 & 3591 &  110 \\
Gl680 & 17:35:13 & -48:40:53 & 363 & -4.88 & 0.03 & -0.22 & -0.19 & 0.09 & 3390 &  110 \\
Gl682 & 17:37:03 & -44:19:11 & 177 & -4.93 & 0.04 & 0.11 & 0.10 & 0.09 & 2912 &  110 \\
Gl686 & 17:37:53 & +18:35:30 & 328 &    - & 0.02 & -0.37 & -0.35 & 0.09 & 3493 &  110 \\
Gl693 & 17:46:35 & -57:19:11 & 133 &    - & 0.03 & -0.30 & -0.28 & 0.09 & 3232 &  110 \\
Gl699 & 17:57:49 & +04:41:36 & 496 &    - & 0.01 & -0.52 & -0.51 & 0.09 & 3338 &  110 \\
Gl701 & 18:05:07 & -03:01:53 & 520 &    - & 0.04 & -0.27 & -0.27 & 0.09 & 3510 &  110 \\
GJ1224$^1$ & 18:07:33 & -15:57:47 & 35 & -3.97 & 0.27 & -0.10 & -0.25 & 0.20 & 2860 &  150 \\
G141-29$^1$ & 18:42:44 & +13:54:17 & 24 &    - & 0.26 & 0.09 & -0.08 & 0.20 & 3011 &  150 \\
Gl729$^1$ & 18:49:49 & -23:50:12 & 135 & -3.77 & 0.31 & -0.10 & -0.40 & 0.20 & 3058 &  150 \\
GJ1232$^1$ & 19:09:51 & +17:40:07 & 24 &    - & 0.05 & 0.14 & 0.03 & 0.20 & 2893 &  150 \\
Gl752A & 19:16:55 & +05:10:05 & 535 &    - & 0.04 & 0.06 & 0.05 & 0.09 & 3339 &  110 \\
Gl754 & 19:20:48 & -45:33:30 & 80 &    - & 0.03 & -0.17 & -0.14 & 0.09 & 3005 &  110 \\
GJ1236 & 19:22:03 & +07:02:36 & 53 &    - & 0.03 & -0.42 & -0.47 & 0.09 & 3280 &  110 \\
GJ1256 & 20:40:34 & +15:29:57 & 32 &    - & 0.06 & 0.10 & 0.06 & 0.09 & 2853 &  110 \\
Gl803$^{1,3}$ & 20:45:10 & -31:20:30 & 202 &    - & 0.40 & 0.32 & - & - & 3430 &  150 \\
LHS3583 & 20:46:37 & -81:43:12 & 69 &    - & 0.03 & -0.18 & -0.22 & 0.09 & 3236 &  110 \\
LP816-60 & 20:52:33 & -16:58:30 & 97 &    - & 0.04 & -0.06 & -0.07 & 0.09 & 2960 &  110 \\
Gl832 & 21:33:34 & -49:00:36 & 925 &    - & 0.03 & -0.19 & -0.17 & 0.09 & 3446 &  110 \\
Gl846 & 22:02:10 & +01:24:00 & 643 & -4.82 & 0.07 & 0.06 & 0.01 & 0.09 & 3588 &  110 \\
LHS3746 & 22:02:29 & -37:04:54 & 71 &    - & 0.03 & -0.15 & -0.13 & 0.09 & 3013 &  110 \\
Gl849 & 22:09:40 & -04:38:30 & 410 &    - & 0.03 & 0.24 & 0.22 & 0.09 & 3143 &  110 \\
GJ1265 & 22:13:42 & -17:41:12 & 28 &    - & 0.04 & -0.09 & -0.20 & 0.09 & 2941 &  110 \\
LHS3799$^1$ & 22:23:07 & -17:36:23 & 25 &    - & 0.36 & 0.18 & 0.10 & 0.20 & 2820 &  150 \\
Gl876 & 22:53:17 & -14:15:48 & 554 &    - & 0.03 & 0.15 & 0.14 & 0.09 & 2954 &  110 \\
Gl877 & 22:55:46 & -75:27:36 & 369 &    - & 0.03 & -0.01 & -0.00 & 0.09 & 3266 &  110 \\
Gl880 & 22:56:35 & +16:33:12 & 351 &    - & 0.06 & 0.07 & 0.03 & 0.09 & 3602 &  110 \\
Gl887 & 23:05:52 & -35:51:12 & 1434 &    - & 0.04 & -0.24 & -0.24 & 0.09 & 3507 &  110 \\
LHS543 & 23:21:37 & +17:17:25 & 81 & -4.94 & 0.04 & 0.25 & 0.23 & 0.09 & 2872 &  110 \\
Gl908 & 23:49:13 & +02:24:06 & 845 &    - & 0.02 & -0.44 & -0.44 & 0.09 & 3511 &  110 \\
LTT9759 & 23:53:50 & -75:37:53 & 168 &    - & 0.07 & 0.21 & 0.17 & 0.09 & 3326 &  110 \\
\label{table:full}
\end{longtable}
}

\onecolumn
\longtab{8}{
\begin{landscape}
\fontsize{8.0}{8.0}{ \selectfont
\begin{longtable}{l c r r r r r r r r r r r | r r r r r r r r }
\caption{Comparison of our parameters with other results from the literature for the stars in common. Sorted by right ascension.} \\ 

\hline
\hline

\vspace{1mm}

star & Sample & \multicolumn{11}{c}{[Fe/H]} & \multicolumn{8}{c}{$T_{eff}$} \\

     &      & This work  &  B05   &  SL10 &  J12  & WW05 &  RA12 &  O12  &   T12  & EN13  &  M13a & M13b  & This work & WW05 & RA12    & O12     &  B12    & M13a    & M13b    & R13 \\

\hline
\endfirsthead
\caption{continued.}\\
\hline

\vspace{1mm}

star & Sample & \multicolumn{11}{c}{[Fe/H]} & \multicolumn{8}{c}{$T_{eff}$} \\
     &        & This work  &  B05   &  SL10 &  J12  & WW05 &  RA12 &  O12  &   T12  & EN13  &  M13a & M13b  & This work & WW05 & RA12    & O12     &  B12    & M13a    & M13b    & R13 \\
\hline
\endhead
\hline
\multicolumn{21}{l}{References: B05 - \citet{Bonfils-2005}; SL10 - \citet{Schlaufman-2010}; J12 - \citet{Johnson-2012}; WW05 - \citet{Woolf-2005}; RA12 - \citet{Rojas-Ayala-2012}; O12 - \citet{Onehag-2012};} \\ 
\multicolumn{21}{l}{T12 - \citet{Terrien-2012}; EN13 - \citet{Newton-2013}; M13a - \citet{Mann-2013a}; M13b - \citet{Mann-2013b}; B12 - \citet{Boyajian-2012}; R13 - \citet{Rajpurohit-2013a}.} \\
\endfoot
Gl1 & Y & -0.45 & -0.49 & -0.46 & -0.3 & - & - & - & - & - & - & - & 3567 & - & - & - & 3541 & - & - & - \\
Gl12 & N & -0.29 & - & - & - & - & - & - & - & -0.17 & - & - & 3239 & - & - & - & - & - & - & - \\
Gl54.1 & Y & -0.38 & -0.11 & -0.55 & 0.05 & - & - & - & - & - & - & - & 3088 & - & - & - & - & - & - & 3100 \\
L707-74 & N & -0.38 & - & - & - & - & - & - & - & - & - & - & 3353 & - & - & - & 3324 & - & - & - \\
Gl87 & Y & -0.32 & -0.37 & -0.3 & -0.52 & - & - & - & - & - & - & - & 3555 & - & - & - & 3584 & - & - & 3600 \\
Gl105B & Y & -0.02 & -0.15 & -0.09 & -0.26 & - & - & -0.06 & - & - & -0.32 & - & 2894 & - & - & 3261 & - & 3505 & - & - \\
HIP12961 & Y & 0.22 & -0.22 & -0.1 & -0.44 & - & 0.01 & - & - & - & - & - & 3823 & - & 3838 & - & 4035 & - & - & - \\
LHS1481 & N & -0.76 & - & - & - & - & - & - & - & - & - & - & 3510 & - & - & - & 3306 & - & - & - \\
LP771-95A & Y & -0.34 & -0.08 & 0.07 & -0.17 & - & - & - & - & - & - & - & 3236 & - & - & - & - & - & - & - \\
GJ1057 & N & -0.1 & - & - & - & - & - & - & - & 0.24 & - & - & 2916 & - & - & - & - & - & - & 2900 \\
Gl145 & N & -0.28 & - & - & - & - & - & - & - & - & - & - & 3270 & - & - & - & 3373 & - & - & - \\
GJ1065 & N & -0.23 & - & - & - & - & - & - & - & - & - & - & 3062 & - & - & - & - & - & - & 3200 \\
GJ163 & Y & 0.07 & -0.06 & 0.07 & -0.13 & - & - & - & - & - & - & - & 3276 & - & - & - & 3475 & - & - & - \\
Gl176 & Y & -0.01 & -0.0 & 0.13 & -0.05 & - & 0.15 & 0.04 & - & - & - & - & 3355 & - & 3581 & 3361 & 3527 & - & - & - \\
GJ179 & Y & 0.12 & 0.05 & 0.27 & 0.14 & - & 0.23 & - & 0.19 & - & - & - & 3086 & - & 3424 & - & - & - & - & - \\
LHS1723 & N & -0.24 & - & - & - & - & -0.06 & - & - & - & - & - & 3167 & - & 3054 & - & - & - & - & - \\
LHS1731 & N & -0.19 & - & - & - & - & - & - & - & - & - & - & 3273 & - & - & - & 3355 & - & - & - \\
Gl191 & Y & -0.85 & -0.97 & -1.07 & -0.5 & -0.99 & - & - & - & - & - & - & 3510 & 3570.0 & - & - & 3716 & - & - & - \\
Gl203 & N & -0.22 & - & - & - & - & - & - & - & -0.21 & - & - & 3138 & - & - & - & - & - & - & - \\
Gl205 & Y & 0.19 & 0.2 & 0.33 & 0.07 & 0.21 & 0.35 & - & - & - & - & 0.49 & 3670 & 3760.0 & 4012 & - & 3709 & - & 3850 & - \\
Gl213 & Y & -0.11 & -0.21 & -0.24 & -0.33 & - & -0.25 & - & - & - & - & - & 3082 & - & 3167 & - & - & - & - & - \\
Gl229 & Y & -0.03 & -0.05 & 0.07 & 0.0 & - & - & - & - & - & - & - & 3633 & - & - & - & 3672 & - & - & 3700 \\
HIP31293 & Y & -0.05 & -0.01 & 0.13 & 0.01 & - & - & - & - & - & - & - & 3288 & - & - & - & 3441 & - & - & - \\
HIP31292 & Y & -0.06 & -0.18 & -0.1 & 0.04 & - & - & - & - & - & - & - & 3184 & - & - & - & - & - & - & - \\
G108-21 & N & -0.02 & - & - & - & - & - & - & - & -0.01 & - & - & 3186 & - & - & - & 3415 & - & - & - \\
Gl250B & Y & -0.08 & -0.14 & -0.04 & -0.04 & - & 0.01 & -0.05 & -0.02 & - & -0.24 & - & 3453 & - & 3569 & 3376 & 3511 & 3459 & - & - \\
Gl273 & Y & -0.01 & -0.16 & -0.07 & 0.08 & - & -0.17 & - & - & -0.04 & - & - & 3090 & - & 3293 & - & - & - & - & - \\
LHS1935 & N & -0.22 & - & - & - & - & - & - & - & - & - & - & 3181 & - & - & - & 3372 & - & - & - \\
Gl299 & N & -0.53 & - & - & - & - & -0.46 & - & - & -0.56 & - & - & 3373 & - & 3021 & - & - & - & - & - \\
Gl300 & Y & 0.13 & -0.03 & 0.17 & 0.25 & - & - & - & - & - & - & - & 2841 & - & - & - & - & - & - & - \\
GJ2066 & Y & -0.17 & -0.15 & -0.05 & -0.04 & - & - & - & - & -0.11 & - & - & 3421 & - & - & - & 3501 & - & - & - \\
GJ1123 & N & 0.15 & - & - & - & - & - & - & - & - & - & - & 2779 & - & - & - & - & - & - & 3100 \\
Gl341 & Y & -0.14 & -0.21 & -0.11 & -0.18 & - & - & - & - & - & - & - & 3575 & - & - & - & 3694 & - & - & - \\
GJ1125 & Y & -0.09 & -0.21 & -0.15 & -0.06 & - & - & - & - & - & - & - & 3112 & - & - & - & - & - & - & - \\
Gl357 & Y & -0.3 & -0.37 & -0.35 & -0.01 & - & - & - & - & - & - & - & 3344 & - & - & - & 3429 & - & - & 3500 \\
Gl358 & Y & -0.01 & 0.01 & 0.17 & -0.02 & - & - & - & - & - & - & - & 3178 & - & - & - & 3425 & - & - & - \\
Gl367 & Y & -0.07 & -0.08 & 0.03 & -0.1 & - & - & - & - & - & - & - & 3394 & - & - & - & 3538 & - & - & - \\
Gl382 & Y & 0.02 & 0.03 & 0.16 & -0.01 & - & - & - & - & - & - & - & 3401 & - & - & - & 3584 & - & - & 3700 \\
Gl393 & Y & -0.2 & -0.15 & -0.06 & -0.02 & - & - & - & - & -0.18 & - & - & 3431 & - & - & - & 3475 & - & - & 3500 \\
Gl402 & N & 0.03 & - & - & - & - & 0.2 & - & - & - & - & - & 2943 & - & 3334 & - & - & - & - & - \\
GJ3634 & Y & -0.07 & -0.08 & 0.04 & 0.1 & - & - & - & - & - & - & - & 3405 & - & - & - & 3495 & - & - & - \\
Gl413.1 & Y & -0.1 & -0.15 & -0.06 & -0.06 & - & - & - & - & - & - & - & 3394 & - & - & - & 3532 & - & - & - \\
Gl433 & Y & -0.17 & -0.21 & -0.12 & -0.12 & - & - & - & - & - & - & - & 3480 & - & - & - & 3560 & - & - & 3600 \\
Gl436 & Y & -0.03 & -0.05 & 0.07 & -0.11 & - & 0.04 & 0.08 & 0.01 & 0.02 & - & 0.01 & 3354 & - & 3469 & 3263 & 3469 & - & 3520 & - \\
Gl438 & Y & -0.36 & -0.39 & -0.34 & -0.24 & - & - & - & - & - & - & - & 3505 & - & - & - & 3562 & - & - & - \\
Gl447 & Y & -0.17 & -0.14 & -0.26 & 0.09 & - & -0.01 & - & - & - & - & - & 3036 & - & 3065 & - & - & - & - & - \\
Gl465 & Y & -0.62 & -0.56 & -0.66 & -0.31 & - & - & - & - & - & - & - & 3472 & - & - & - & 3395 & - & - & 3500 \\
Gl479 & Y & 0.01 & 0.01 & 0.16 & -0.03 & - & - & - & - & - & - & - & 3218 & - & - & - & 3449 & - & - & - \\
Gl480.1 & N & -0.48 & - & - & - & - & - & - & - & - & - & - & 3211 & - & - & - & 3257 & - & - & - \\
Gl486 & N & 0.03 & - & - & - & - & - & - & - & 0.03 & - & - & 2941 & - & - & - & - & - & - & 3300 \\
Gl514 & Y & -0.16 & -0.16 & -0.06 & -0.11 & - & - & - & - & - & - & - & 3526 & - & - & - & 3624 & - & - & - \\
Gl526 & Y & -0.22 & -0.22 & -0.13 & -0.26 & -0.1 & -0.3 & - & - & - & - & -0.31 & 3515 & 3650.0 & 3642 & - & 3585 & - & 3646 & - \\
Gl536 & Y & -0.14 & -0.21 & -0.12 & -0.22 & - & - & - & - & - & - & - & 3525 & - & - & - & 3647 & - & - & - \\
Gl555 & Y & 0.14 & 0.0 & 0.23 & 0.26 & - & 0.22 & - & - & - & - & - & 2839 & - & 3288 & - & - & - & - & - \\
Gl569A & Y & -0.06 & -0.01 & 0.12 & 0.04 & - & - & - & - & - & - & - & 3289 & - & - & - & 3495 & - & - & - \\
Gl581 & Y & -0.2 & -0.22 & -0.17 & 0.08 & - & -0.1 & -0.15 & -0.06 & - & - & -0.15 & 3248 & - & 3534 & 3308 & - & - & 3487 & - \\
Gl588 & Y & 0.06 & 0.02 & 0.16 & 0.11 & - & - & - & - & - & - & - & 3291 & - & - & - & 3517 & - & - & - \\
Gl618A & Y & -0.06 & -0.11 & 0.01 & -0.03 & - & - & - & - & - & - & - & 3200 & - & - & - & 3431 & - & - & - \\
Gl628 & Y & -0.02 & -0.11 & 0.02 & 0.14 & - & -0.02 & - & - & - & - & - & 3057 & - & 3380 & - & - & - & - & - \\
Gl643 & N & -0.26 & - & - & - & - & -0.22 & - & - & - & - & - & 3102 & - & 3376 & - & - & - & - & - \\
GJ1214 & Y & 0.05 & -0.01 & 0.13 & 0.51 & - & 0.2 & - & - & 0.28 & - & - & 2817 & - & 3245 & - & - & - & - & - \\
Gl667C & Y & -0.5 & -0.59 & -0.64 & -0.23 & - & - & - & - & - & - & - & 3445 & - & - & - & 3500 & - & - & - \\
Gl674 & Y & -0.23 & -0.25 & -0.19 & -0.0 & - & - & -0.11 & - & - & - & - & 3334 & - & - & 3305 & 3408 & - & - & - \\
GJ676A & Y & 0.26 & 0.12 & 0.21 & -0.11 & - & - & - & - & - & - & - & 4071 & - & - & - & 3931 & - & - & - \\
Gl678.1A & Y & -0.14 & -0.2 & -0.1 & -0.3 & - & - & - & - & - & - & - & 3591 & - & - & - & 3712 & - & - & - \\
Gl680 & Y & -0.19 & -0.09 & 0.03 & -0.12 & - & - & - & - & - & - & - & 3390 & - & - & - & 3475 & - & - & - \\
Gl682 & Y & 0.1 & 0.0 & 0.23 & 0.39 & - & - & - & - & - & - & - & 2912 & - & - & - & - & - & - & - \\
Gl686 & Y & -0.35 & -0.38 & -0.32 & -0.38 & - & -0.28 & - & - & - & - & - & 3493 & - & 3693 & - & 3578 & - & - & - \\
Gl693 & Y & -0.28 & -0.29 & -0.31 & -0.02 & - & - & - & - & - & - & - & 3232 & - & - & - & - & - & - & - \\
Gl699 & Y & -0.51 & -0.29 & -0.68 & -0.41 & - & -0.39 & - & - & - & - & -0.4 & 3338 & - & 3266 & - & - & - & 3238 & - \\
Gl701 & Y & -0.27 & -0.26 & -0.18 & -0.13 & -0.2 & - & - & - & - & - & - & 3510 & 3630.0 & - & - & 3580 & - & - & - \\
Gl752A & Y & 0.05 & -0.02 & 0.1 & 0.17 & - & - & - & - & - & - & - & 3339 & - & - & - & 3551 & - & - & - \\
GJ1236 & N & -0.47 & - & - & - & - & - & - & - & -0.21 & - & - & 3280 & - & - & - & 3282 & - & - & - \\
GJ1256 & N & 0.06 & - & - & - & - & 0.2 & - & - & 0.26 & - & - & 2853 & - & 3080 & - & - & - & - & - \\
LHS3583 & N & -0.22 & - & - & - & - & - & - & - & - & - & - & 3236 & - & - & - & 3370 & - & - & - \\
LP816-60 & N & -0.07 & - & - & - & - & 0.06 & - & - & - & - & - & 2960 & - & 3405 & - & - & - & - & - \\
Gl832 & Y & -0.17 & -0.23 & -0.15 & 0.04 & - & - & - & - & - & - & - & 3446 & - & - & - & 3544 & - & - & - \\
Gl846 & Y & 0.01 & -0.12 & -0.0 & -0.13 & - & - & - & - & - & - & - & 3588 & - & - & - & 3768 & - & - & - \\
Gl849 & Y & 0.22 & 0.21 & 0.42 & 0.25 & - & 0.31 & 0.35 & 0.26 & - & - & - & 3143 & - & 3601 & 3196 & 3530 & - & - & - \\
Gl876 & Y & 0.14 & 0.04 & 0.28 & 0.33 & - & 0.19 & - & 0.18 & - & - & - & 2954 & - & 3473 & - & - & - & - & - \\
Gl877 & Y & -0.0 & -0.06 & 0.06 & -0.18 & - & - & - & - & - & - & - & 3266 & - & - & - & 3467 & - & - & - \\
Gl880 & Y & 0.03 & 0.06 & 0.19 & -0.18 & - & - & - & - & - & - & 0.21 & 3602 & - & - & - & 3626 & - & 3731 & - \\
Gl887 & Y & -0.24 & -0.33 & -0.24 & -0.31 & -0.22 & - & - & - & - & - & -0.06 & 3507 & 3680.0 & - & - & 3654 & - & 3695 & - \\
Gl908 & Y & -0.44 & -0.5 & -0.45 & -0.41 & - & -0.59 & - & - & - & - & - & 3511 & - & 3995 & - & 3602 & - & - & - \\
LTT9759 & Y & 0.17 & 0.13 & 0.28 & 0.14 & - & - & - & - & - & - & - & 3326 & - & - & - & 3593 & - & - & - \\
\label{table:compfull}
\end{longtable}}
\end{landscape}
}

\appendix

\section{Using the method}

The code of our technique is written in python 2.7 and can be downloaded at \url{http://www.astro.up.pt/resources/mcal}. The program is very simple to use. The first step is to write the filenames of your spectra into \textit{stars.txt}, replacing the two demonstration filenames, \textit{Gl105B\_S1D.fits} and \textit{Gl849\_S1D.fits}. Then, one just needs to change the startup options, described in the startup section of the file \textit{runallv1.py}. Depending on the resolution and S/N of the spectra, one should use the values of Table \ref{table:snres} as the reference of precision of [Fe/H] and $T_{eff}$.

The compressed zip file \textit{calibrationv3.zip} contains all the necessary files needed to run the program, as described in the following list:
\begin{itemize}
\item \textit{runallv1.py} - script to run all the other programs. In the startup section one can choose to use FFT to filter high frequency noise, the file type of the input spectra (FITS or text file), and the name of the file with the full path of the spectra.
\item \textit{fft\_filterv1.py} - function that performs the FFT filtering of the spectra.  
The default setting of the filter in \textit{runallv1.py} is 'off'.
\item \textit{int\_calc\_stars.py} - function to calculate the pseudo EWs of the relevant lines. It uses lines.rdb as input. An output file, ew\_out.npz, is also created. The function also estimates the H$\alpha$ index described by \citet{Gomes_da_Silva-2011} and warns if the star is too active. It takes 3-5 minutes per star to calculate the EWs.
\item \textit{mcalv1.npz} - function that calculates the [Fe/H] and $T_{eff}$ of each star using the calibration matrix file \textit{coef\_cal.npz}. The output will be displayed on the screen and can also be optionally saved to a file (check the startup section of \textit{runallv1.py} for details). 
\item \textit{stars.txt} - text file with the full path of the spectra. This file should have all the spectra files for analysis.
\item \textit{Gl105B\_S1D.fits} and \textit{Gl849\_S1D.fits} are two HARPS spectra that can be used to demonstrate how the program works. Their full file names appear in the file stars.txt. One should remove them from \textit{stars.txt} before calibrating new stars.
\end{itemize}

\bibliographystyle{aa}
\bibliography{mylib}

\end{document}